\documentclass[aps,prb,superscriptaddress,amsmath,amssymb,floats,twocolumn,longbibliography,floatfix
]{revtex4-1}

\usepackage{graphicx}
\usepackage{placeins}
\usepackage{dcolumn}
\usepackage{bm}
\usepackage{color}
\usepackage{comment}
\usepackage{svg}
\usepackage{booktabs}
\usepackage{siunitx}
\usepackage{makecell}
\usepackage{float}
\usepackage{appendix}
\usepackage{subcaption}

\begin{document}

\preprint{APS/123-QED}

\title{Refined spin Hamiltonian on the Cairo pentagonal lattice of Bi$_2$Fe$_4$O$_9$}

\author{Emma Y. Lenander}
\affiliation{Nanoscience Center, Niels Bohr Institute, University of Copenhagen, DK-2100 Copenhagen \O , Denmark}
\affiliation{Institute  of  Physics,  \'Ecole  Polytechnique  F\'ed\'erale  de  Lausanne  (EPFL),  CH-1015  Lausanne,  Switzerland}

\author{Frida B. Nielsen }
\affiliation{Nanoscience Center, Niels Bohr Institute, University of Copenhagen, DK-2100 Copenhagen \O , Denmark}

\author{Jakob Lass}
\affiliation{Paul Scherrer Institute; Center for Neutron and Muon Sciences, CH-5232 Villigen, Switzerland}

\author{Ursula B. Hansen} 
\affiliation{Institut Laue-Langevin, 71 Avenue des Martyrs, CS20156, 38042 Grenoble Cedex 9, France}

\author{Kristine M. L. Krighaar} 
\affiliation{Nanoscience Center, Niels Bohr Institute, University of Copenhagen, DK-2100 Copenhagen \O , Denmark}

\author{Asbjørn Preuss}
\affiliation{Nanoscience Center, Niels Bohr Institute, University of Copenhagen, DK-2100 Copenhagen \O , Denmark}
\affiliation{Department of Physics, Technical University of Denmark, DK-2800 Kongens Lyngby, Denmark}

\author{Tobias Weber} 
\affiliation{Institut Laue-Langevin, 71 Avenue des Martyrs, CS20156, 38042 Grenoble Cedex 9, France}

\author{Mechthild Enderle}
\affiliation{Institut Laue-Langevin, 71 Avenue des Martyrs, CS20156, 38042 Grenoble Cedex 9, France}

\author{Henrik Jacobsen}
\affiliation{Data Management and Software Centre, Asmussens Allé 305, 2800 Kongens Lyngby, Denmark}
\affiliation{Nanoscience Center, Niels Bohr Institute, University of Copenhagen, DK-2100 Copenhagen \O , Denmark}

\author{Uwe Stuhr} 
\affiliation{Paul Scherrer Institute; Center for Neutron and Muon Sciences, CH-5232 Villigen, Switzerland}

\author{Ryoichi Kajimoto} 
\author{Mitsutaka Nakamura}
\affiliation{Materials and Life Science Division, J-PARC Center, Tokai, Ibaraki 319-1195, Japan}

\author{Manfred Burianek} 
\affiliation{University Bremen, Faculty 5 Geosciences, 28359 Bremen, Germany}

\author{Andrea Kirsch}
\email{a.kirsch@ruhr-uni-bochum.de}
\affiliation{Nanoscience Center, Department of Chemistry, University Copenhagen, 2100 Copenhagen \O, Denmark}
\affiliation{Research Center Future Energy Materials and Systems of the Research Alliance Ruhr, 44801 Bochum, Germany}
\affiliation{Faculty of Chemistry and Biochemistry, Ruhr University Bochum, 44801 Bochum, Germany}

\author{Henrik M. Rønnow} 
\affiliation{Institute  of  Physics,  \'Ecole  Polytechnique  F\'ed\'erale  de  Lausanne  (EPFL),  CH-1015  Lausanne,  Switzerland}
\affiliation{Nanoscience Center, Niels Bohr Institute, University of Copenhagen, DK-2100 Copenhagen \O , Denmark}

\author{Kim Lefmann} 
\email{lefmann@nbi.ku.dk}
\affiliation{Nanoscience Center, Niels Bohr Institute, University of Copenhagen, DK-2100 Copenhagen \O , Denmark}

\author{Pascale P. Deen}  
\email{pascale.deen@ess.eu}
\affiliation{European Spallation Source ESS ERIC, SE 21100 Lund, Sweden}
\affiliation{Nanoscience Center, Niels Bohr Institute, University of Copenhagen, DK-2100 Copenhagen \O , Denmark}
\date{\today}

\begin{abstract}
The frustrated magnet Bi$_2$Fe$_4$O$_9$ has been reported to exhibit complex spin dynamics coexisting with conventional spin wave excitations.
The magnetic Fe$^{3+}$ ($S=5/2$) ions are arranged into a distorted two-dimensional Cairo pentagonal lattice with weak couplings between the layers, developing long-ranged non-collinear antiferromagnetic order below 245 K.  
In order to enable studies and modelling of the complex dynamics close to $T_N$, we have reexamined the magnetic excitations across the complete energy scale ($0 < \hbar \omega < 90$~meV) at 10~K. We discover two distinct gaps, which can be explained by introducing, respectively, easy axis and easy plane anisotropy on the two unequivalent Fe-sites. We develop a refined spin Hamiltonian that accurately accounts for the dispersion of essentially all spin-wave branches across the full spectral range, except around 40~meV, where a splitting and dispersion are observed. We propose that this mode is derived from phonon hybridization. Polarisation analysis shows that the system has magnetic anisotropic fluctuations, consistent with our model. A continuum of scattering is observed above the spin wave branches and is found to principally be explained by an instrumental resolution effect. 
The full experimental mapping of the excitation spectrum and the refined spin Hamiltonian provides a foundation for future quantitative studies of spin waves coexisting with unconventional magnetic fluctuations in this frustrated magnet found at higher temperatures.
\end{abstract}

\pacs{Valid PACS appear here}

\maketitle


\section{Introduction}
Magnetic frustration derived from competing exchange interactions or geometric symmetries often result in exotic emergent states of matter\cite{Banerjee2016, Broholm2020_QSL, Savary_2017_QSL, Balents_2010_QSL, Ramirez_1999_spin_ice, Castelnovo_2008_spin_ice, Poree_2025}. Of these exotic states, we will here concentrate on the classically frustrated materials. 
These systems, frequently, have broad correlated magnetic scattering features, where spin waves coexist with complex spin dynamics beyond simple spin wave excitations. This may signify novel emergent magnetic states beyond current theories and is exemplified by magnetic monopoles in the spin ice material Ho$_2$Ti$_2$O$_7$ \cite{fennell2009}, hidden order in the spin liquid Gd$_3$Ga$_5$O$_{12}$ \cite{Paddison2015, Ambrumenil_2015}, and as clusters of emerging order of toroidal moments in the frustrated material h-YMnO$_3$ \cite{Janas_2021, Lass_2024, Tosic_2024}.

One particularly rare geometric setup, which generates magnetic frustration, is the Cairo pentagonal lattice. This pattern is realized in Bi$_2$Fe$_4$O$_9$. 
Here, classical Fe$^{3+}$ ions ($S = 5/2$)\cite{high_spin_bfo_park} occupy two distinct sites, forming a network of corner-sharing pentagons. These pentagons induce a fairly unexplored geometric frustration in the quasi-two-dimensional system with a non-collinear antiferromagnetic (AF) order\cite{Beauvois2020, LeDuc2021}. 

We have chosen to study the dynamics of this system with inelastic neutron scattering (INS) that directly measures magnetic excitation.
Our aim is to study whether there is coexistence between semi-classical spin waves from an ordered ground state and more complex fluctuations originating from the frustration. 
Exotic dynamics have previously been reported in Bi$_2$Fe$_4$O$_9$ with a paramagnetic state consisting of uncorrelated dimers\cite{Beauvois2020} and cooperative paramagnetic state\cite{Singh_2019} above the AF transition temperature ($T_N=245$ K). 
Even in the ordered phase, T $<$  $T_N$, complex two-magnon dynamics have been proposed to coexist with more conventional magnon excitations\cite{Iliev_2010}. A more recent study further reports weak spin-phonon coupling\cite{Roy_2025}.

We extend and improve the parameters of the magnetic Hamiltonian presented previously by Duc Le et al.\cite{LeDuc2021} and Beauvois et al.\cite{Beauvois2020} by mapping the full spin wave spectrum and using linear spin wave theory (LSWT). Our Hamiltonian describes almost all features of the excitation spectrum and is compatible with the experimentally observed ground state. We observed a continuum of scattering and find that this is likely a result of instrument resolution, but we cannot exclude that part of the signal could be caused by more complex dynamics, possibly of quantum origin.\\

\begin{figure}[ht!]
    \centering
    \includegraphics[width=0.9\linewidth]{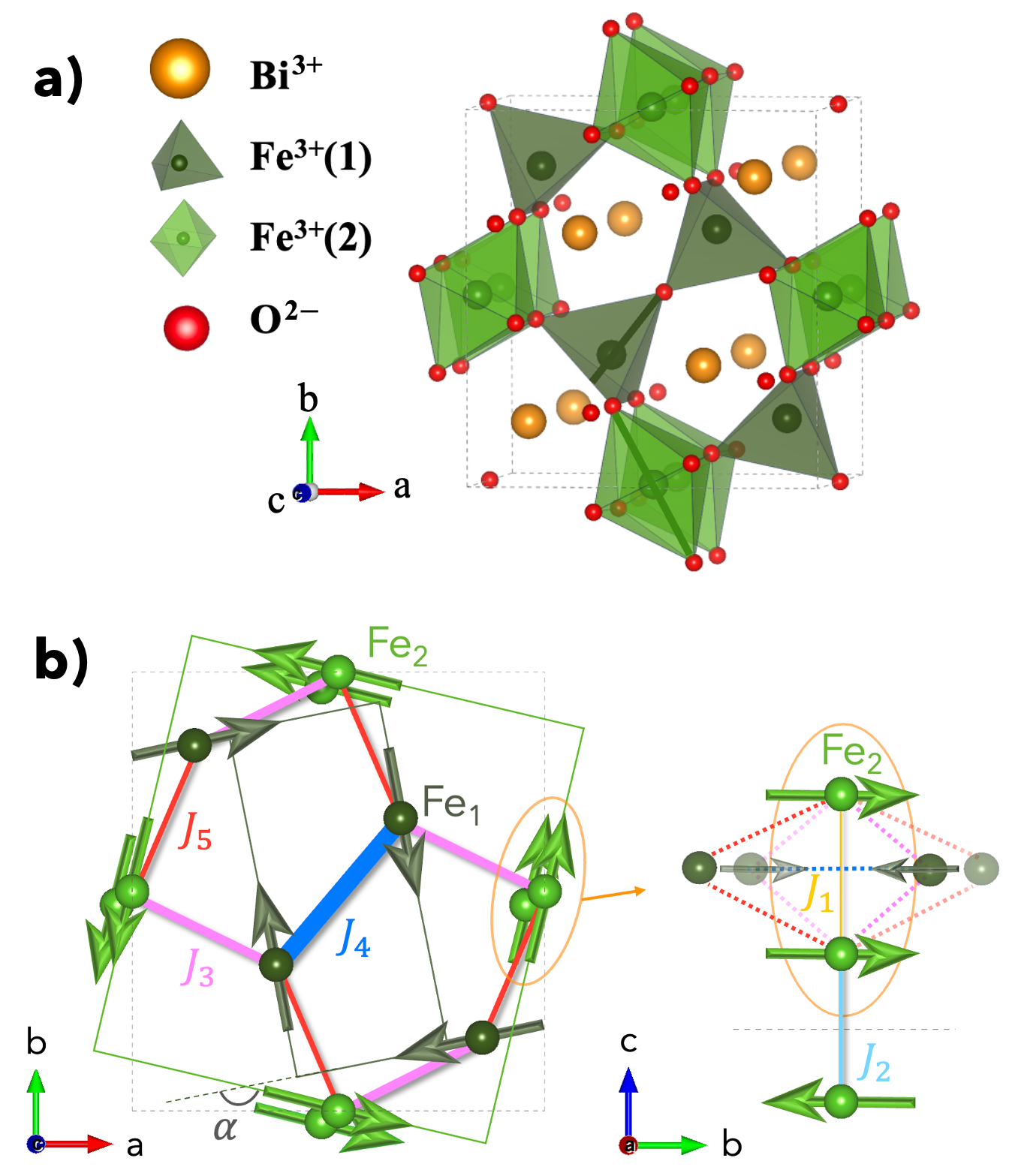}
    \caption{\textbf{a)}: Crystal structure of Bi$_2$Fe$_4$O$_9$. The local uniaxial axes are indicated respectively for Fe$_1$ (tetrahedral) and Fe$_2$ (octahedral) as a line. \textbf{b)}: Magnetic structure of Bi$_2$Fe$_4$O$_9$ proposed by Ref. \onlinecite{Ressouche2009}. The material has five nearest neighbour exchange interactions $J_1-J_5$, with two sublattices for the different iron sites; Fe$_1$ (dark green) and Fe$_2$ (bright green) rotated $\alpha=155^\circ$. The magnitude of the interactions are indicated by the width of the bonds.
    }
    \label{fig:structure}
\end{figure}

The crystal structure of Bi$_2$Fe$_4$O$_9$ is shown in Fig.~\ref{fig:structure}a. It is orthorhombic, and each unit cell contain two formula units. The eight magnetic Fe$^{3+}$ ions ($3d^5$) are equally distributed on the $4h$ and $4e$ Wyckoff sites of the \textit{Pbam} space group. These sites have different connectivities and oxygen coordination, making them nonequivalent; tetrahedral (Fe$_1$) and octahedral (Fe$_2$) both with $S=5/2$. 

Columns of edge-sharing Fe$_2$ octahedra along the $c$-axis are linked by corner-sharing Fe$_1$ tetrahedra and Bi atoms in the $ab$-plane. Viewed along the $c$-axis, the Fe-atoms form distorted pentagons, see Fig.~\ref{fig:structure}a-b. This geometry is closely related to the Cairo pentagonal lattice, except that the site with fourfold connectivity (Fe$_2$) is constituted by a pair of Fe$_2$ atoms sandwiching the pentagonal plane, resulting in slightly different bond lengths and angles. The lattice parameters are $7.9745(7)$, $b=8.4449(9)$, and $c=6.0067(4)$ Å$^{-1}$ found from single crystal neutron diffraction\cite{Murshed2013} at 300~K. 

Upon cooling below $T_N$,
Bi$_2$Fe$_4$O$_9$ transitions from a cooperative paramagnetic state to long-range antiferromagnetic (AFM) order characterized by a propagation vector\cite{Ressouche2009} $\textbf{k}=(1/2, 1/2, 1/2)$. In the AFM phase, the spins are arranged non-collinearly with two sublattices of Fe$_1$ and Fe$_2$ with an angle of $\alpha=155^\circ$. The four magnetic moments of Fe$_1$ are oriented $90^\circ$ to each other in a rectangle (see the rectangle connecting the Fe$_1$ spins in Fig.~\ref{fig:structure}b). The magnetic moments of the two Fe$_2$ pairs are also oriented $90^\circ$ to each other\cite{Ressouche2009}. \\
The Curie-Weiss temperature ($\theta_{CW}$) has been reported on single crystals to be $\theta_{CW} \approx -1670$ K and $-1468$ K found by Ref. \onlinecite{Ressouche2009} and Ref. \onlinecite{Zatsiupa2013}, respectively. The corresponding index of frustration, $f=\theta_{CW}/T_N$, are 7 and 5.7. The frustration is derived from the pentagonal coordination, resulting in competing AFM exchange interactions.

We consider five main exchange interactions $J_1$ - $J_5$, see Fig.~\ref{fig:structure}b. The interactions along the $c$-direction involve only the Fe$_2$ atoms in the octahedron columns and are achieved through two interactions; $J_1$, which is between the atoms within the unit cell (the pair), and $J_2$, which is the interaction between the atoms of adjacent cells. Within the pentagonal layer in the $ab$-plane, each Fe$_1$ interacts with its nearest neighbour Fe$_1$ via $J_4$ (180$^\circ$ angle of the Fe$_1$-O-Fe$_1$ bond) and with two nearest neighbours of Fe$_2$ pairs via $J_3$ and $J_5$. The difference between $J_3$ and $J_5$ stems from the different locations of the oxygen atoms in the bonds.

There are two previous INS reports 
of the magnetic excitations in Bi$_2$Fe$_4$O$_9$. Beauvois et al.\cite{Beauvois2020} reports spin waves in the energy transfer range $0<\hbar \omega <35$~meV measured on a small crystal ($\approx 0.5$ g),
while Duc Le et al.\cite{LeDuc2021} measured the excitations $2<\hbar \omega <30$~meV on $\approx 0.6$~g co-aligned single crystals. Additionally, they\cite{LeDuc2021} performed INS on a 20~g powder up to $\hbar \omega =100$~meV, showing that the spin waves extend up to 90~meV. They both report a dispersionless band, but at different energies: 19~meV (Ref. \onlinecite{Beauvois2020}) or 40~meV (Ref. \onlinecite{LeDuc2021}). From linear spin wave theory, both works suggest a magnetic Hamiltonian with five exchange interactions and an easy-plane anisotropy of the system, reported in Table \ref{tab:interactions}. The easy-plane anisotropy for all iron-sites in the $ab$-plane results in one mode (involving precession out of the plane) being gapped ($\approx 5$~meV in Ref.~\onlinecite{LeDuc2021}), while the other modes remain gapless. They both find $J_4$ to be the dominating interaction. However, the two studies differ in whether the small $J_1$ interaction is FM or AFM. In our study, we conclude the $J_1$ is FM. 

\begin{table}[h!]
    \centering
    \setlength{\tabcolsep}{12pt}
    \renewcommand{\arraystretch}{1.2}
    \begin{tabular}{lccc}
        \toprule[0.1pt]
        & This work & Ref. \onlinecite{Beauvois2020} & Ref. \onlinecite{LeDuc2021} \\
        \midrule[0.1pt]  
        $J_1$  & -0.2(1)  & 3.7(2)   & -0.22(3)  \\
        $J_2$  & 1.40(7)   & 1.3(2)   & 1.39(5)   \\
        $J_3$  & 6.4(1)   & 6.3(2)   & 6.5(2)    \\
        $J_4$  & 27.9(8)   & 24.0(8)  & 27.6(6)   \\
        $J_5$  & 3.1(1)   & 2.9(1)   & 3.1(2)    \\
        $A$    &          & 0.03     & 0.096(5)  \\
        $A_1$  & 0.034(2)   &          &           \\
        $A_2$  & -0.046(2)   &          &           \\
        \bottomrule[0.1pt]
    \end{tabular}
    \caption{Spin wave exchange parameters fitted to experimental data given in~meV. Positive values indicate antiferromagnetic exchange. $A$ denotes easy-plane anisotropy in $ab$-plane on all sites, while $A_1$ indicates the easy-plane for the Fe$_1$ sites and $A_2$ is easy-axis anisotropy for the Fe$_2$ sites (as described in the text).}
    \label{tab:interactions}
\end{table}

\section{Methods}
\begin{figure}[ht!]
    \centering
    \includegraphics[width=0.9\linewidth]{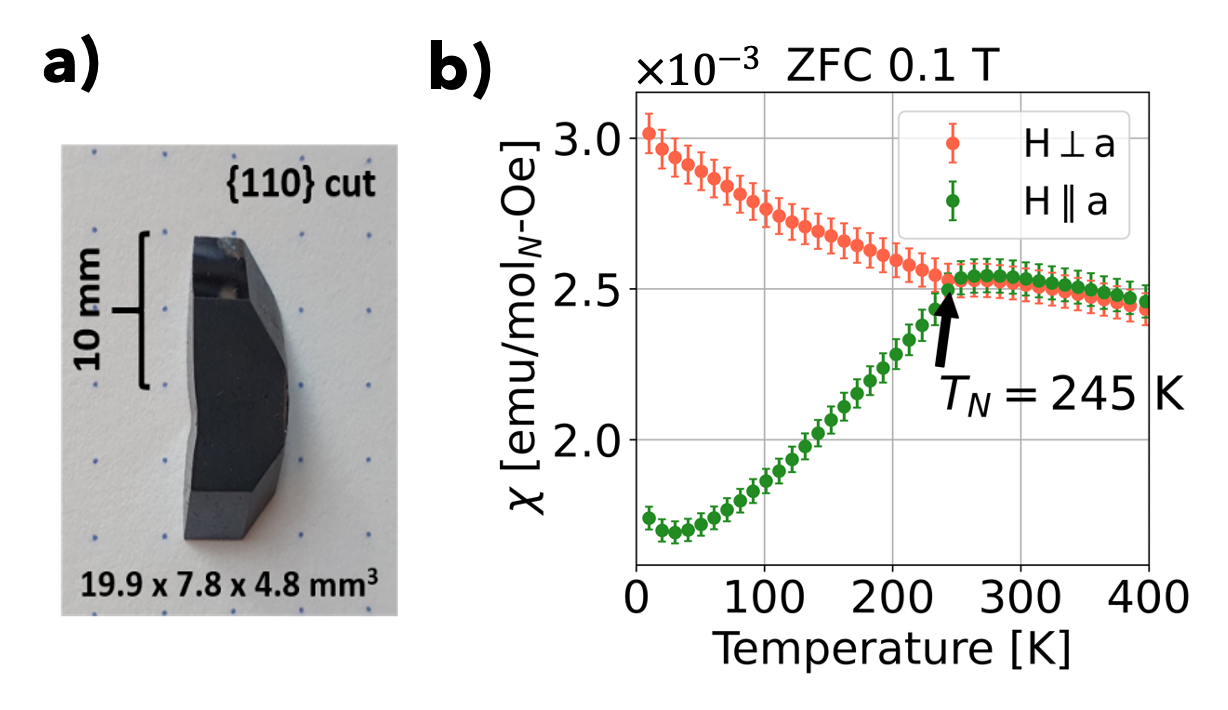}
    \caption{\textbf{a)}: Picture of the 2.35 g single crystal used in this study. 
    \textbf{b)}: Magnetic susceptibility of a small crystal of the same sample growth, showing $T_N=245$ K.}
    \label{fig:crystal}
\end{figure}

The single crystal used in the present study was grown along [001] using the top seeded solution growth technique. It is a 2.35~g crystal piece (Fig.~\ref{fig:crystal}a) of $20\times20\times8$ mm$^3$ of a much larger crystal\cite{Burianek_2009_growth}. 
Our X-ray diffraction data shows high crystalline quality, and neutron diffraction shows a mosaic spread narrower than the detection limit ($<0.5^\circ$). \\
The magnetic susceptibility in Fig.~\ref{fig:crystal}b was measured (PPMS DynaCool in VSM mode) on a small single crystal piece from the same growth as the one used for INS. The data is almost completely identical to that of Ref. \onlinecite{Ressouche2009} and we also obtain $T_N = 245$~K. A slight upturn is seen at low temperatures, which we assign to the crystal coming from the outside of the growth, so it is likely not to have the same quality as the one used for INS. 

\subsection{Inelastic neutron scattering (INS)}
We have used multiple different inelastic neutron spectrometers to measure the excitations; the cold-neutron spectrometer CAMEA (at the Paul-Scherrer-Institut (PSI), Switzerland) was used to cover 0-9.2~meV energy transfer, the thermal spectrometers EIGER (PSI), covered 7-24~meV energy transfer, and IN20 (at the Institut-Laue-Langevin, France) covered 20-45~meV energy transfer. To reach the upper bands (3-95~meV), the high-intensity spectrometer 4SEASONS (at J-PARC, Japan) was utilized. All experiments were performed at 10~K on the same single crystal in orange helium-flow cryostats, except at 4SEASONS where a closed-cycle refrigerator was used. The sample was mounted with the [HH0] and [00L] directions in the horizontal scattering plane.

The triple-axis-spectrometer (TAS)-like CAMEA instrument has a quasi-continuous coverage in two dimensions in \textbf{Q}-space 
and the energy axis.\cite{Lass2023} Data were acquired by performing 120$^\circ$ sample rotation scans in 0.5$^\circ$ steps for 5 different incoming energies, (5.0, 6.8, 8.6, 10.4, and 12.2~meV plus 0.13~meV offsets) and using a monitor of 125\,000 counts. We measured at the highest possible scattering angles at all energies, i.e. $-79^\circ$ at 5~meV, $-78^\circ$ at 6.8~meV, $-64^\circ$ at 8.6~meV, and at $-49^\circ$ for 10.4 and 12.2~meV, as well as $+4^\circ$ for dark angle interlacing. 
This data were converted and treated using the dedicated software package MJOLNIR (version 1.3.1.post4).\cite{Lass2020MJOLNIR,jakob_lass_2023_8183140} 

Both EIGER\cite{stuhr_2017_EIGER} and IN20 are thermal TAS. In both setups, the measurements were performed with a constant final energy of 14.688~meV with a double focused pyrolytic graphite PG(002) monochromator. At IN20 a double focused PG(002) analyzer was used, while at EIGER we used a horizontal focusing analyzer setup with open collimation. 
At EIGER, a 37~mm thick PG filter was placed between the sample and the analyzer in order to suppress higher order neutrons, while at IN20 a velocity selector and two PG filters were used (lengths 2 and 5~cm, respectively).\\
To access the high energy bands of $60<\hbar\omega < 85$~meV, we measured INS on the time-of-flight (TOF) spectrometer 4SEASONS\cite{Kajimoto_2011_4SEASONS}. The data were collected with a Fermi chopper frequency of 300~Hz using the repetition-rate-multiplication technique\cite{Nakamura_2009_multiE} with incident neutron energies of $E_i=389$, 110 and 51.0~meV (with respective energy resolutions (FWHM) of 46.3, 7.6, and 2.7~meV at the elastic position). Energies lower than $E_i=51.0$~meV were suppressed to obtain better background. The crystal was rotated by $180^\circ$ in steps of $0.5^\circ$.
We used the Utsusemi\cite{Inamura_2013_Utsusemi} and Horace\cite{Ewings_2016_Horace} software for the data analysis.

As a complement to the unpolarised measurements of the magnetic dispersion, we employed full XYZ polarisation analysis with PASTIS-3 at IN20\cite{Jullien_2021_Pastis3} to separate nuclear and magnetic scattering, and to distinguish between magnetic amplitudes in the horizontal scattering plane and perpendicular to it. The setup consisted of a PG(002) monochromator and the FlatCone multi-analyser detector\cite{Kempa_2006_flatcone} unit of 31 Si(111) analyzers with final neutron wavelength of 3.0 Å$^{-1}$. We used the velocity selector in front of the monochromator and had open collimation. PASTIS-3 contains two 3He-cells to prepare the incident and analyse the outgoing neutron polarisation in a common magnetic guide field of 16 G, applied vertical or in the horizontal scattering plane. An RF-flipper allowed to flip the incident polarisation. All polarised measurements were performed in sample rotation scans in 1.25 degree steps to cover the AFM zone centres (-1.5 -1.5 0.5) and (-0.5 -0.5 1.5), at the energy transfers 10, 12, 15.5, 24 and 28 meV. The data were corrected for polarisation and transmission decay and separated into nuclear, magnetic in-plane ($M_{yy}$) and out-of-plane amplitude scattering ($M_{zz}$) with the nplot MATLAB toolset\cite{nplot_ill}.

\begin{figure*}[hb]
    \includegraphics[width=\textwidth]{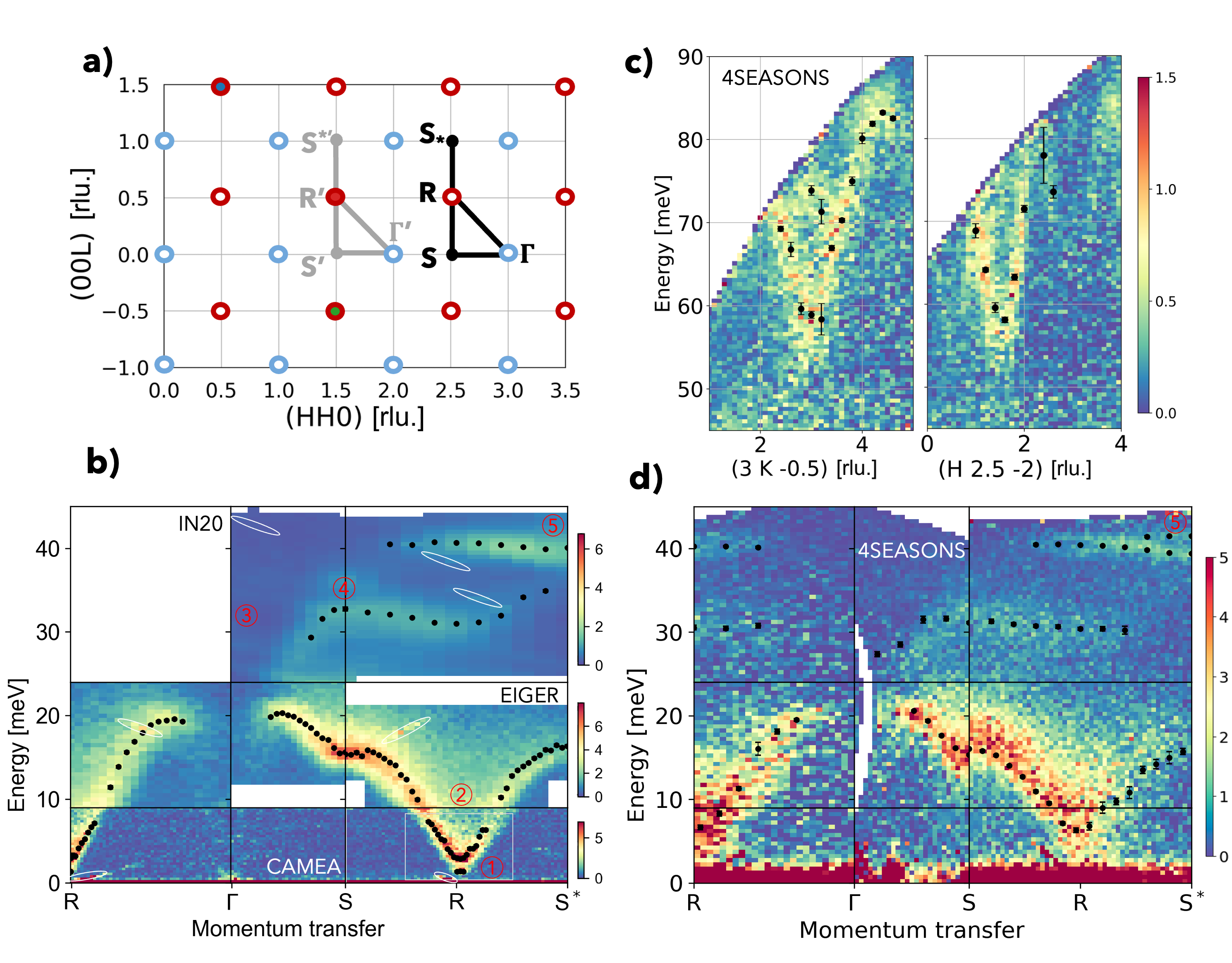} 
    \caption{
    \textbf{a)}: Sketch of the scattering plane (HHL) investigated by INS, with the nuclear (blue circles) and magnetic (red circles) Bragg peak positions. The measurement cuts are shown in black lines cutting through R=(2.5 2.5 0.5), $\Gamma$=(3 3 0), S=(2.5 2.5 0) and S$^*$=(2.5 2.5 1). The grey path indicates the same cut in another Brillouin Zone (-1 in H). The colored magnetic Bragg peaks are the cut positions in Fig. \ref{app:double_gap_fit}b.  
    \textbf{b)}: INS intensity shown as a function of energy transfer at \textbf{Q} given by the path illustrated in a). Data from CAMEA (0-9~meV, grey path), EIGER (9-24~meV, black path) and IN20 (24-45~meV, black path). The white ellipsis indicate spurious signals. The red numbers are discussed in the text.
    \textbf{c)} INS intensity collected on 4SEASONS showing the upper branch of the excitation spectrum of 45-90~meV energy transfer, with an integration of $\pm 0.4$ in L. 
    \textbf{d)}: INS intensity collected on 4SEASONS with $E_i=51$~meV (black path). The plot is 
    integrated $\pm0.1$~rlu. in the out of plane directions, except the cut $\Gamma$-$S$, which is integrated $\pm0.2$~rlu. along L to improve statistics. The black lines match the corresponding lines in panel b). \textbf{b-d} The intensity on the colour scales are in arbitrary units and the black points show fitted peak positions as described in the text. All data is at 10~K. 
    }
    \label{fig:data_colorplots}
\end{figure*}

\section{Results}
We show an overview of $S(\mathbf{Q},\hbar\omega)$ for Bi$_2$Fe$_4$O$_9$ as collected across the various instruments over the energy- and \textbf{Q}-range in Fig.~\ref{fig:data_colorplots}. The data from the different instruments were scaled to each other to provide a visual overview. 
Since different instruments cover different ranges of energy transfer, we measured in different Brillouin zones to be able to close the scattering triangle. Fig.~\ref{fig:data_colorplots}a shows the path in reciprocal space for the CAMEA\cite{CAMEA_data} data in grey and for EIGER\cite{EIGER_data}, IN20\cite{IN20}, and 4SEASONS\cite{JPARC_data} in black. 
The energy mode positions and the integrated intensity at constant \textbf{Q}-cuts have been fitted with a Gaussian lineshape with a linear background, shown in the figures as black points with errorbars. White ellipses indicate spurious signals that we will ignore in the following.

\subsection{Excitations 0-9~meV: Gapped dispersion} \label{sect:data_double_gap}

\begin{figure}[H]
    \centering
    \includegraphics[width=1\linewidth]{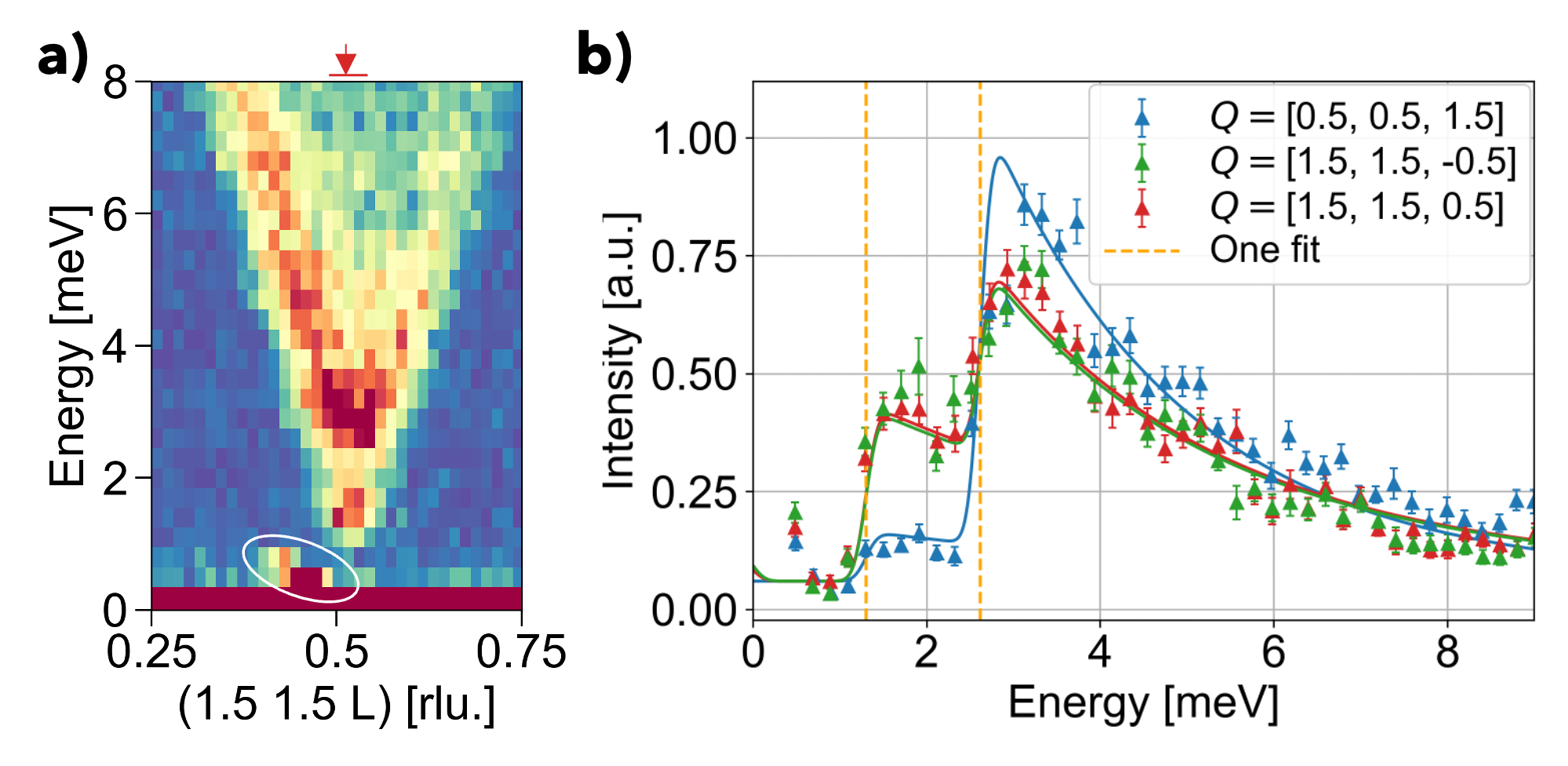}
    \caption{\textbf{a)}: Zoom in at position R'=(1.5 1.5 0.5) (region (1) marked with a grey box in Fig.~\ref{fig:data_colorplots}b) of CAMEA data, showing the double gap in the data. \textbf{b)}: Constant Q-cuts through the data at three different magnetic Bragg peak positions. The integration width in both directions are 0.03~rlu. in \textbf{Q}. The Bragg peak positions are shown in Fig. \ref{fig:data_colorplots}a), additionally for the red cut, an arrow in a) shows the cut position and the line is the integration width. The yellow dashed vertical lines indicate the simultaneously fitted gap positions of all three \textbf{Q}-positions.}
    \label{fig:double_peak_1Dcut}
\end{figure}

We first look at the low energy data from CAMEA. Fig.~\ref{fig:double_peak_1Dcut}a shows a zoom of the data indicated by the region labelled (1) in Fig.~\ref{fig:data_colorplots}b.
In this region, the spin wave dispersion is linear, as expected for an AFM. Interestingly, the excitations at the magnetic zone centre R'=(1.5 1.5 0.5) are clearly gapped. Furthermore, we see a splitting of the lowest modes, resulting in a double gap. Constant-$Q$ cuts at three magnetic Bragg peak positions; (1.5 1.5 $\pm$0.5) and (0.5 0.5 1.5) are shown in Fig.~\ref{fig:double_peak_1Dcut}b, where the double gap is observed in all three cases. To determine the size of the gaps, the peaks have been fitted with an approximate convolution function\cite{fit_gap_paper} (described in appendix \ref{app:double_gap_fit_function}) to accommodate the resolution tail and get a more precise value of the gap. The FWHM of the fit is fixed to the energy resolution of CAMEA. Fitting all three \textbf{Q}-positions simultaneously (see appendix \ref{app:double_gap_fit_final}), yields gaps of $1.30(1)$~meV and $2.62(1)$~meV, marked with dashed yellow lines in the figure. 
We show below that these gaps originate from single-ion anisotropy, which lifts the magnon degeneracy and stabilizes the magnetic order (section \ref{sec:aniso}). 

\subsection{Excitations 9-35~meV: Continuum of scattering}
\label{sec:data_continuum}
The dispersion from CAMEA continues into the EIGER data, where the spin wave dispersions are in agreement with previously published data\cite{Beauvois2020,LeDuc2021}.

In Fig.~\ref{fig:data_colorplots}b (2) a pronounced continuum of scattering is present above the spin wave dispersion. Such a continuum has previously not been reported. This could indicate a two-magnon dispersion or a potential quantum continuum. At R'=(1.5 1.5 0.5) and R=(2.5 2.5 0.5), this scattering extends from 1.6~meV up to the 30~meV mode, and is thus observed on three different TAS instruments. However, the continuum looks less pronounced in the 4SEASONS data, Fig.~\ref{fig:data_colorplots}d. This opens the question, whether the observed continuum stems from the instrument resolution rather than from a physical effect. Such a resolution tail could arise if the spin wave would disperse strongly within the span of the resolution function. Around a gapped minimum, this effect results in intensity in energies above the dispersion, but not below. To investigate this, we compare data from all four instruments at two different \textbf{Q}-values; R=(2.5 2.5 0.5) and S=(2.5 2.5 0). 

The four instruments have different resolution functions. If the observed continuum is a resolution effect, it should vary with the resolution. On the other hand, if the effect is intrinsic to the sample, it should be mostly independent of the instrument resolution. With this in mind, we now look at the data with varying resolutions.

\begin{figure}[H]
    \centering
    \includegraphics[width=1\linewidth]{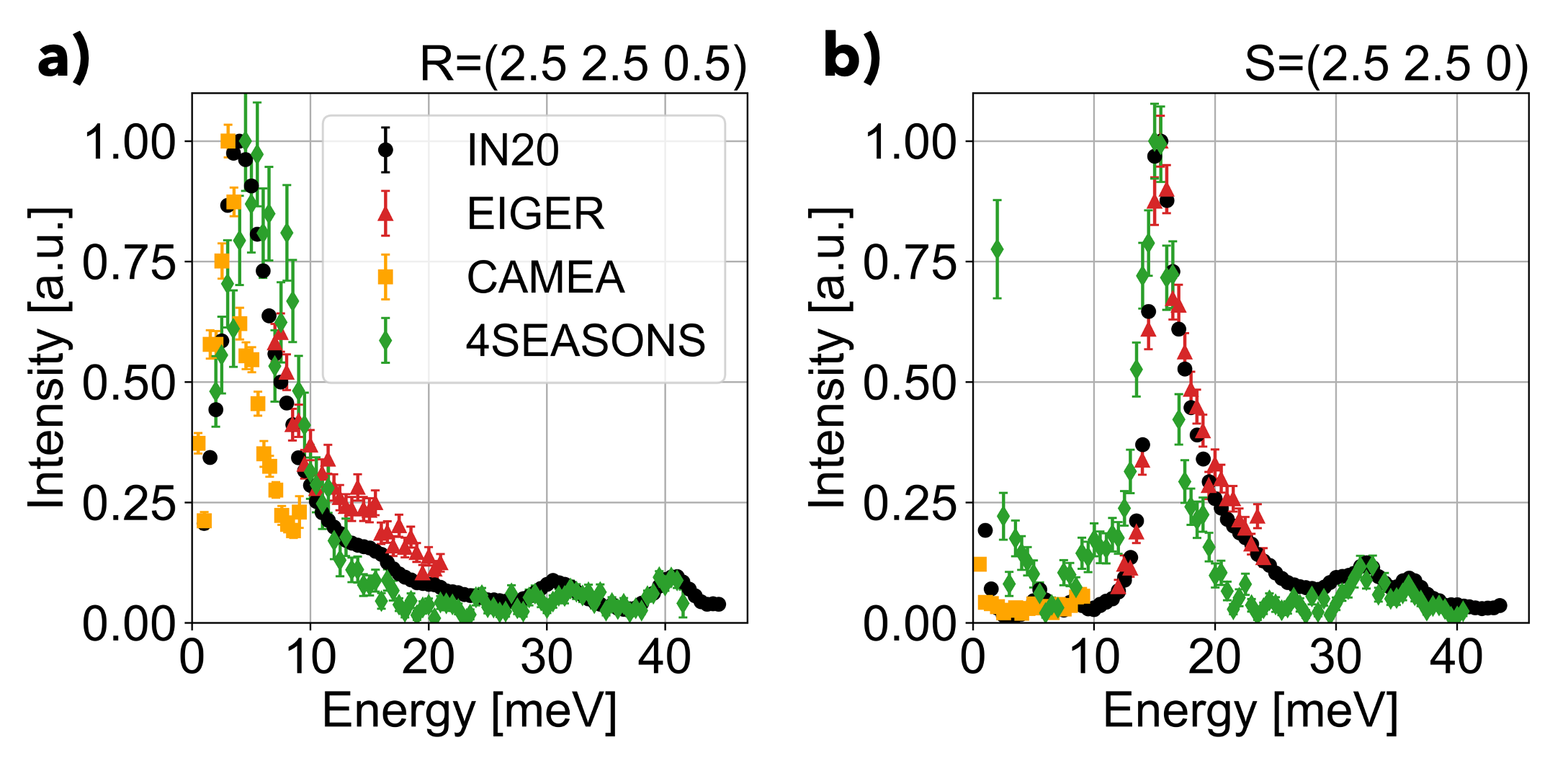}
    \caption{Comparison of data from all four instruments at two different \textbf{Q}-values; \textbf{a)}: R=(2.5 2.5 0.5) and \textbf{b)}: S=(2.5 2.5 0). The data are normalised to be on the same scale. 
    The 4SEASONS data are integrated by $\pm 0.1$~Å$^{-1}$ in all directions. }
    \label{fig:continuum}
\end{figure}

\begin{figure}[ht!]
    \centering
    \includegraphics[width=0.9\linewidth]{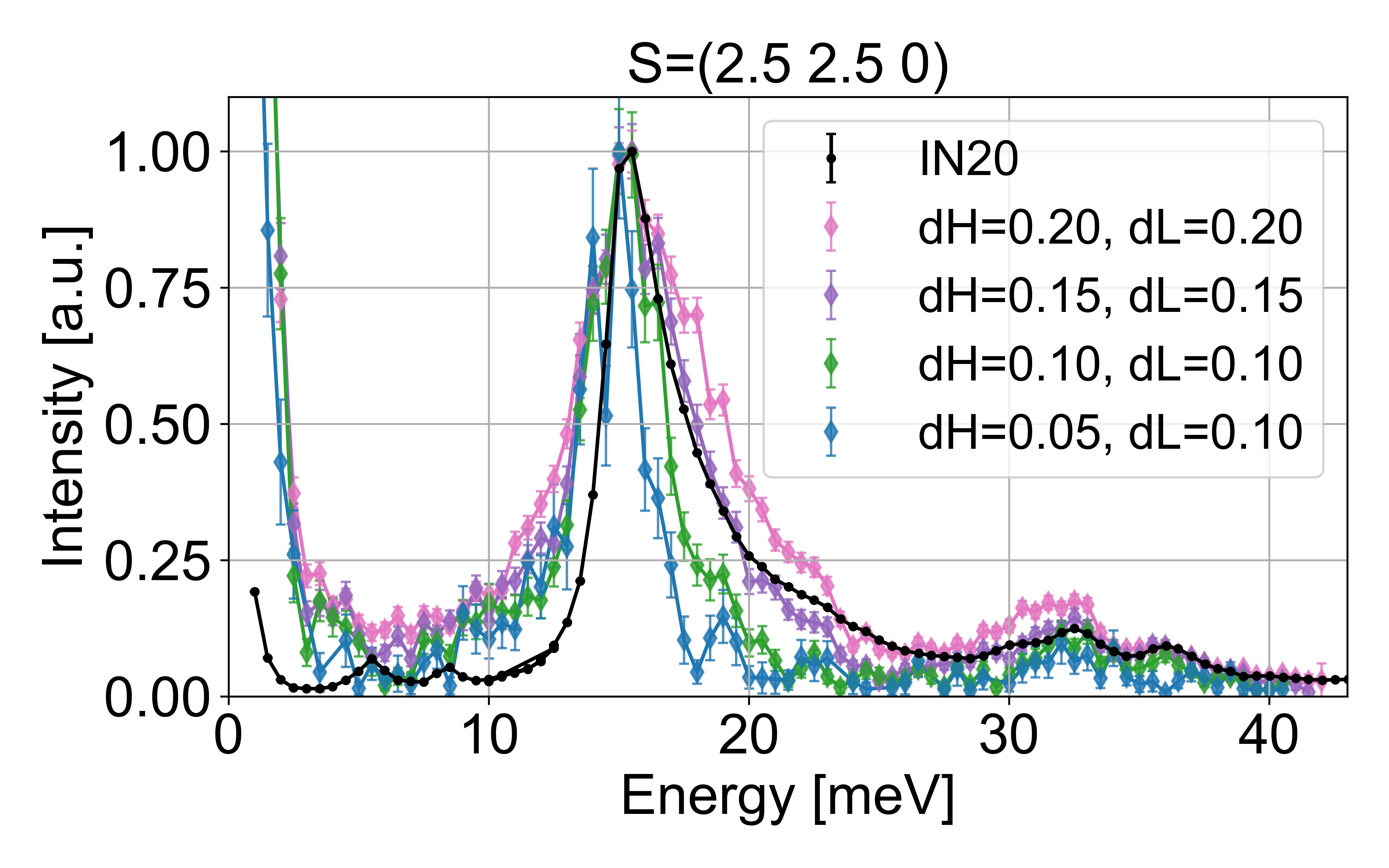}
    \caption{4SEASONS constant \textbf{Q}-cut at S=(2.5 2.5 0) with varying integration width compared to IN20. The $\pm0.1$ Å$^{-1}$ width data (green) is also presented in Fig.~\ref{fig:continuum}b. With increasing integration width, the intensity of the high energy peak tail increases rapidly. }
    \label{fig:diff_integration_4SEASONS}
\end{figure}

At low energies, CAMEA has the best resolution, whereas those of IN20, EIGER and 4SEASONS ($E_i=51$ meV) are roughly equal, see Fig.~\ref{fig:continuum}a. Here, the peak at 3~meV seen with EIGER, IN20 and 4SEASONS have roughly the same peak tail, while the tail of the peak measured at CAMEA is more narrow. 
At higher energies, the TAS resolution broadens, while the 4SEASONS resolution narrows as $E_{\rm f} \rightarrow 0$. Hence, in Fig.~\ref{fig:continuum}b, the peak at 16~meV in the 4SEASONS data is narrower than the TAS data. The resolution of 4SEASONS (TOF) is often elongated in the direction corresponding to the TOF scan trajectory. In Fig. \ref{fig:data_colorplots}d, you can see a slope extending from the R position at $E = 0$ toward the upper left. This may be the tail of the Bragg peak caused by the resolution function (Bragg tail). However, the
\textbf{Q}-resolution in TOF instruments is often dominated by the  cutting and integration procedures used when analysing the data, which can be varied post-experiment. By varying the integration width, we can therefore vary the resolution of our 4SEASONS data.
Decreasing the integration width of the 4SEASONS data should show a narrowing of the tail if the peak is produced by resolution, but be unchanged if there is a genuine continuum of scattering. Looking at S=(2.5 2.5 0) in Fig.~\ref{fig:diff_integration_4SEASONS}, we see the tail diminishing with decreasing integration width.

These results indicate that the observed continuum primarily is an instrumental resolution effect. We return to this analysis in section \ref{sect:model_continuum} to determine if a part of the broadening is caused by more exotic origins.

The IN20 data, Fig.~\ref{fig:data_colorplots}b top, shows weaker modes near 30~meV and 40~meV energy transfer.
At 30~meV, 4SEASONS (1.0~meV) has a much better energy resolution than IN20 (5.4~meV). 
We may pick up signals from other \textbf{Q}-values due to the wider out of plane coverage of IN20. In the cut $\Gamma$=(3 3 0) to S=(2.5 2.5 0) at 25-40~meV, Fig.~\ref{fig:data_colorplots}b (3), multiple weakly dispersive signals are observed, which are likely optical phonons. At S=(2.5 2.5 0) scattering is observed at both 33~meV and at 36~meV, Fig. \ref{fig:data_colorplots} (4), where only the 33~meV follows the shape of the spin wave dispersion. We have compared equal \textbf{Q}-cuts, for IN20 and 4SEASONS data ($E_i=51$~meV), from Fig.~\ref{fig:data_colorplots}b and \ref{fig:data_colorplots}d, to exclude that the additional scattering is part of the spin wave spectrum. An example is shown in Fig.~\ref{fig:diff_integration_4SEASONS}, where the signal at 36~meV disappears as the integration width is reduced. This indicates that the additional signals are not part of the spin wave excitation.


\subsection{Excitations 35-45~meV: Dispersionless mode}

\begin{figure*}[h]
    \centering
    \includegraphics[width=1\linewidth]{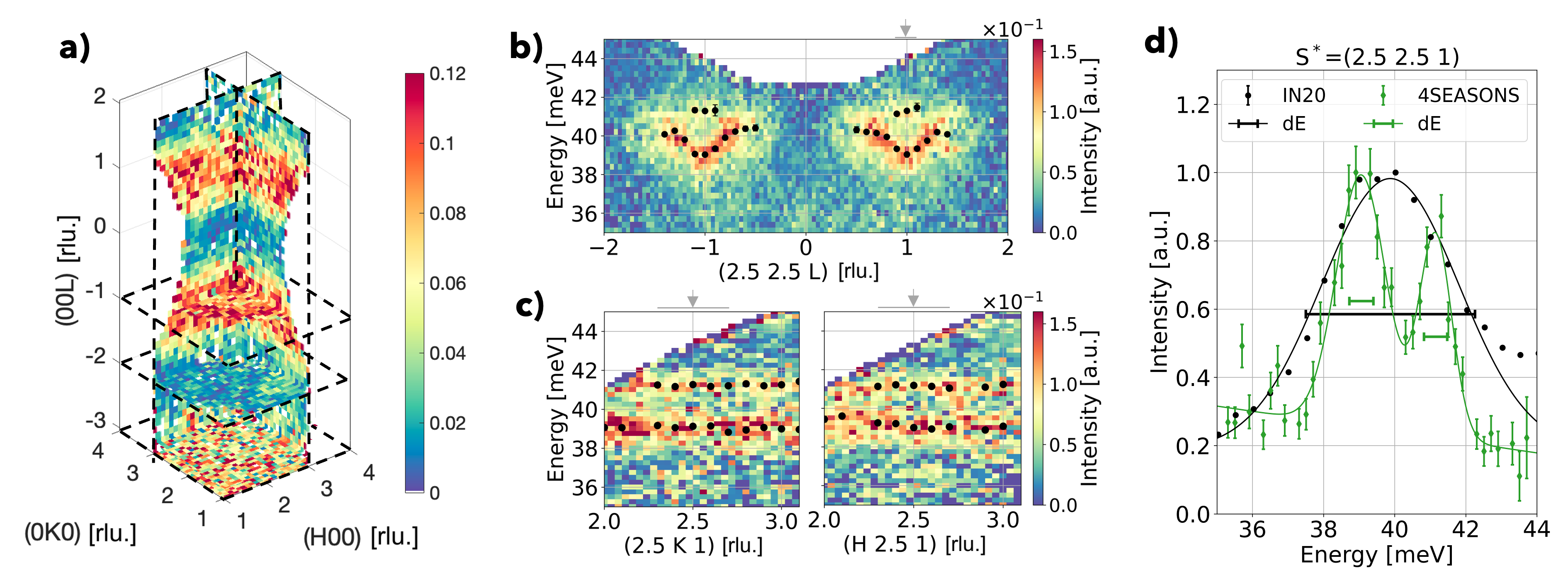}
    \caption{Zoom-in on the magnetic mode close to 40~meV of the 4SEASONS $E_i=51$~meV data. \textbf{a)}: 3D figure of HKL at $40\pm1.5$~meV energy transfer. The highlighted planes are at (H3L), (3KL) and planes in HK at L=$-3,-2$ and $-1$. \textbf{b)}: Dispersion along L with H=K=2.5, showing that the mode is dispersive and splits. \textbf{c)}: Dispersion at L=1 along K (left panel) and H (right panel), showing a splitting of the 40~meV mode. \textbf{d)}: Intensity as a function of energy transfer at S$^*$=(2.5 2.5 1) for both IN20 (black) and 4SEASONS (green) with integration $\pm0.1$ rlu. in L and $\pm0.2$ rlu. in H and K (indicated with the grey arrow in b) and c) ). Their respective resolutions are drawn at half maximum of each fitted peak and the data is normalised to 1.}
    \label{fig:40meV_mode}
\end{figure*}

A 40~meV magnetic mode, previously measured in a powder sample, was proposed to be dispersionless\cite{LeDuc2021}. However, our data, Fig.~\ref{fig:data_colorplots}b and \ref{fig:data_colorplots}d (5), indicate that the mode has a weak dispersion. In addition, the mode splits in two close to S$^*$=(2.5 2.5 1) (see Fig.~\ref{fig:data_colorplots}d). \\
From the 4SEASONS data with $E_i=51$~meV, the 40~meV mode reveals sheets of scattering in the HK-plane at odd integers of L, shown in 3D \textbf{Q}-space integrated over $\hbar \omega= 40\pm1.5$~meV in Fig.~\ref{fig:40meV_mode}a. The figure shows the sheets of scattering in the HK-plane at L$=-3$ and L$=-1$, which are not present at L$=-2$. The \textbf{Q}-point S$^*$=(2.5 2.5 1) is studied through cuts along the high symmetry directions in H, K and L as a function of energy, Fig.~\ref{fig:40meV_mode}b-c. From these cuts, it is apparent that the 40~meV mode is dispersive along L with a maximum splitting at odd integers. At L$=1$ the mode is split into two modes that are flat in H and K. Taking the average for both cuts the lower mode is found at $39.1\pm0.2$~meV and the upper mode at $41.3\pm0.2$~meV. Fig.~\ref{fig:40meV_mode}d at S$^*$=(2.5 2.5 1) shows the split of the 40~meV mode seen in the 4SEASONS data (green) compared to the single peak in the IN20 data (black). At 40~meV energy transfer, IN20 has an energy resolution of 4.8~meV, while 4SEASONS with $E_i=51$~meV has a much more narrow resolution of 0.7~meV, enabling the observation of this splitting. The resolutions are drawn on the figure for both instruments at the respective peak position(s). In the 4SEASONS data, the calculated resolution is much narrower than the FWHM of the observed peaks. This we assign to the finite integration width in {\bf Q}.

\subsection{Excitations 50-90~meV: Upper branch}
The excitation spectrum was shown to extend up to 90~meV energy transfer on powder samples\cite{LeDuc2021}, but has not been studied on a single crystal before. We have mapped this part of the spectrum at 4SEASONS with $E_i=110$~meV. We show two selected cuts in Fig.~\ref{fig:data_colorplots}c; along K and along H. Along K, the dispersion is more shallow, compared to the one along H, and we see another mode at around 74~meV.
The energy transfer of the spin wave modes extends 58-83~meV, which is indicative of the strength of the exchange coupling constants, especially the exchange interaction $J_4$, as will be elaborated below. 

\subsection{Polarised neutron scattering}
\begin{figure*}[h]
    \centering
    \includegraphics[width=1\linewidth]{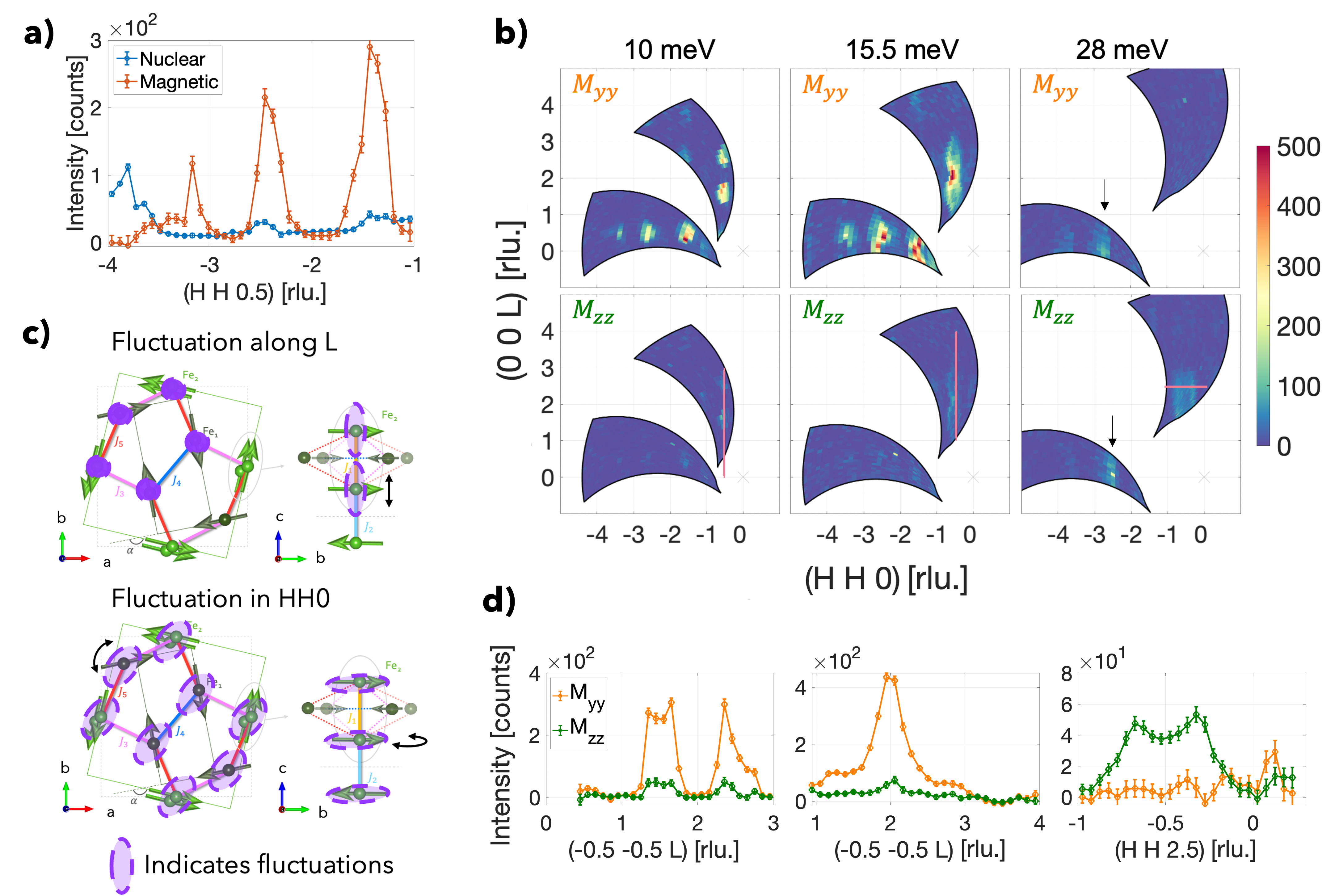}
    \caption{Polarised IN20 data. \textbf{a)}: Separation of the different contributions; nuclear and magnetic scattering at 10~meV energy transfer. Integration is 0.02 Å$^{-1}$. 
    \textbf{b}: Polarised constant energy maps of intensity in the different channels at different energy transfer; 10, 15.5 and 28~meV in the scattering plane [HH0]-[00L]. Top row shows magnetic scattering perpendicular to \textbf{Q} in the scattering plane; $M_{yy}$. The bottom row shows scattering from magnetic scattering perpendicular to the scattering plane; channel $M_{zz}$. 
    \textbf{c)}: The purple ellipse illustrates the plane of the magnetic amplitudes in real space (cf. Fig.~\ref{fig:structure}b). Intensity $M_{yy}$ at $\textbf{Q}$ mainly parallel to (HH0) indicate magnetic amplitudes along the $c$-axis ($L$-axis) (top), while $M_{yy}$-intensity at \textbf{Q} mainly parallel to (00L) indicate magnetic amplitudes along (HH0), in the $ab$-plane (bottom). 
    \textbf{d)}: Constant energy cuts (pink lines in b) comparing $M_{yy}$ and $M_{zz}$. Integration width is 0.02 Å$^{-1}$.}
    \label{fig:pol_maps}
\end{figure*}

The INS polarised data\cite{IN20_pol} (from the PASTIS setup at IN20) allows us to separate nuclear and magnetic scattering, following the method of Ref.~\onlinecite{Stewart_2009_XYZpol}. An example of the separation is shown in Fig.~\ref{fig:pol_maps}a, taken at 10~meV energy transfer. Here, the magnetic scattering is well separated from 
nuclear scattering at the magnetic Bragg peak positions (H H 0.5) for H=$-3.5, -2.5, -1.5$. At H=$-4$ a peak is seen in the nuclear channel, which we identify as a phonon. Neutron scattering only probes magnetic components perpendicular to \textbf{Q}, with the magnetic scattering further split into two contributions; magnetic scattering perpendicular to \textbf{Q} in the scattering plane ($M_{yy}$) and magnetic scattering perpendicular to the scattering plane ($M_{zz}$). With the scattering plane being [HH0]-[00L], the direction of the fluctuating amplitudes giving rise to intensity in $M_{yy}$ will vary for each \textbf{Q}-position, see Fig.~\ref{fig:pol_maps}b. For example at (-1.5 -1.5 0.5) (equivalent to R', but with opposite sign of H), the main contribution to $M_{yy}$ are amplitudes along the $c$-axis, equivalent to 75\% of the expected scattering, compared to 25\% in the $ab$-plane. The direction of these amplitudes are illustrated in Fig.~\ref{fig:pol_maps}c top. In contrast, at (-0.5 -0.5 1.5) the main contribution to $M_{yy}$ is from amplitudes along (HH0), in the $ab$-plane. The intensity in $M_{zz}$ will always come from magnetic amplitudes parallel to (H -H 0), perpendicular to the scattering plane, and within the $ab$-plane, but orthogonal to those shown in Fig.~\ref{fig:pol_maps}c (bottom).\\

In Fig.~\ref{fig:pol_maps}b, we show $M_{yy}$ and $M_{zz}$ as function of (00L) and (HH0) at three energy transfers $\hbar \omega = 10$, 15, 28~meV. We notice that the inelastic magnetic scattering is very anisotropic. The low energy modes (10~meV and 15.5~meV) show strong scattering in the $M_{yy}$-component, but weak scattering in $M_{zz}$. Thus, the low-energy modes fluctuate mainly along the $c$ axis. In contrast, the 28~meV mode shows scattering only present in the $M_{zz}$ component. This is quantified in Fig.~\ref{fig:pol_maps}d, showing relevant cuts in the data as indicated by the pink lines in Fig.~\ref{fig:pol_maps}b. \\
At 28~meV energy transfer at H$=-2.5$ a straight line of intensity is seen along L in both channels with the same intensity and shape (see black arrows in Fig. \ref{fig:pol_maps}b). This could resemble scattering that is not well separated, or something else interfering with the signal. This is also seen at 24~meV (data not shown). We believe it not to be a magnetic signal, and at this time we do not investigate it further. 

\section{Modelling}
We use linear spin wave theory to model our data. The Hamiltonian is given by
\begin{equation}
    H = \sum_{ij} J_{ij} \textbf{S}_i \cdot \textbf{S}_j + \sum_{i} A_{i} S_{\alpha_i}^2,
    \label{eq:Hamiltonian}
\end{equation}
where $J_{ij}$ are the exchange constants and $J > 0$ indicates AFM interactions. $A_i$ are the single-ion anisotropies with uniaxial axis $\alpha_i$ for the $i$th Fe atom. We show below that the anisotropy for Fe$_1$ is positive ($A_1 > 0$) giving an easy-plane anisotropy, thus perpendicular to the uniaxial axis, while the anisotropy for Fe$_2$ is negative ($A_2<0$), giving an easy-axis anisotropy along the uniaxial axis of Fe$_2$.

We refine the model parameters based on the experimentally determined mode positions, the uncertainty on the positions and the integrated peak intensities using SpinW\cite{Toth_2015,SpinW_program}, which is a software implementation of linear spin-wave theory.

In this section, we first specify the impact of the interaction parameter on the spectrum, then the anisotropies and gaps are discussed. 
Finally, we optimize the parameters and compare the calculations to the experimental data. More visual information about testing the parameters in the model is given in appendix \ref{app:test_spinW}.

\subsection{Effects of the individual exchange parameters}
Each of the exchange parameters affects the spin wave spectrum differently. This information is important to build an understanding of the material. Below, we describe the direct effect of the exchange parameters in a scenario with zero anisotropy. 

The $J_1$ interaction generates an out-of-phase precession of the FM aligned Fe$_2$ pair of spins, yielding a flat mode at an intermediate energy. The position of this dispersionless band was first proposed by Ref. \onlinecite{Beauvois2020} to be at 19~meV, but was later found by Ref. \onlinecite{LeDuc2021} to be at 40~meV, while the 19~meV mode was explained to be a spurion. In the model, the band position depends on the size and magnitude of $J_1$; the stronger FM, the higher the mode will increase in energy. A greater AFM $J_1$ will decrease the mode energy. 

The $J_2$ interaction, connecting Fe$_2$ spins in different layers, modulates the lower bands together with $J_3$ and $J_5$. $J_1$ and $J_2$ do not affect the upper bands, since these mostly depend on fluctuations on the Fe$_1$ sites. $J_2$ influences the dispersion along the L-direction; the stronger the AFM interaction, the larger the spin wave velocity at R=(2.5 2.5 0.5) and the more the 30~meV band disperses.

The two different interactions connecting the Fe$_1$ and Fe$_2$ sites, $J_3$ and $J_5$, differ due to an asymmetry in the positions of the oxygen ligands. From previous studies, $J_3$ is a factor of two larger than $J_5$. Switching the magnitude of the interactions results in slight differences in the intensities and no noticeable difference in energy. This means that the two interactions have very similar effect on the spin waves. Both interactions determine the bandwidths of the lower (0-35~meV) and upper (60-85~meV) bands, thus the greater AFM $J_3$ and $J_5$, the larger the bandwidth. When removing one of these interactions, there is no splitting of the degenerate lower, flat, and upper bands. When increasing one of the interactions, the degeneracy of the lower and upper bands are broken. For the lower bands, the previously mentioned 30~meV mode splits from the acoustic mode at R=(2.5 2.5 0.5). 

The dominating in-plane $J_4$ interaction splits the dispersive modes into two branches, such that a stronger AFM coupling lifts the upper bands. Hence, $J_4$ mostly involves precession of the spins on Fe$_1$ sites. The lower bands are not affected. The observation of the upper bands being between 58 and 83~meV points to a remarkably large antiferromagnetic $J_4$, as also proposed in previous studies\cite{Beauvois2020, LeDuc2021}.

\subsection{Anisotropies and gaps}
\label{sec:aniso}
Following Hund's rules, Fe$^{3+}$ in the high-spin configuration ($S=5/2$) has orbital moment $L=0$. However, the spin-orbit coupling $H_{\rm SOC}=\lambda \textbf{L} \cdot \textbf{S}$, where $\lambda$ is the spin-orbit coupling constant, can still act as a perturbation to the system, thus giving the spins a weak anisotropy. This mechanism can lead to both Single-Ion Anisotropy (SIA) and Dzyaloshinskii-Moriya (DM) interactions.
SIA depends on the crystal field splitting ($\Delta$): $A\propto\lambda^2/\Delta$ and it dictates the preferred orientation of individual spins. DMI is an exchange-driven effect and depends on the symmetric exchange interaction ($J$) and the Hubbard repulsion ($U$): $D\propto\lambda/U J$. It induces canted or noncollinear spin configurations. If DMI is strong, it can overcome SIA and drive weak ferromagnetism, which has previously been observed for Bi$_2$Fe$_4$O$_9$ nano-crystals\cite{Udod2024}, where the smaller the particle, the larger the FM tendencies. In contrast, larger single crystals do not exhibit weak FM\cite{Zatsiupa2013}. 

\paragraph{Single-ion anisotropies (SIA).}

From the CAMEA data, we see a double gap at low energies, suggesting that there are two anisotropies in the system. This could mean that the two Fe sites have different anisotropies. 
We estimate the direction of the anisotropy of the Fe sites by considering the individual complexes and whether they contain a symmetry axis. For the octahedral Fe$_2$ we propose the uniaxial axis to be along the longest distorted direction. For the tetrahedral complex, the direction is less clear. However, from the crystal structure of Bi$_2$Fe$_4$O$_9$, we propose the uniaxial axis to be where the two Fe$_1$ (tetrahedral) face each other and are coupled through $J_4$. Both are drawn in Fig.~\ref{fig:structure}a.

\begin{figure}[H]
    \centering
    \includegraphics[width=1\linewidth]{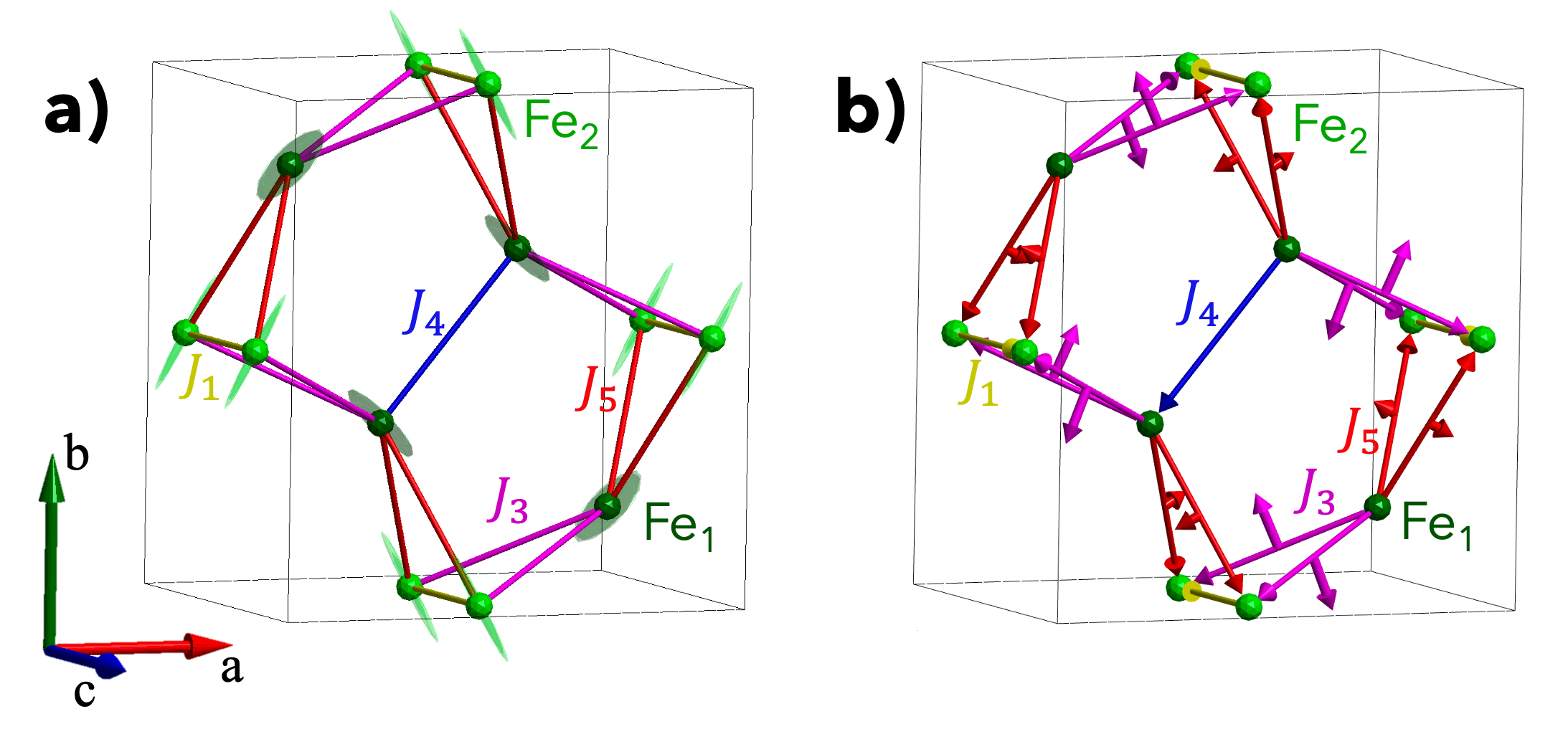}
    \caption{Anisotropies discussed in the text. \textbf{a)} Shows SIA, easy-plane for Fe$_1$ (dark green) and easy-axis for Fe$_2$ (light green). \textbf{b)} Shows implemented DMI for $J_3$ (pink) and $J_5$ (red). The magnitude for \textbf{D} is 10\% of the exchange interaction, thus the DMI on $J_3$ is almost twice as big as that on $J_5$.}
    \label{fig:DMI_SIA}
\end{figure}

We tested all nine combinations of axial, planar or zero anisotropy for each of the two different iron sites. We found that if one of the Fe sites has an axial anisotropy, the dispersion becomes gapped, but the bands do not split. If one Fe site has a planar anisotropy, the bands split, but the lowest band is not gapped. Since we see both effects, we conclude that a combination of the two anisotropies is required. The combination, Fe$_1$ being easy-plane and Fe$_2$ being easy-axis anisotropic, gives an optimized spin structure close to the one found by neutron diffraction\cite{Ressouche2009}. It is stable against rotation of all spins around the $c$-axis and with respect to a small rotation of the Fe$_1$ against the Fe$_2$. The dispersion is gapped, and the modes are split as observed at position R=(2.5 2.5 0.5). These anisotropies are indicated in Fig.~\ref{fig:DMI_SIA}a. 

\paragraph{Dzyaloshinskii–Moriya interaction (DMI).}
The DMI is allowed on some bonds by symmetry and we investigate the effect of adding the DMI term, $\textbf{D}_{ij}\cdot(\textbf{S}_i \times \textbf{S}_j)$, to the Hamiltonian in eq. \eqref{eq:Hamiltonian}. However, including the DM interaction does not improve our model, which we hereby explain.  

In Bi$_2$Fe$_4$O$_9$, DMI is only allowed by symmetry on the $J_3$ and $J_5$ interactions. 
The DMI-vector (\textbf{D}) is defined as the vector perpendicular to the plane created by the bonds in the super-exchange interaction (Fe$_1$-O-Fe$_2$), see Fig.~\ref{fig:DMI_SIA}b. We observe that for both $J_3$ and $J_5$, \textbf{D} mainly lies in the $ab$-plane, with a small component along $c$. Common for both $J_3$ and $J_5$, is that they connect Fe$_1$ with one of the spins in the pair of Fe$_2$ spins, where the pair is connected by $J_1$. For the pair, \textbf{D} are in opposite directions in the $ab$-plane, but in the same direction for the $c$-axis. The net DMI contribution (the sum of all \textbf{D}'s) for the unit cell always equals zero, as expected for a centrosymmetric space group. In a mean field consideration, one would thus expect DMI to have a relatively small influence on the overall dispersion, which is in agreement with our modelling. We can fit the data both with or without DMI. 
When having five exchange interactions and DMI on $J_3$ and $J_5$ in the model and no SIA, the spin configuration in the ground state does not converge if the magnitude of $D$ is too large ($>7$\% of the magnitude of the exchanges). Below this threshold, the dispersion is extremely similar to the one for $D = 0$ and no gap opens at R=(2.5 2.5 0.5). The DMI perturbation in the Hamiltonian adds zero energy, due to the weak FM $J_1$ between the Fe$_2$ pair. From this, we conclude that the gap opens due to the SIA on the two iron sites and not due to the DMI. 

One observation that we cannot explain with the model of the five exchange interactions and the SIA model is the splitting of the 40~meV modes at S$^*$=(2.5 2.5 1) from the 4SEASONS data (Fig.~\ref{fig:40meV_mode}).
Adding the SIA to the model with DMI stabilizes the structure, and larger DMI can be allowed for the spin configuration in the ground state to converge. However, due to the FM cancellation, even for large values, e.g. 50\% of $J_3$ and $J_5$ magnitude, DMI also here has minimal influence on the dispersion. DMI can create a splitting of the 40~meV band, however, the dispersion is opposite the data along L. In H and K, where in the data the dispersion is flat, the model is also dispersing. 

To summarize, from the experimental results we find no reason to add the DMI in the final model, thus, we have chosen to operate with the five exchange interactions and two SIA. We cannot exclude that a combination of other parameters could reproduce the experimental features, but from the extensive tests performed in this study (see appendix \ref{app:40meV_tests}), no such combination was found.

\subsection{Parameters of the Hamiltonian}
\begin{figure*}[h]
    \includegraphics[width=\textwidth]{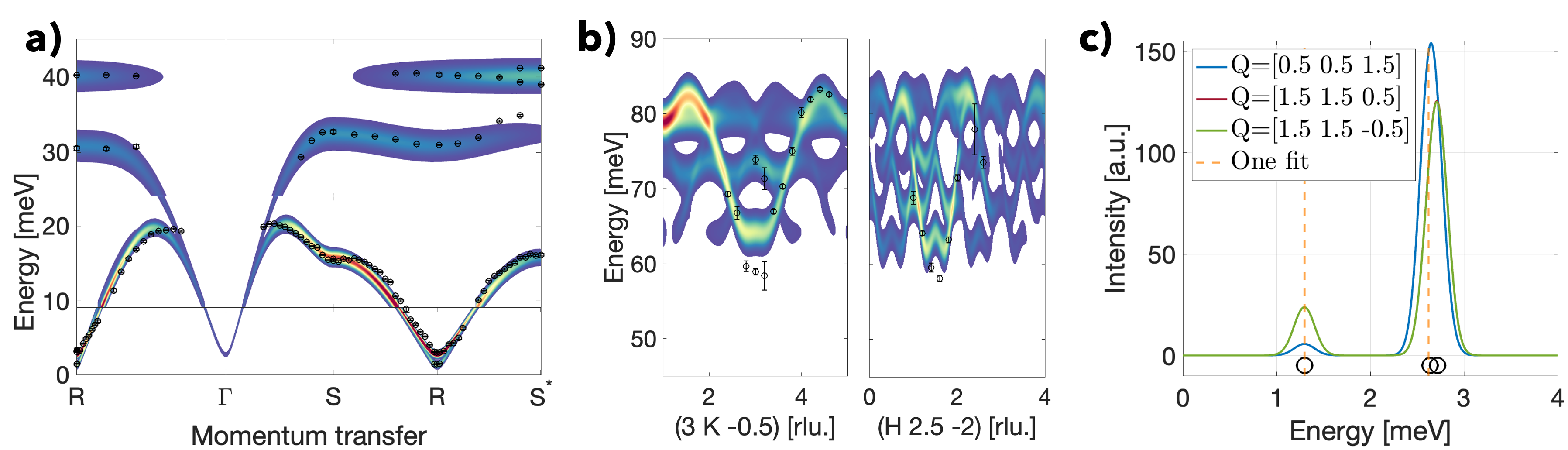} 
    \caption{The final spin wave model. \textbf{a)}: The model with the same cuts as the data in Fig.~\ref{fig:data_colorplots}b and the same black errorbars, except the 30 meV in R-$\Gamma$ and the 40~meV mode, which comes from Fig.~\ref{fig:data_colorplots}d. The energy ranges 0-9, 9-24, 24-45~meV have been plotted with respective energy resolutions of 0.5, 1 and 2~meV. \textbf{b)}: The model of the upper branch with the same cuts shown in the data in Fig.~\ref{fig:data_colorplots}c and the same black errorbars. \textbf{c)}: Constant Q-cut at three different magnetic Bragg peak positions, similar to the data in Fig.~\ref{fig:double_peak_1Dcut}b (fitted experimental gap values are plotted in yellow dashed lines), plotted with an energy resolution, $\Delta E=0.11$~meV. The mode positions from the model are shown in empty black rings below the peaks. }
    \label{fig:spinW}
\end{figure*}

\begin{figure}[h]
    \includegraphics[width=0.85\linewidth]{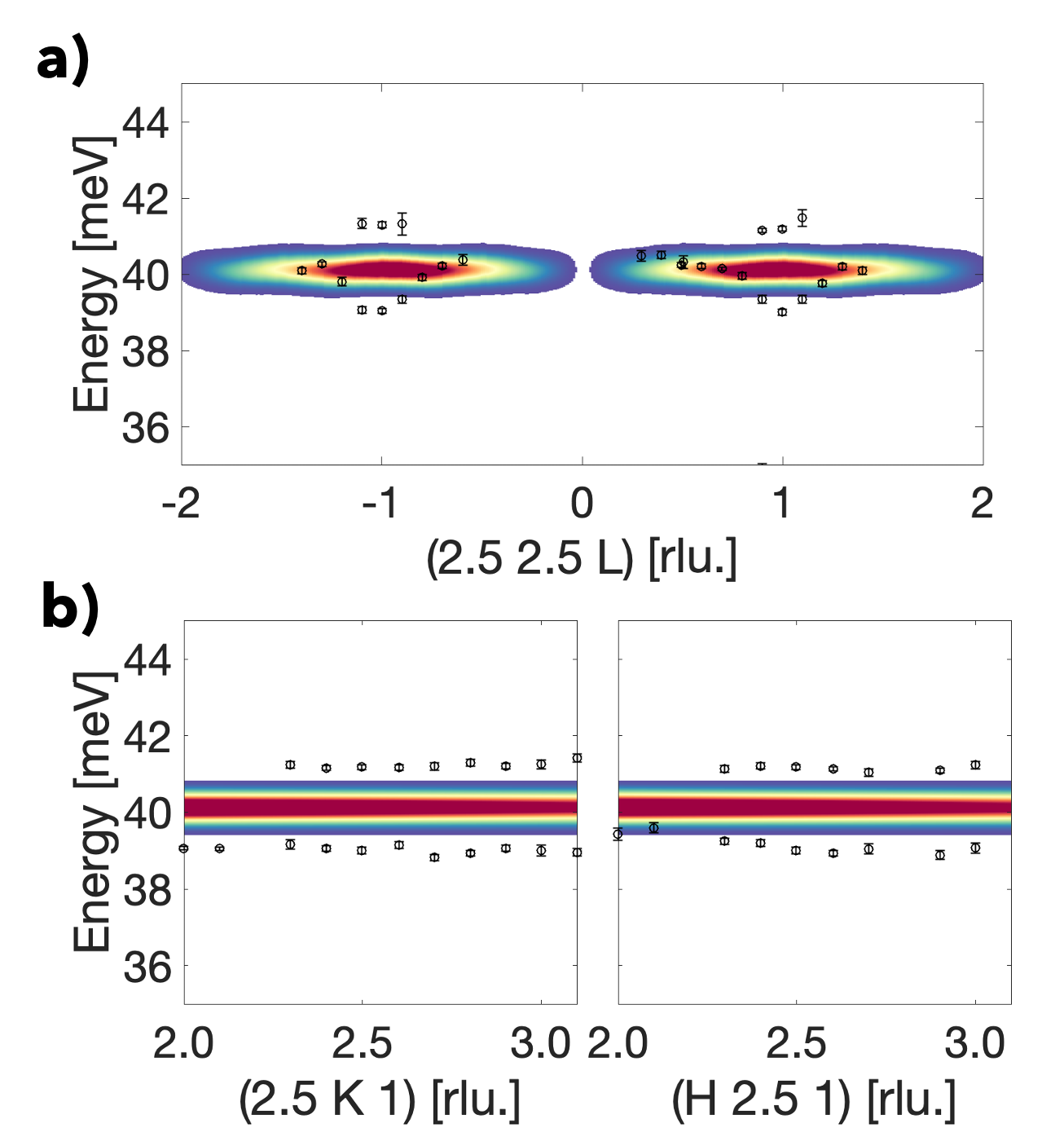} 
    \caption{Spin wave model description of the 40~meV mode \textbf{a)}: Model along L with H=K=2.5, like Fig.~\ref{fig:40meV_mode}b, with the fitted data in black error bars. \textbf{b)}: Model at L=1 along K (left panel) and H (right panel), similar to the data in Fig.~\ref{fig:40meV_mode}c.}
    \label{fig:spinW_40meV}
\end{figure}

We now optimize the spin wave model to the data shown in Fig.~\ref{fig:data_colorplots} and Fig. \ref{fig:40meV_mode} by varying $J_1-J_5$ and the magnitude of the SIA, $A_1$ and $A_2$. The final spin wave dispersion is shown in Fig.~\ref{fig:spinW} and Fig.~\ref{fig:spinW_40meV}. We have iteratively fitted the parameters and optimized the ground state magnetic structure to the new parameters. 

We have performed the following fitting procedure: We first fitted all parameters until convergence, then we used these parameters to optimize the magnitude of the anisotropies to get the experimental gap sizes. Lastly, with the fixed anisotropies, the exchange interactions were fitted until convergence. This yielded the final model presented here. The optimized exchange and anisotropy parameters are given in Table \ref{tab:interactions}.

In Fig.~\ref{fig:spinW}a-b, we show the spin wave model along with the experimental data points and their statistical uncertainty. We find a very good overall agreement across the whole energy range 0-85~meV.
In Fig.~\ref{fig:spinW}a the model accurately reproduces the dispersion, but small differences are observed. The most prominent is that the slope of the dispersion around R=(2.5 2.5 0.5) is higher in the fitted experimental data than in the spin wave model. Looking at the individual energy scans of the relevant data (the EIGER data), asymmetric resolution effects artificially cause an increase of the fitted mode position. 
The high energy modes in Fig.~\ref{fig:spinW}b match the experimental data (Fig. \ref{fig:data_colorplots}c) well, however, looking at the left panel it slightly overestimates the energy of the bottom of the higher branch, around 60~meV. The wide \textbf{Q}-integration range used to obtain this data could very well cause the bottom of the mode to be more diffuse and thus less precisely fitted. 

Fig.~\ref{fig:spinW}c shows the gaps at three different values of \textbf{Q} (like shown in Fig.~\ref{fig:double_peak_1Dcut}b); the experimental found positions are shown with vertical yellow dashed lines. The mode positions found in the model are at 1.30, 2.64 (doubly degenerate), and 2.72~meV (plotted as black empty circles). Comparing the experimental and modelled peak positions, the model finds the energy of the low energy mode, but is slightly overestimating the energy of the high energy modes. The intense peak in the data at 2.64~meV turns out to represent three close lying modes. 
For the two peaks in the model, the intensity ratio is 5.3 for red/green (intensity of 2.64 meV mode divided by intensity of 1.30 meV mode) and 27.6 for the blue.
This is much larger ratios than what is observed in the data, 1.7 for the red/green and 6.3 for blue.
However, since SpinW does not use the intensity in the fitting routine, inconsistencies of this type are expected. A small adjustment of the anisotropies can recover the observed intensity ratio at the expense of a slight shift of the lowest gap value. 

The 40~meV mode, which in the data is shown to soften and split, is not well reproduced in the spin wave model, seen in Fig.~\ref{fig:spinW_40meV}. Our attempts to do so have included SIA, DMI (mentioned above), higher order SIA, introducing further interactions, dipolar interaction, and allowing for asymmetric exchange interactions, see appendix \ref{app:Frida}. None of these were able to reproduce the 40~meV mode split correctly. It has been possible to split the mode, but not to obtain the correct dispersion. In our final model, the 40~meV mode is degenerate and essentially flat, with a small splitting of approximate 0.1~meV. 
   
\subsection{Polarisation}
\begin{figure}
    \centering
    \includegraphics[width=1\linewidth]{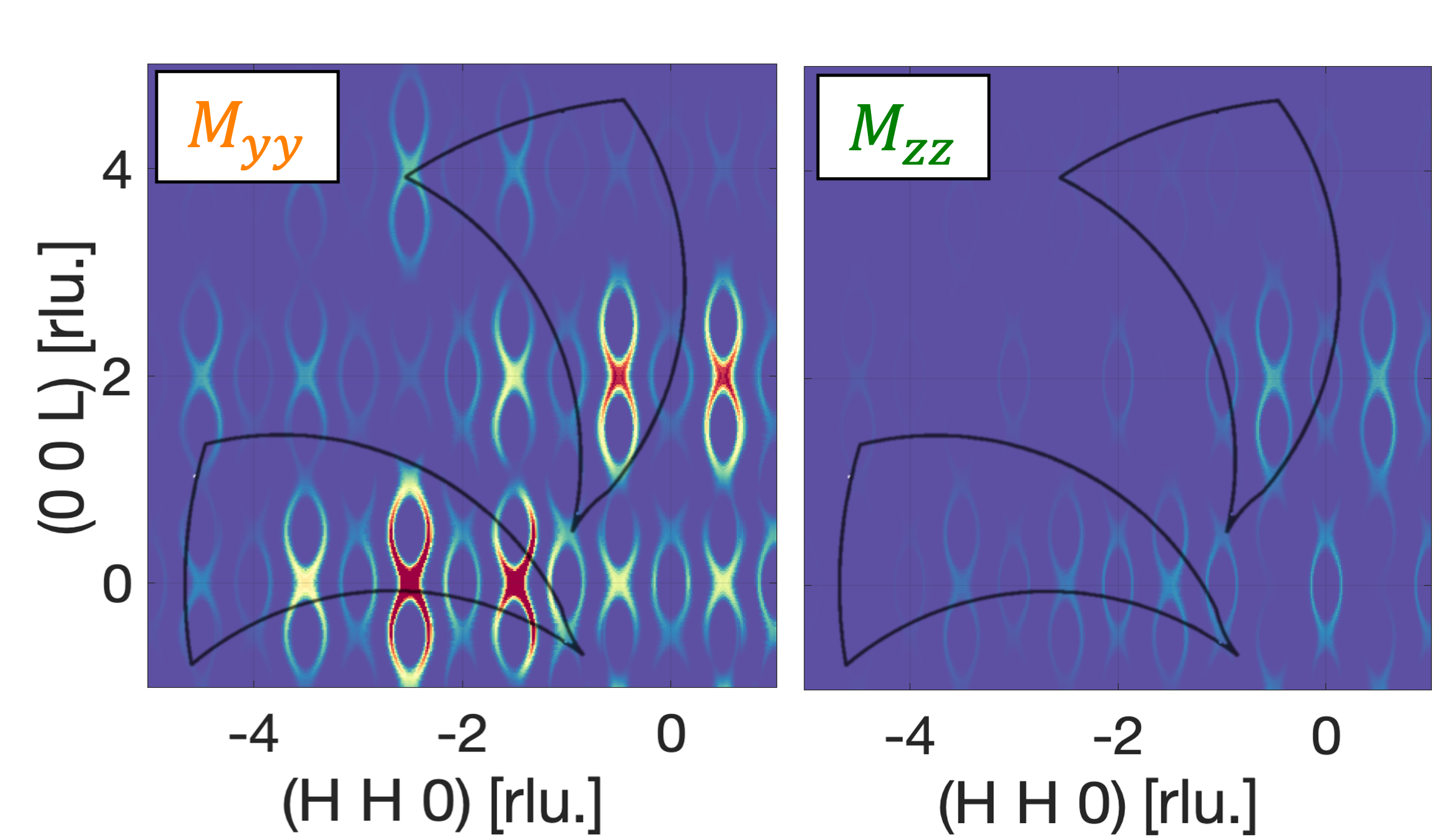}
    \caption{Spin wave model of polarised constant energy maps at 15.5~meV in scattering plane [HH0]-[00L] for $M_{yy}$ and $M_{zz}$. The experimental covered area, from Fig.~\ref{fig:pol_maps}b middle panel, are indicated by the black contours.}
    \label{fig:pol_maps_spinW}
\end{figure}

The behaviour of the polarised experimental data in Fig.~\ref{fig:pol_maps}a is well captured by the spin wave model, e.g. at 15.5~meV seen in Fig.~\ref{fig:pol_maps_spinW}. At lower energies $M_{yy}$ dominates, but at higher energies this signal is weaker and $M_{zz}$ dominates. \\
By exploring the directions of the fluctuating magnetic amplitudes along the $a$, $b$ and $c$-axis in the spin wave calculations, we have observed that the low energy spin waves have magnetic amplitudes that are orthogonal to all ordered moments simultaneously, i.e., perpendicular to the $ab$-plane, see appendix \ref{app:pol}. Summing, the calculated intensities coming from amplitudes in the $ab$-plane at 10~meV energy transfer (in the low-energy range), we find that they are half the size of the amplitudes along $c$. This is as expected. We see, however, a large difference between amplitudes along $a$ and $b$, with $a$ being very dominant at most \textbf{Q}-positions at low energies. Given the easy-axis anisotropy of Fe$_2$ has a main contribution along $b$, the Fe$_2$ spins prefer to fluctuate in $a$ and $c$ for the low energies. 

Additionally, the model shows that not all spins need to share $c$ as one of their easy-plane directions in order to reproduce an intensity pattern observed in Fig. \ref{fig:pol_maps}b. Thus, the anisotropic amplitudes are seemingly related to the co-planarity of the canted spin-structure. An anisotropy that generates this type of co-planar canted spin structure will share the same pattern of $M_{yy}$ being dominant for low energies and $M_{zz}$ for high energies. 

\subsection{A potential continuum} \label{sect:model_continuum}
In section \ref{sec:data_continuum} we showed that at least some of the apparent continuum in the 9-35 meV range (see Fig.~\ref{fig:data_colorplots}b, Fig.~\ref{fig:double_peak_1Dcut}, Fig.~\ref{fig:continuum} and Fig. \ref{fig:diff_integration_4SEASONS}) is caused by resolution effects. However, it was unclear if any of the scattering was intrinsic to the system. We now return to approach this question more quantitively.

\begin{figure}
    \centering
    \includegraphics[width=1\linewidth]{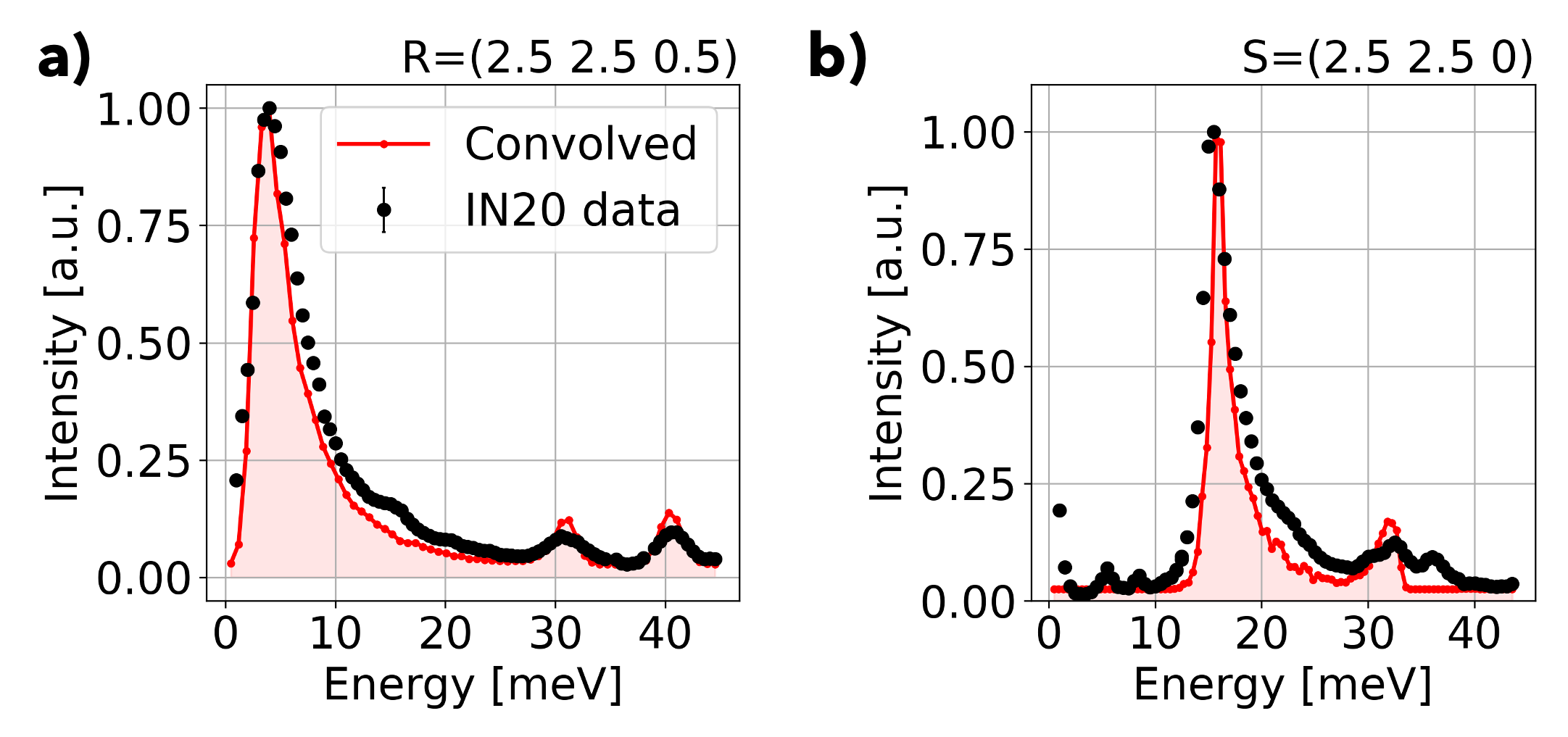}
    \caption{Convolution of the spin wave models with instrument resolution of IN20 at two different \textbf{Q}-values; \textbf{a)}: R=(2.5 2.5 0.5) and \textbf{b)}: S=(2.5 2.5 0). The figures are comparable to the cuts shown in Fig.~\ref{fig:continuum}.}
    \label{fig:convoluted}
\end{figure}

We use Takin\cite{Takin2023} to calculate the convolution of our spin wave model with the resolution of IN20 at two different \textbf{Q}-values; R=(2.5 2.5 0.5) and S=(2.5 2.5 0) in the energy range 2-45~meV, see Fig.~\ref{fig:convoluted}. The convoluted spectra describe most of the tail of scattering present in the apparent continuum.
The exact width of the tail is not reproduced, implying that there may be weak additional scattering. However, this result, combined with the comparison of the different instruments (Fig.~\ref{fig:continuum}), varying the integration width of the 4SEASONS data (Fig. \ref{fig:diff_integration_4SEASONS}) and the fitting of the gaps (Fig.~\ref{fig:double_peak_1Dcut}b), indicate that the majority of the apparent continuum scattering is simply a resolution effect. 

\section{Discussion and conclusion}

\subsection{Spin wave model}
As previously mentioned, two groups; Duc Le et al.\cite{LeDuc2021} and Beauvois et al.\cite{Beauvois2020}, studied the magnetic excitations of Bi$_2$Fe$_4$O$_9$. Both groups performed INS measurements of $\sim 0.6$ g single crystals in the range 0-35~meV and Duc Le et al. also performed powder INS in the energy range 0-90~meV. They report a flat magnetic mode in \textbf{Q} at different energies, resulting in 
different signs on their values of the $J_1$ exchange interaction. We have measured the whole excitation spectrum on a 2.35 g single crystal. Comparing the model parameters found in previous papers\cite{LeDuc2021, Beauvois2020} (see Table \ref{tab:interactions}), we find very good agreement with especially Ref.~\onlinecite{LeDuc2021}. We confirm that $J_1$ is FM and produces a flat band at 40~meV, and that $J_4$ is the dominant exchange interaction. 

With the reported exchange parameters, we can analytically calculate the Curie Weiss temperature\cite{Curie_Weiss} for Bi$_2$Fe$_4$O$_9$:
\begin{align}
    \theta_{CW} & = \frac{S(S+1)}{3k_B} (4J_1 + 4 J_2 + 16 J_3 + 4J_4 + 16 J_5) \\
     & \approx -1137\text{ K}. 
    \label{eq:curie_weiss_eq}
\end{align}
    
This is significantly smaller than the previous reported experimental values\cite{Ressouche2009, Zatsiupa2013}. In both these cases, the susceptibility data appears to have been fitted below the Curie Weiss temperature, $\mid \theta_{CW} \mid$, thus, in a range where the Curie Weiss law is not valid. It is not possible to measure the susceptibility above, because the crystal will start melting for $T>900$ K. Therefore, estimating $\theta_{CW}$ from the spin wave model is required. Our values imply a frustration index of $f=4.6$, somewhat lower than previously reported.

With polarised INS, we observed the fluctuations to be anisotropic. Using the spin wave model in eq.~\eqref{eq:Hamiltonian}, we find that the low-lying modes fluctuate mainly along the out-of-plane ($c$) direction, with weaker anisotropic fluctuations in the $ab$-plane.

In contrast to previous reports, we have discovered that the dispersion is doubly gapped. This cannot be described by the previously proposed $ab$-planar anisotropy for all sites. We determine a planar SIA of the Fe$_1$ and an axial SIA of the Fe$_2$ sites. From the resulting spin wave model, we have 3 modes instead of the two we report from the data. However, since two of them are very close in energy, we were not able to convincingly fit 3 modes to the data. 

While our spin wave model describes the overall dispersion on all energy scales, it does not match the dispersion extracted from the data perfectly. Likewise, while the model reproduces the overall momentum dependence of intensity across the different modes, the intensity ratios are not exactly matched. This indicates that further refinement of the model could be possible by allowing additional free parameters. For instance, the approximation of having just a planar or axial SIA may be too simplistic. Allowing the SIA to be ellipsoidal might yield better agreement. 

Previously, a single dispersionless mode\cite{LeDuc2021} was reported around 40~meV. In contrast, we uncover that this mode is split and dispersive.
We have not been able to reproduce this observation in our spin wave model. Based on extensive tests of different additional terms in the Hamiltonian (NNN exchange interactions, DMI, dipolar interactions and introducing symmetric and asymmetric exchange interactions on the allowed matrix elements given by symmetry), we exclude that any of these give rise to the observed magnon behaviour. Another possibility is that the dispersion is caused by hybridisation between the magnon mode and a phonon mode. From a polarised Raman-scattering study of Bi$_2$Fe$_4$O$_9$ single crystals, first-order Raman phonon lines were identified and assigned to definite atomic motions, e.g. the mode $A_g(5)$ at 322 cm$^{-1}$ (39.9~meV) at 12 K, which corresponds to atomic motion of Fe$_2$ along the $c$-axis\cite{Iliev_2010}. These magnetic ions are connected through the $J_1$ exchange, which determines the energy of the 40~meV modes. Thus, it is very likely that the eigenmodes for this phonon are relevant for the magneto electric coupling giving rise to a magnon-phonon hybridization. Such magneto-electric coupling may be relevant to account for the possible magneto-dielectric and multiferroic behaviour of this compound, reported in Refs. \onlinecite{Singh_2008_magnetoelectric, Roy_2025, Ressouche2009}.

\subsection{Complex dynamics}
To summarize, we have investigated the observed scattering above the low-lying spin wave branches and studied whether it is a physical effect, such as a continuum, or a resolution effect of the instruments. We have compared the different instruments (TAS and TOF) at constant-Qs, we have studied different integrations on the TOF data, convoluted the IN20 (TAS) resolution function onto our spin wave model and fitted the energy gaps with an approximate resolution function. All these results describe most of the large tail of scattering above the spin waves. For all the tests, however, there seem to be a little more scattering present than what is described by instrumental resolution. It seems there could be a weak continuum, which is also what we expect from two-magnon scattering. 
This type of scattering will be polarised differently than spin wave scattering and could be probed through polarised neutron scattering. We attempted this, by measuring at 10~meV in two polarisation channels, but did not have sufficient resolution nor statistics to conclude anything regarding the continuum. Details can be found in 
appendix \ref{app:pol_con}.

There is a large interest in the determination of non-trivial quantum states of matter expected for geometrically frustrated compounds. 
Experimental signatures often reported as potential evidence for such state and their concomitant quantum fluctuations include absence of signatures of order in susceptibility and neutron diffraction and broad excitation continua in INS\cite{Han_2012, Lake_2005}. 
However, several of these signatures can also be observed in classical systems\cite{Lane_2025}, and it is, therefore, important to build an understanding of them.

In Bi$_2$Fe$_4$O$_9$, it is not yet clear whether more complex dynamics are co-existing with magnon modes at 10 K. For the paramagnetic phase, one could expect the dimers ($J_4$ interaction) to be strongly correlated well above the ordering temperature in a dimerised paramagnetic state\cite{Beauvois2020, LeDuc2021}.
Indeed, neutron diffraction from powder samples is indicative of short range correlations above $T_N$\cite{Singh_2019}, and analysis of polarised neutron Bragg scattering is consistent with dimerization on the Fe$_1$ sites\cite{Beauvois2020}. 
Armed with a detailed magnetic Hamiltonian it will be possible to explore theoretically and experimentally how the system evolves from such local correlations to a complex non-collinear ordered state and whether the resulting spin waves are accompanied by other types of fluctuations of a more diffuse nature.

\acknowledgments
We thank Mahn Duc Le for valuable discussions. We would like to acknowledge technical assistance from the 3He group at ILL: D. Julien, P. Chevalier, S. Batisse, and N. Thiery. \\
This project was supported by the Danish Committee for Research Infrastructure through DANSCATT and the ESS Lighthouse QMAT.  This work was supported by the Swiss National Science Foundation (SNSF) project 200021-228473 - Quantum Magnetism and Spectroscopy. Andrea Kirsch gratefully acknowledges the Deutsche Forschungsgemeinschaft (DFG, German science foundation) for funding of the project Ki 2427/1-1 (\#429360100).

Data from SINQ were acquired on CAMEA (proposal ID: 20212721) and EIGER (20221533). Data from ILL were acquired on IN20 (TEST-3263, 4-05-908) and from J-PARC on 4SEASONS (2023B0095). The data that support the findings of this article are openly available\cite{CAMEA_data, EIGER_data, IN20, JPARC_data, IN20_pol}, embargo periods may apply.

\bibliography{Bi2Fe4O9}
\clearpage
\twocolumngrid  
\clearpage
\onecolumngrid

\appendix
\newpage
\section*{Appendix}

\section{Fitting a double gap}
\label{app:double_gap_fit}

\subsection{Approximated resolution function for fitting gap value of AFM}
\label{app:double_gap_fit_function}
When performing a Triple-Axis Spectrometer (TAS) neutron scattering experiment, we do not measure an ideal delta-function signal, because the signal is being broadened by the instrument. This is described by the resolution function $G(\textbf{Q}-\textbf{Q}', \omega-\omega')$, which for a TAS instrument is a multi-dimensional Gaussian that accounts for instrumental contributions to the measured signal in momentum and energy space. The measured intensity in a TAS experiment is a convolution of the sample's intrinsic scattering function ($S(\textbf{Q},\hbar\omega)$) with the instrument resolution function:
\begin{equation}
    I(\textbf{Q},\hbar\omega) = \int G(\textbf{Q}-\textbf{Q}', \hbar\omega-\hbar\omega')S(\textbf{Q}',\hbar\omega') \text{d}\textbf{Q}' \text{d}\omega'.
    \label{eq:intensity}
\end{equation}
The resolution of an instrument depends on multiple factors; the incident and final neutron wave vectors $\textbf{k}_i$ and $\textbf{k}_f$, the mosaic spreads of the monochromator and analyzer, the collimation settings (pre-monochromator, between monochromator/sample, etc.), the instrument geometry (distances, scattering angles). The energy resolution is also affected by the spread in wavelength of the neutrons. 

A gapped AFM magnon dispersion can often be approximated by
\begin{equation} 
    E(\textbf{Q})=\sqrt{(a(\textbf{Q}-\textbf{Q}_0))^2+\Delta^2},
    \label{eq:AFM_dispersion}
\end{equation}
where $a$ is the spin-wave velocity (the slope of the dispersion at \textbf{Q} values away from the gap region) and $\Delta$ is the energy gap at the bottom of the dispersion centred at $\textbf{Q}_0$. Thus, at the bottom, the dispersion follows a parabola; $E(\textbf{Q})=\alpha(\textbf{Q}-\textbf{Q}_0)^2+\Delta$, where $\alpha=a^2/(2\Delta)>0$ determines the shape of the parabola. 

We are interested in fitting the gaps of the dispersion, and for this an approximate function can be used, described in Ref. \onlinecite{fit_gap_paper}. Here, it is assumed that the resolution in one \textbf{Q}-direction is very good and can be described by a delta-function, while in the other two, we have a Gaussian resolution with the same width, $\sigma_Q$. The energy resolution, $\sigma_E$, is also assumed to be Gaussian:
\begin{equation} \label{eq1}
    I(\mathbf{Q}, \hbar \omega) = \int \exp \left( -\frac{|\hbar\omega' - \Delta| \Delta}{a^2 \sigma_Q^2} - \frac{(\hbar\omega - \hbar\omega')^2}{2 \sigma_E^2} \right) \cdot \frac{\pi}{\alpha} \, \text{d}\omega'
\end{equation}
This numerical function can be used to fit a constant \textbf{Q}-scan, where intensity has been measured as a function of energy transfer, $\hbar \omega$.

For Bi$_2$Fe$_4$O$_9$ we observe 2 gaps in the CAMEA data, see Fig.~\ref{fig:double_peak_1Dcut}. Thus, the approximated function becomes: 
\begin{equation}
    f(\hbar \omega) = 
\begin{cases}
A_1 \exp\left( -\dfrac{|\hbar \omega - \Delta_1| \Delta_1}{(a \sigma_Q)^2} \right), & \hbar\omega > \Delta_1 \\
0, & \hbar\omega \leq \Delta_1
\end{cases}
+ 
\begin{cases}
A_2 \exp \left( -\dfrac{|\hbar\omega - \Delta_2| \Delta_2}{(a \sigma_Q)^2} \right), & \hbar\omega > \Delta_2 \\
0, & \hbar\omega \leq \Delta_2, 
\end{cases}
\end{equation}
where $A_1$ and $A_2$ are normalisation constants. The function is constrained not to contribute to intensity below the gaps at $\Delta_1$ or $\Delta_2$, respectively. 
This function is convoluted with the energy resolution, resulting in the function used for fitting the gaps: 
\begin{equation}
    I(\textbf{Q},\hbar\omega) = \int f(\hbar\omega')  \exp \left( - \frac{(\hbar\omega - \hbar\omega')^2}{2 \sigma_E^2} \right) \text{d}\omega'+B,
\end{equation}
where $B$ is a constant background. 

\subsection{Fitting the energy gaps in Bi$_2$Fe$_4$O$_9$}
\label{app:double_gap_fit_final}

\begin{figure}[h]
    \centering
    \includegraphics[width=0.8\linewidth]{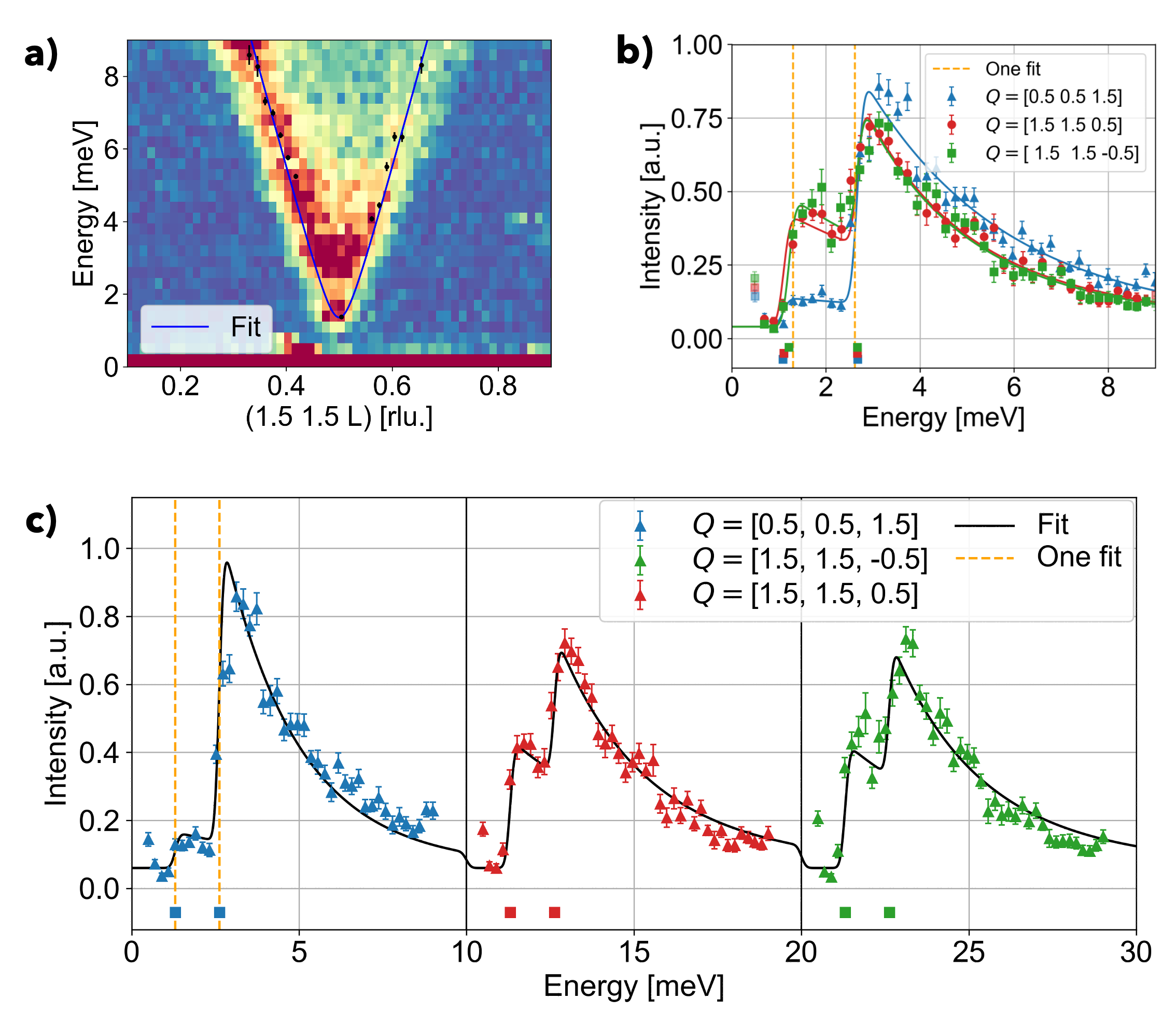}
    \caption{Fitting the two energy gaps: \textbf{a}) Zoom in of the CAMEA data in Fig. \ref{fig:double_peak_1Dcut}a and the slope $a=54$ meV/Å$^{-1}$ plotted on top (blue) at $\textbf{Q}_0$=(1.5 1.5 0.5). \textbf{b}) Fitting the three \textbf{Q}-positions separately. \textbf{c}) Fitting the three \textbf{Q}-positions simultaneously, used in Fig \ref{fig:double_peak_1Dcut}b. Both b-c, has the fitted gap positions plotted as filled squares below zero. The yellow dashed lines, are the gap positions found by the simulaneous fit. }
    \label{fig:2_gaps_1_array}
\end{figure}

The slope of the dispersion, $a$, is found by fitting a gapped AFM magnon dispersion given eq. \eqref{eq:AFM_dispersion}. We use the experimentally fitted low energy gap position and only the mode positions on the linear part of the slope (black errorbars) are used for the fit, see Fig. \ref{fig:2_gaps_1_array}a. Thus, we find $a=54$ meV/Å$^{-1}$. 
For the fitting of the energy gaps, $a$ is fixed and the energy resolution is fixed at 0.11~meV, which is found using MJOLNIR\cite{Lass2020MJOLNIR} (which is, in turn, using Takin\cite{Takin2023}). 

If there are two energy gaps, then they should be at the same energy for all three \textbf{Q}-positions; (0.5 0.5 1.5) and (1.5 1.5 $\pm$0.5). Thus, we can fit the three \textbf{Q}-positions simultaneously, assuming they all have the same \textbf{Q}-resolution, $\sigma_Q$. 
This gives the fit shown in Fig.~\ref{fig:2_gaps_1_array}c. Here $\Delta_1=1.30 \pm 0.01$~meV and $\Delta_2=2.62 \pm 0.01$~meV. The \textbf{Q}-resolution is $0.044\pm0.001$~rlu., which approximately fits with the \textbf{Q}-resolution of 0.05~rlu., which is the largest FWHM found by fitting over Bragg peaks for Bi$_2$Fe$_4$O$_9$.

We also tested fitting the three \textbf{Q}-positions separately, see in Fig.~\ref{fig:2_gaps_1_array}b. Here, the mean of the gap sizes are $\Delta_1=1.13\pm0.06$~meV and $\Delta_2=2.67\pm0.01$~meV. There is a spread on $\Delta_1$ for the different \textbf{Q}-positions, while $\Delta_2$ is very stable across the fits. Compared to the fit, where the gap sizes are linked (described above and plotted as dashed vertical yellow lines), $\Delta_1$ is underestimated and $\Delta_2$ is slightly overestimated. The mean of $\sigma_q=0.044\pm0.005$ rlu. We chose the simultaneous fit to implement in SpinW. 

\newpage
\section{Test of spin wave model}\label{app:test_spinW}

\subsection{Test of exchange parameters}
\begin{figure}[ht!]
    \centering
    \includegraphics[width=0.6\textwidth]{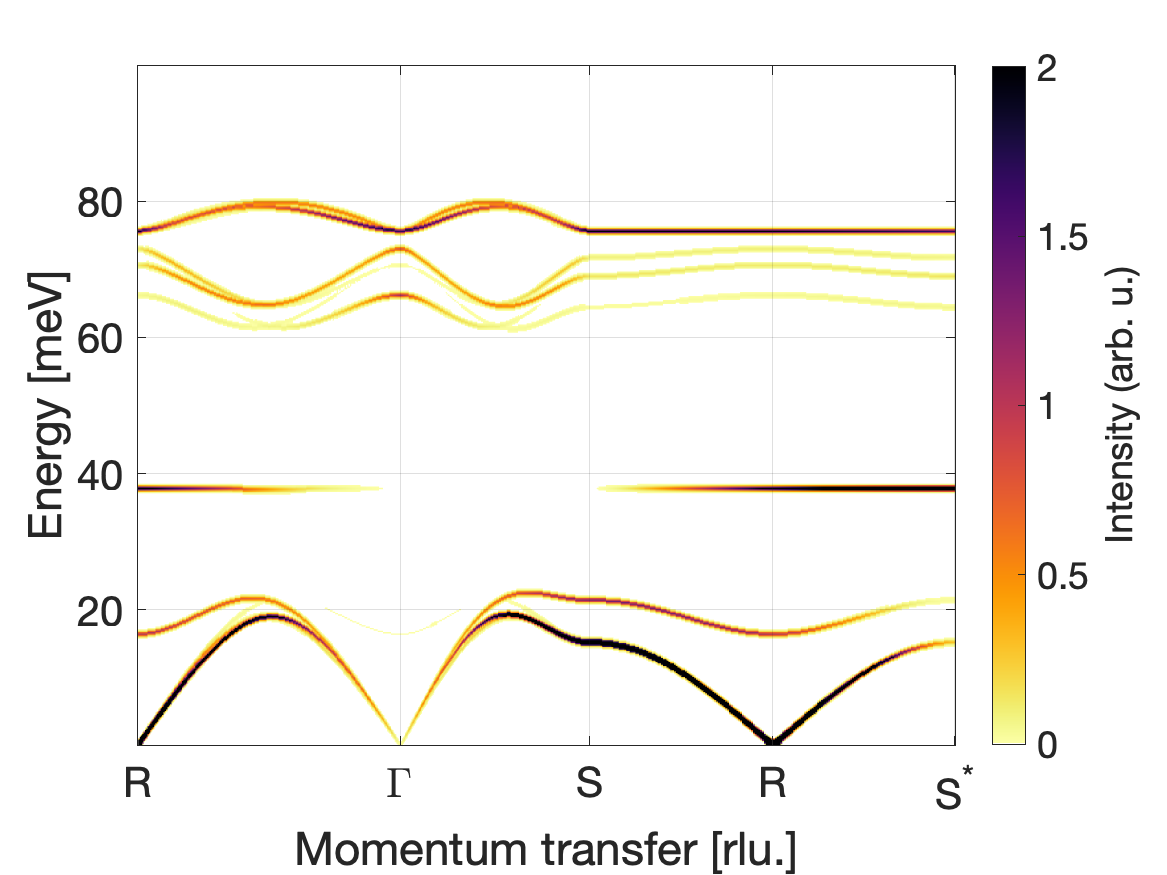}
    \caption{Spin waves in Bi$_2$Fe$_4$O$_9$ calculated using the parameters found by Duc Le et al.\cite{LeDuc2021}, but with zero anisotropy.}
    \label{fig:Duc_zero_anisotropy}
\end{figure}

In this section, we use the exchange interactions found by Duc Le et al. Ref. \onlinecite{LeDuc2021} in Table \ref{tab:interactions} as a starting point, labelled $J_{1D}-J_{5D}$ for this section. The goal is to develop an intuition about how each parameter influences the calculated spin wave dispersion before we begin fitting. The anisotropy is set to zero, such that we only look at the exchange parameters. One parameter is varied at a time, keeping all the others constant, to test how the specific exchange influence the spin wave spectrum.  The spin wave spectrum with these parameters is shown in Fig.~\ref{fig:Duc_zero_anisotropy}. Since the spin waves are present up to 80~meV energy transfer, each plot is on the energy scale 0 to 100~meV with an energy resolution of $dE=0.5$~meV. All figures are plotted on the same colour scale going from 0-2 a.u. 

\subsubsection{Influence of $J_1$}
This is the interaction between the pair of Fe$_2$ spins, one Fe just above and the other just below the pentagonal plane, within the unit cell along the $c$-axis. 

\begin{figure}[ht!]
     \centering
     \begin{subfigure}[t]{0.32\textwidth}
         \centering
         \includegraphics[width=\textwidth]{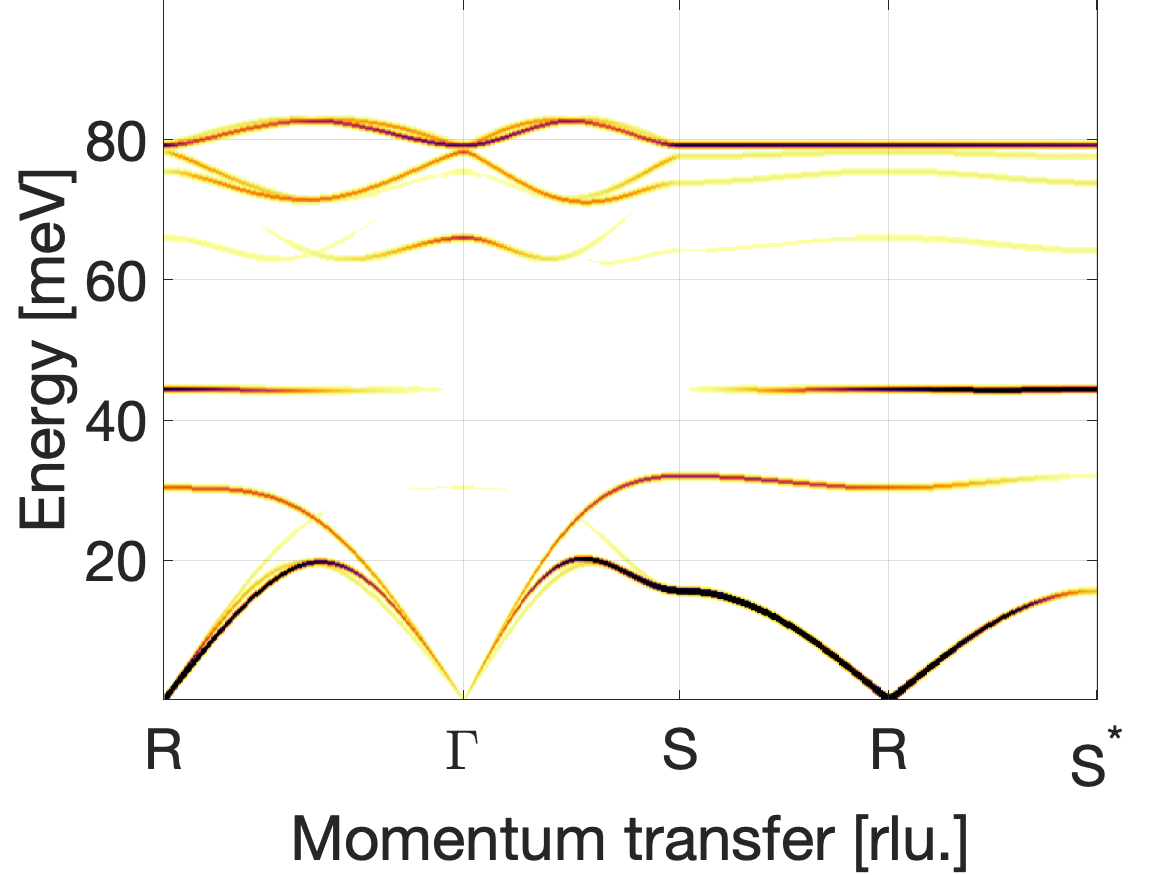}
         \caption{$J_1=-1$~meV}
     \end{subfigure}
     \hfill
     \begin{subfigure}[t]{0.32\textwidth}
         \centering
         \includegraphics[width=\textwidth]{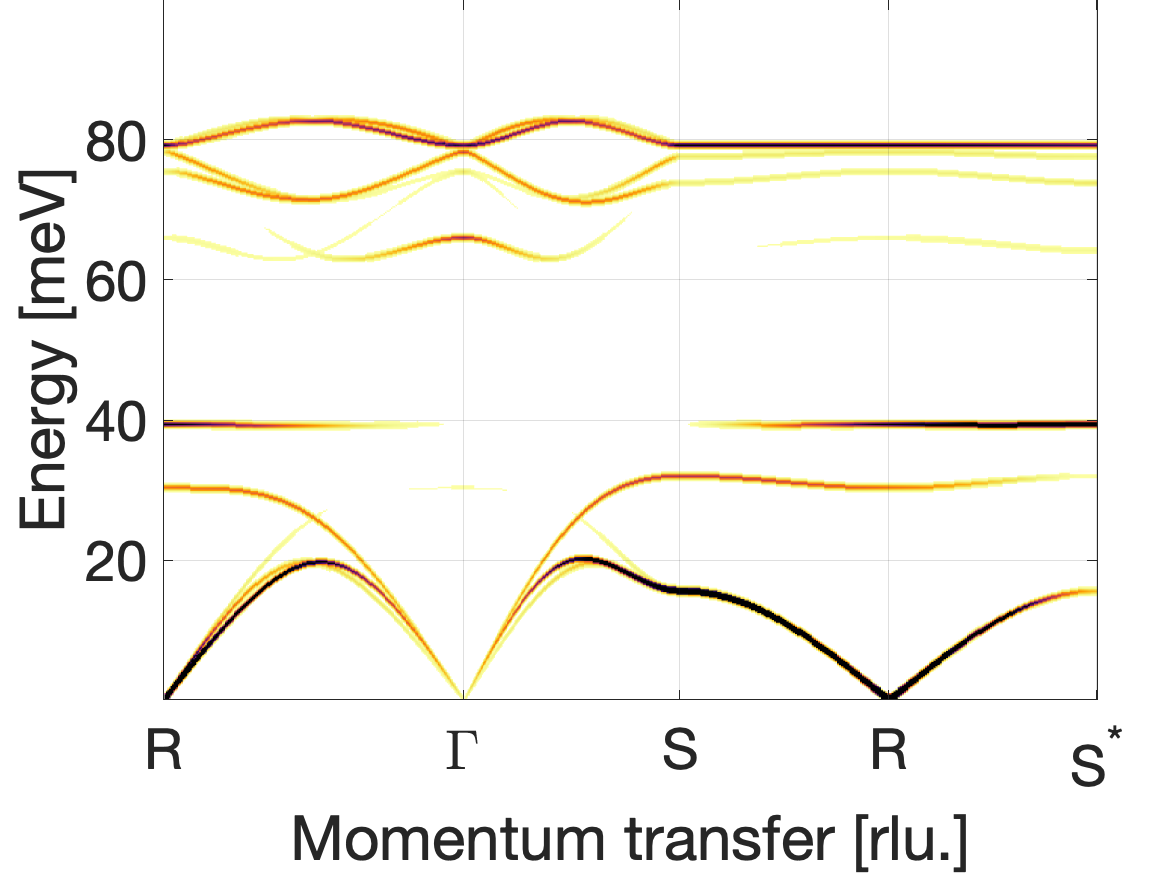}
         \caption{$J_1=0$~meV}
     \end{subfigure}
     \hfill
     \begin{subfigure}[t]{0.32\textwidth}
         \centering
         \includegraphics[width=\textwidth]{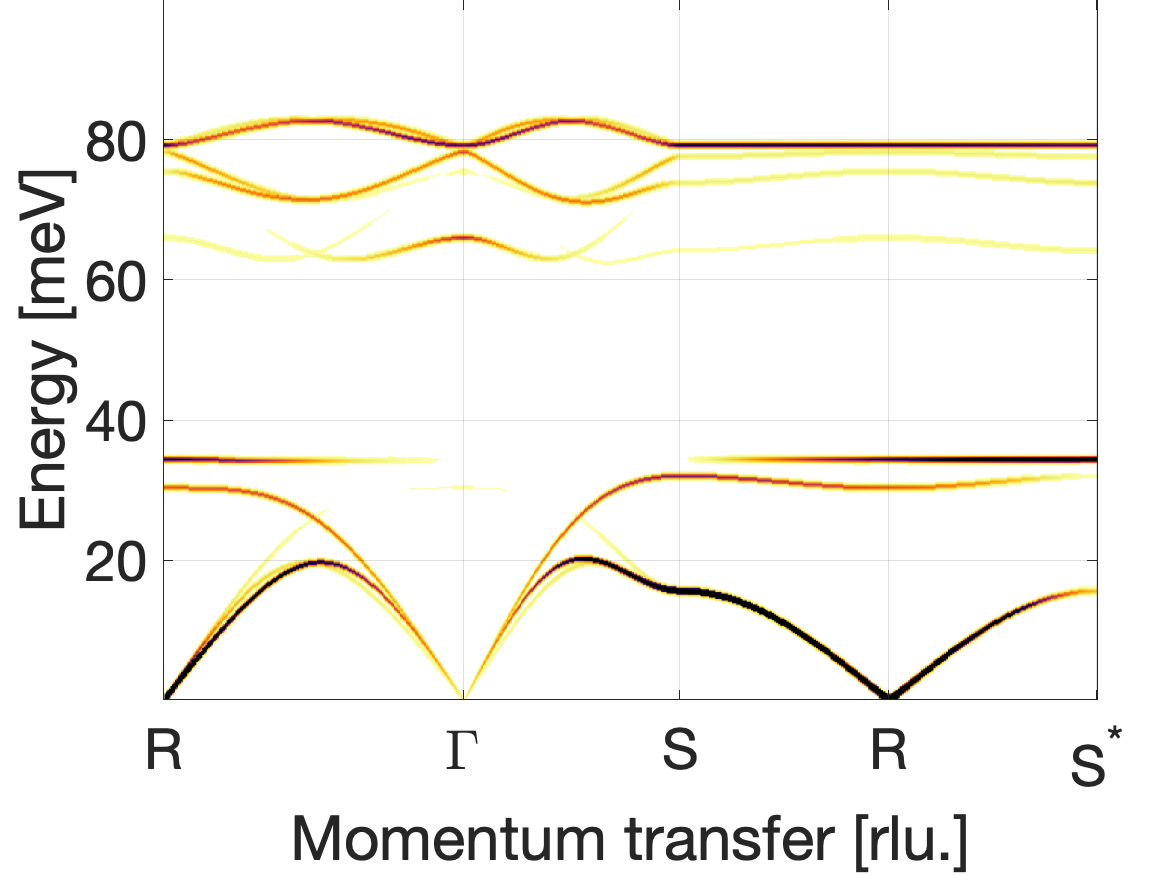}
         \caption{$J_1=1$~meV}
     \end{subfigure}
        \caption{Varying the exchange interaction $J_1$, keeping all other parameters constant.}
        \label{fig:parameter_J1}
\end{figure}

The out-of-phase precession of the FM aligned pair of spins yields a flat band at intermediate energy\cite{Beauvois2020}. Fig.~\ref{fig:parameter_J1} shows the impact of varying $J_1$: A larger FM $J_1$ exchange will cost more energy and hence push the mode up in energy. While a more AFM $J_1$ will push it down in energy.

Duc Le et al. reported a flat band at 40~meV, while Beauvois et al. reports a flat band at 19~meV. This results in them having different signs on the value of the $J_1$ exchange interaction. 
\FloatBarrier

\subsubsection{Influence of $J_2$}
\begin{figure}[ht!]
     \centering
     \begin{subfigure}[t]{0.3\textwidth}
         \centering
         \includegraphics[width=\textwidth]{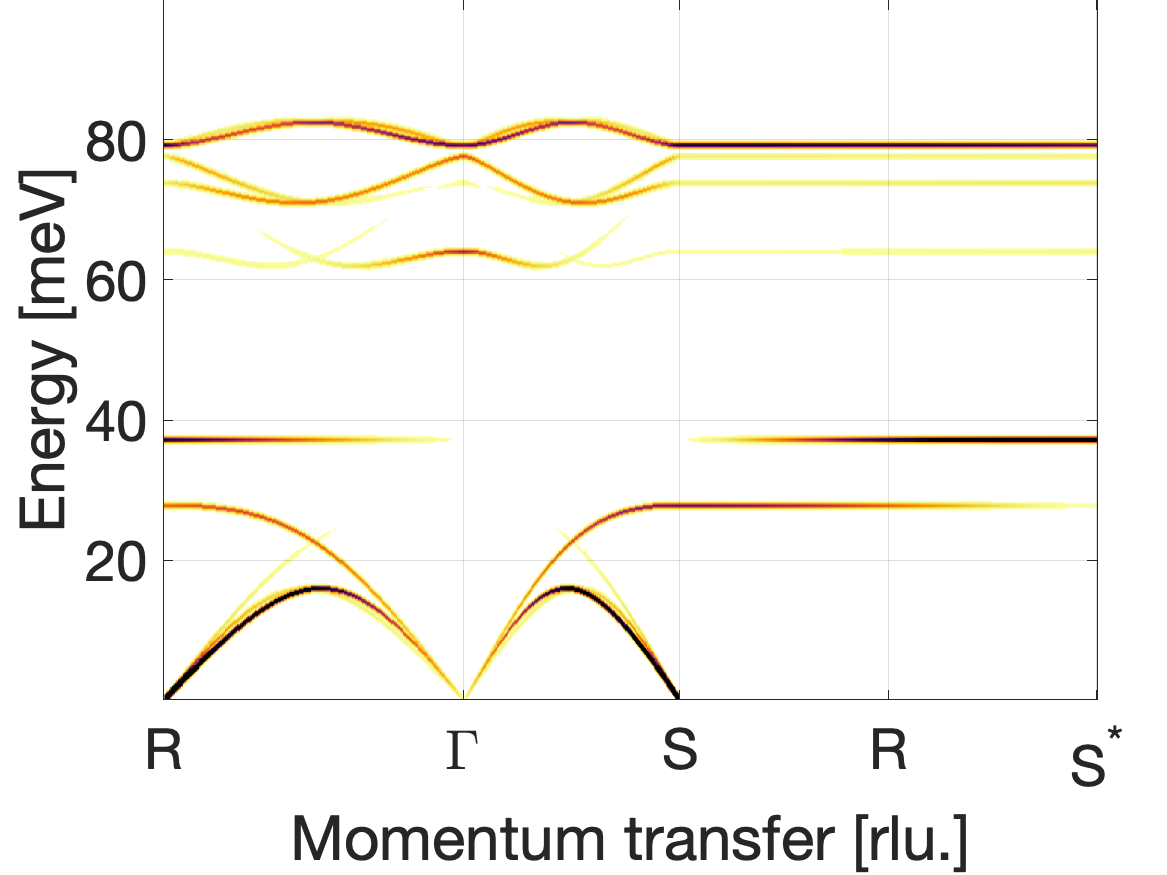}
         \caption{$J_2=0$~meV}
     \end{subfigure}
     \hfill
     \begin{subfigure}[t]{0.3\textwidth}
         \centering
         \includegraphics[width=\textwidth]{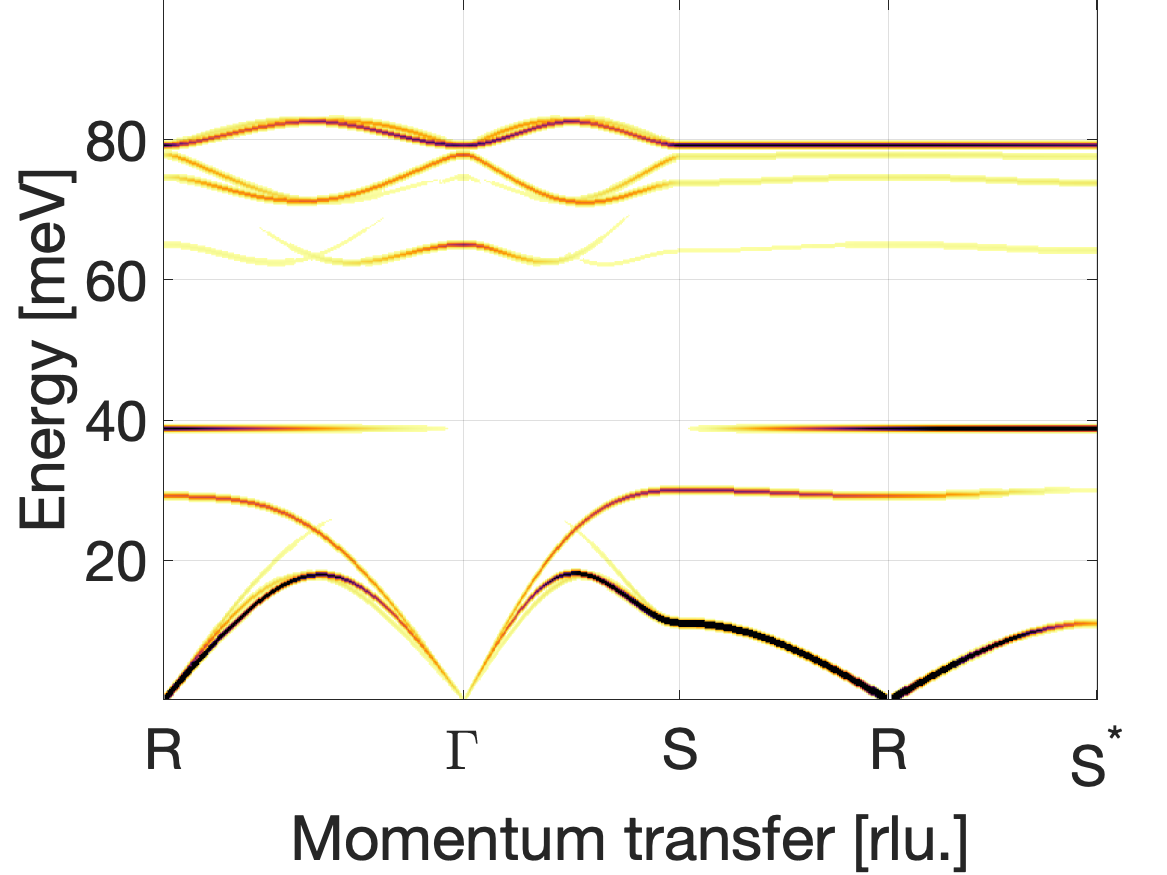}
         \caption{$J_2 = \frac{1}{2} J_{2D} = 0.7$~meV}
     \end{subfigure}
     \hfill
     \begin{subfigure}[t]{0.3\textwidth}
         \centering
         \includegraphics[width=\textwidth]{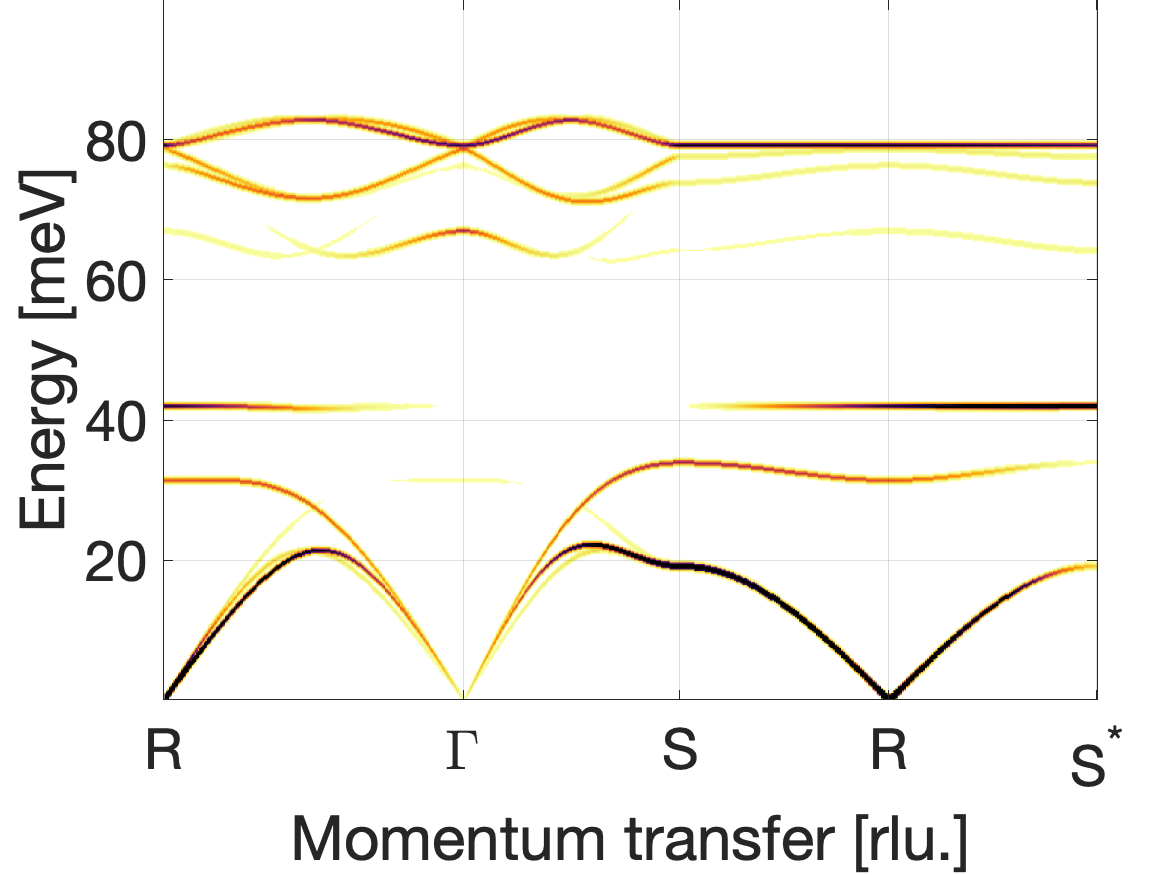}
         \caption{$J_2=\frac{3}{2} J_{2D} = 2.1$~meV}
     \end{subfigure}
        \caption{Varying the exchange interaction $J_2$, keeping all other parameters constant.}
        \label{fig:parameter_J2}
\end{figure}

The exchange interaction $J_2$ is between one spin in a pair of Fe$_2$ spins with another spin in a pair of Fe$_2$ spins in the neighbouring unit cell, thus the interaction along $c$ is to the adjacent cell.\\
This interaction is, together with $J_3$ and $J_5$ described in the next section, modulating the lower bands. It does not affect the upper bands, since they are mostly dependent on the Fe$_1$ sites. Looking at Fig.~\ref{fig:parameter_J2}, the exchange interaction $J_2$ mostly influences the dispersion in the $L$-direction. When $J_2$ is zero, in S to S$^*$ (varying $L$), the symmetry of the magnetic Bragg peak is broken and the lower band has disappeared (non-physical), additionally the 30~meV band is flat. When increasing the value of $J_2$, the lower band appears and the 30~meV band starts dispersing. The stronger the AFM interaction is, the higher is spin wave velocity at R and the more the 30~meV band disperses. 

\FloatBarrier
\subsubsection{Influence of $J_3$ and $J_5$}
\begin{figure}[ht!]
     \centering
     \begin{subfigure}[t]{0.32\textwidth}
         \centering
         \includegraphics[width=\textwidth]{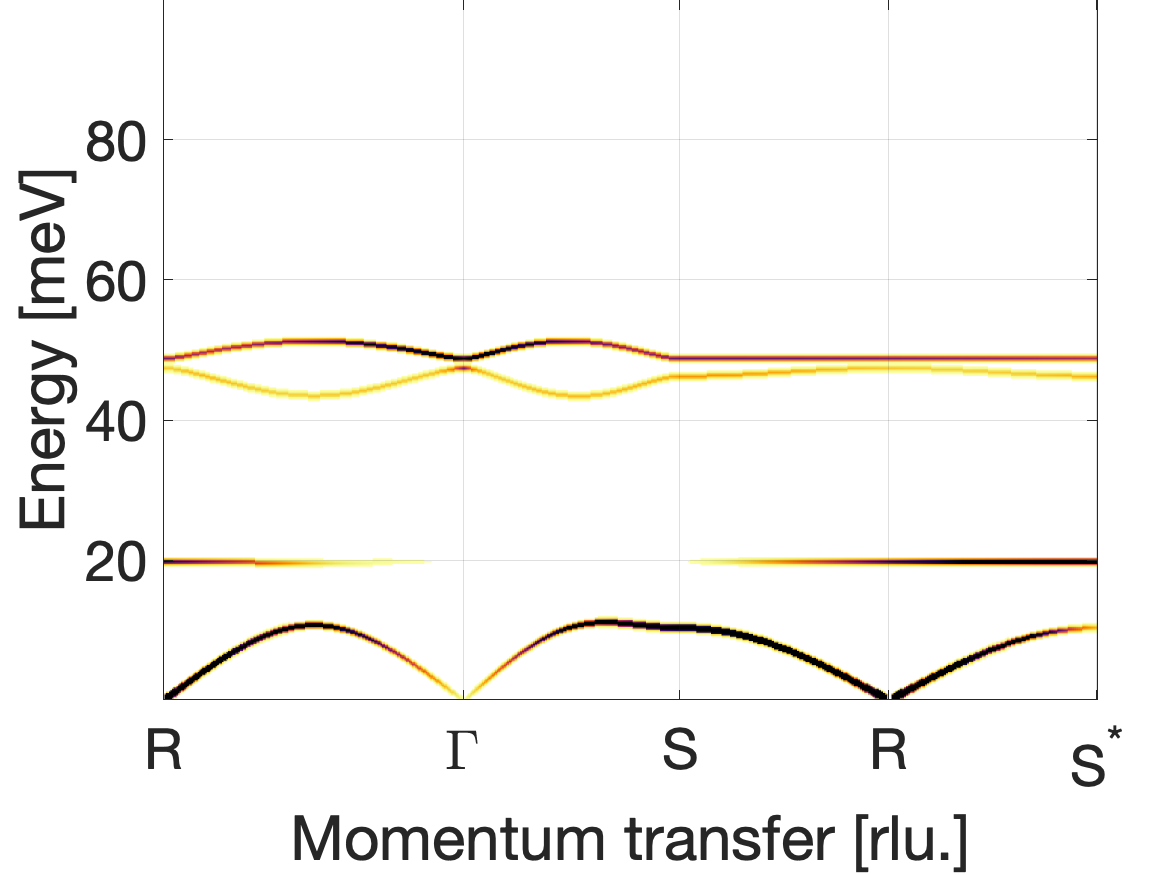}
         \caption{$J_3=0$~meV}
     \end{subfigure}
     \hfill
     \begin{subfigure}[t]{0.32\textwidth}
         \centering
         \includegraphics[width=\textwidth]{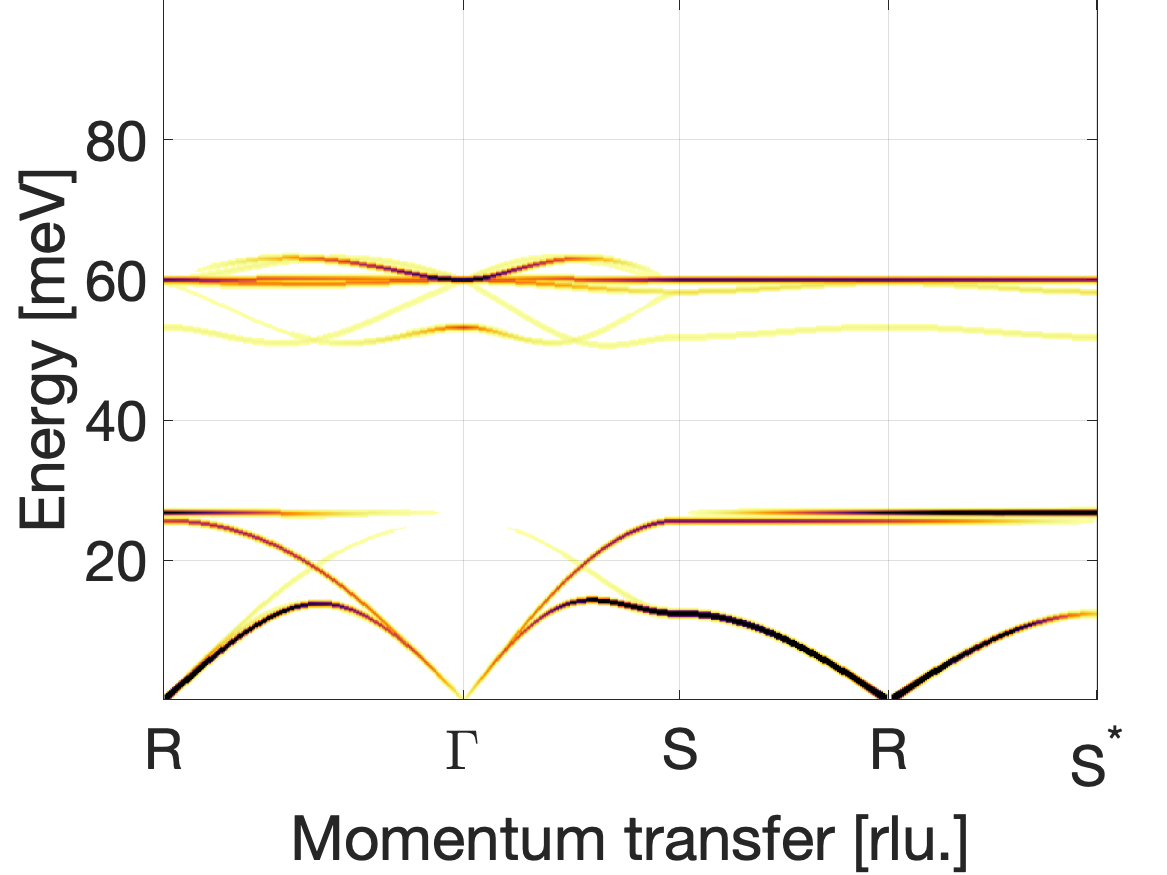}
         \caption{$J_3=\frac{1}{2} J_{3D} = 3.25$~meV}
     \end{subfigure}
     \hfill
     \begin{subfigure}[t]{0.32\textwidth}
         \centering
         \includegraphics[width=\textwidth]{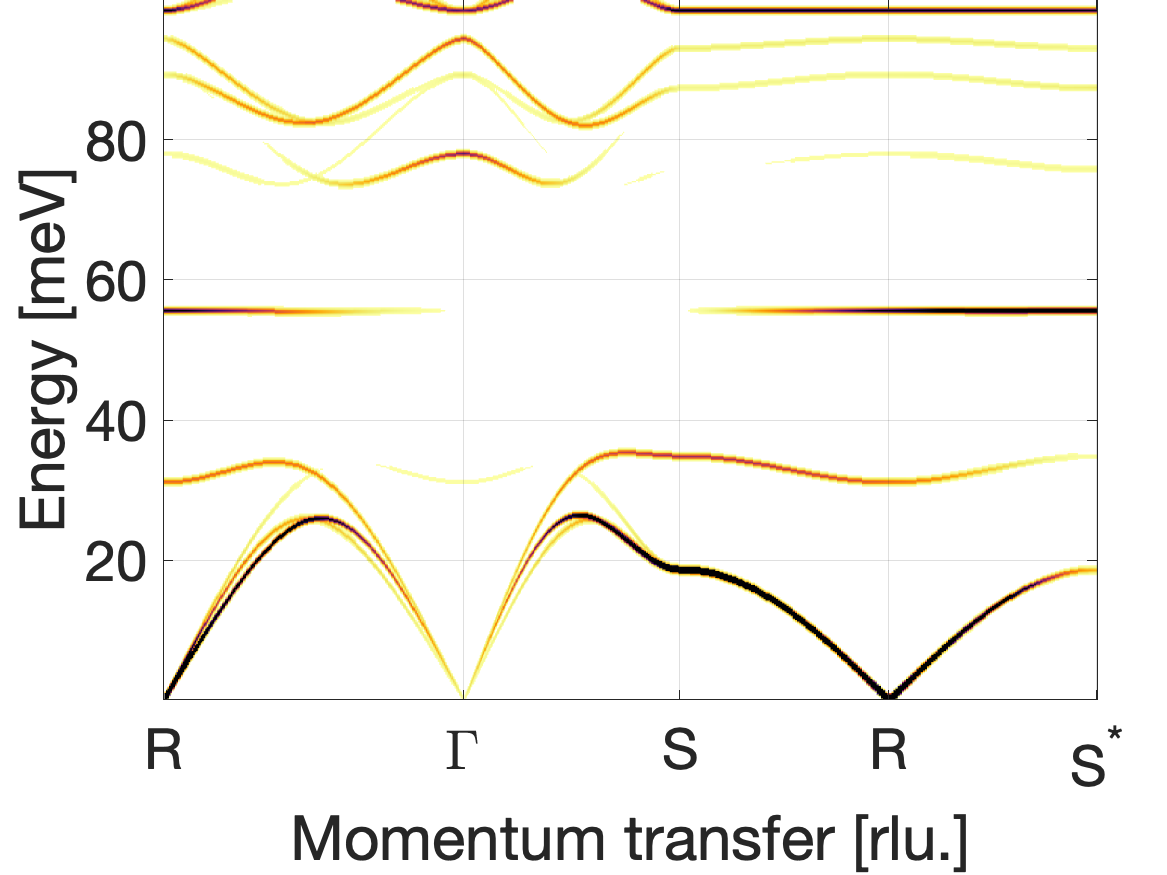}
         \caption{$J_3=\frac{3}{2} J_{3D}=9.75$~meV}
     \end{subfigure}
        \caption{Varying the exchange interaction $J_3$, keeping all other parameters constant.}
        \label{fig:parameter_J3}
\end{figure}

\begin{figure}[ht!]
     \centering
     \begin{subfigure}[t]{0.32\textwidth}
         \centering
         \includegraphics[width=\textwidth]{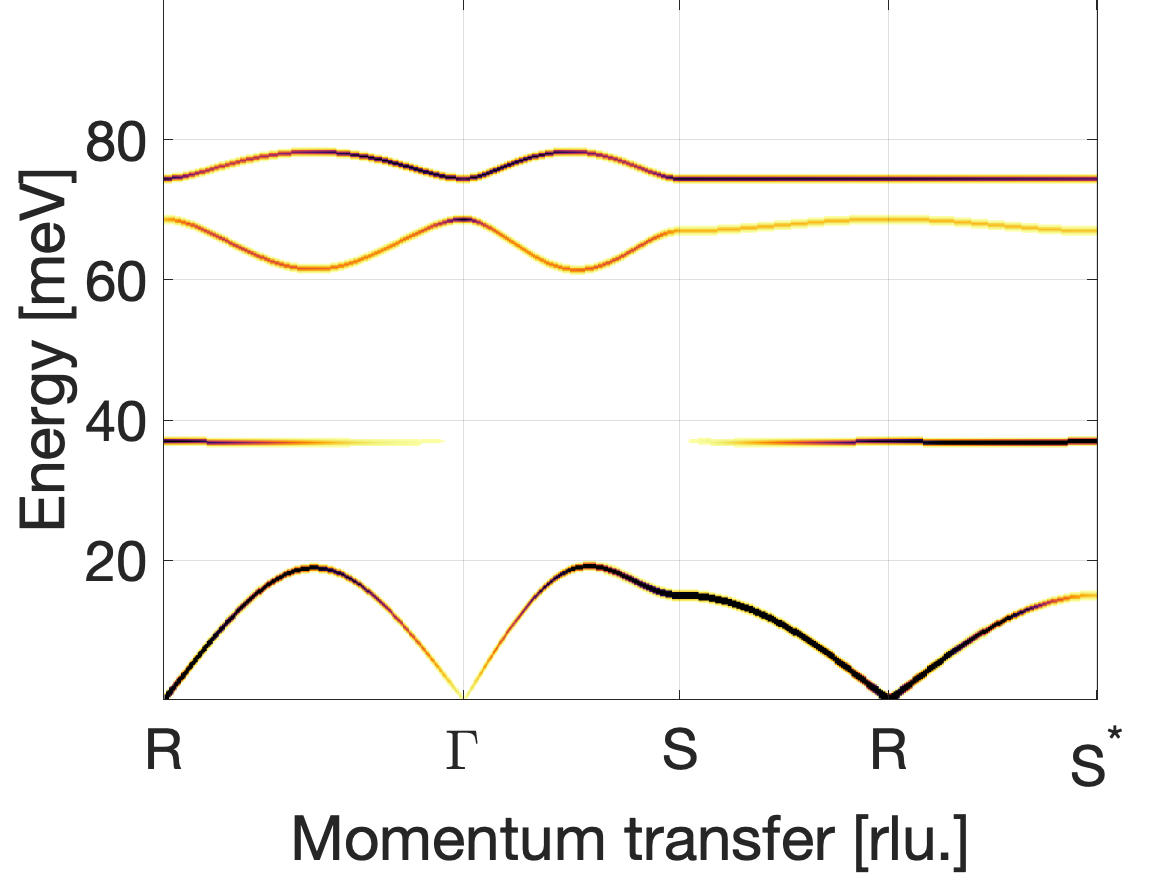}
         \caption{$J_5=0$~meV}
     \end{subfigure}
     \hfill
     \begin{subfigure}[t]{0.32\textwidth}
         \centering
         \includegraphics[width=\textwidth]{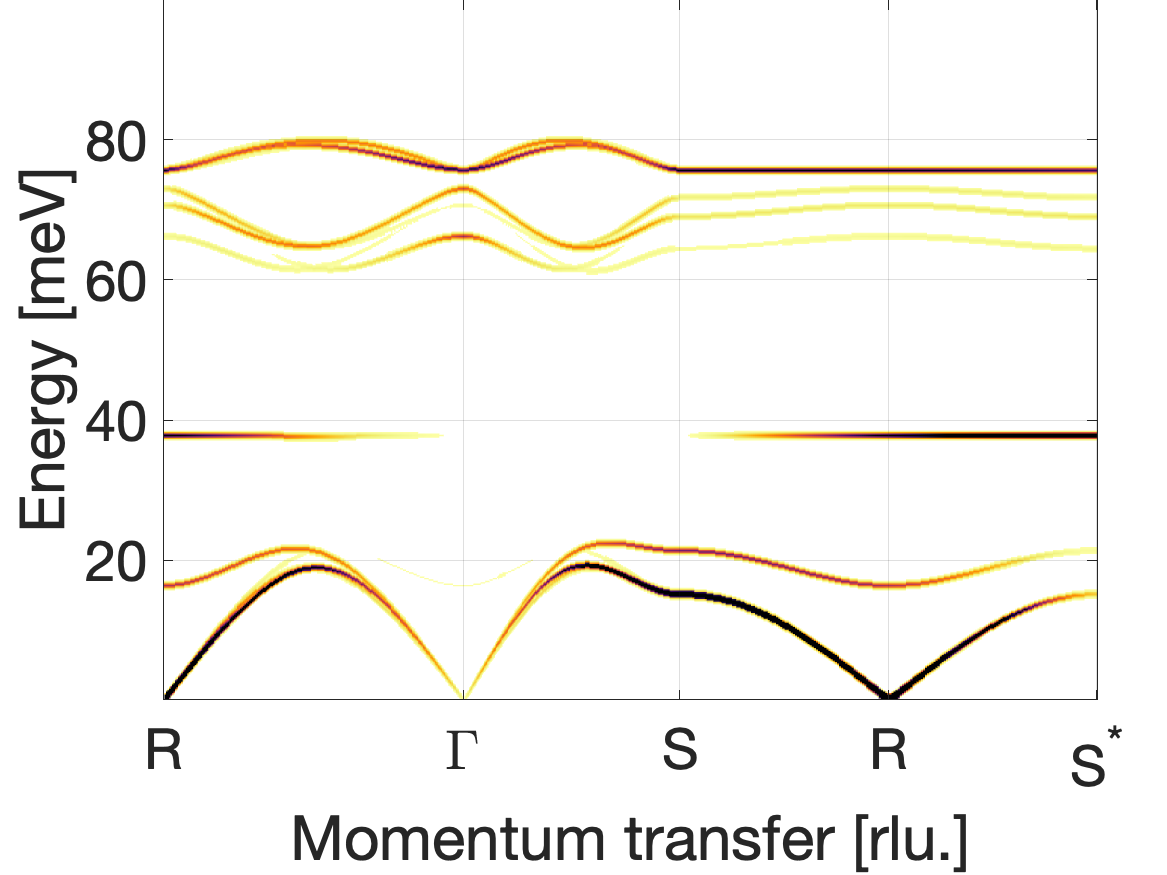}
         \caption{$J_5=\frac{1}{2} J_{5D}=1.55$~meV}
     \end{subfigure}
     \hfill
     \begin{subfigure}[t]{0.32\textwidth}
         \centering
         \includegraphics[width=\textwidth]{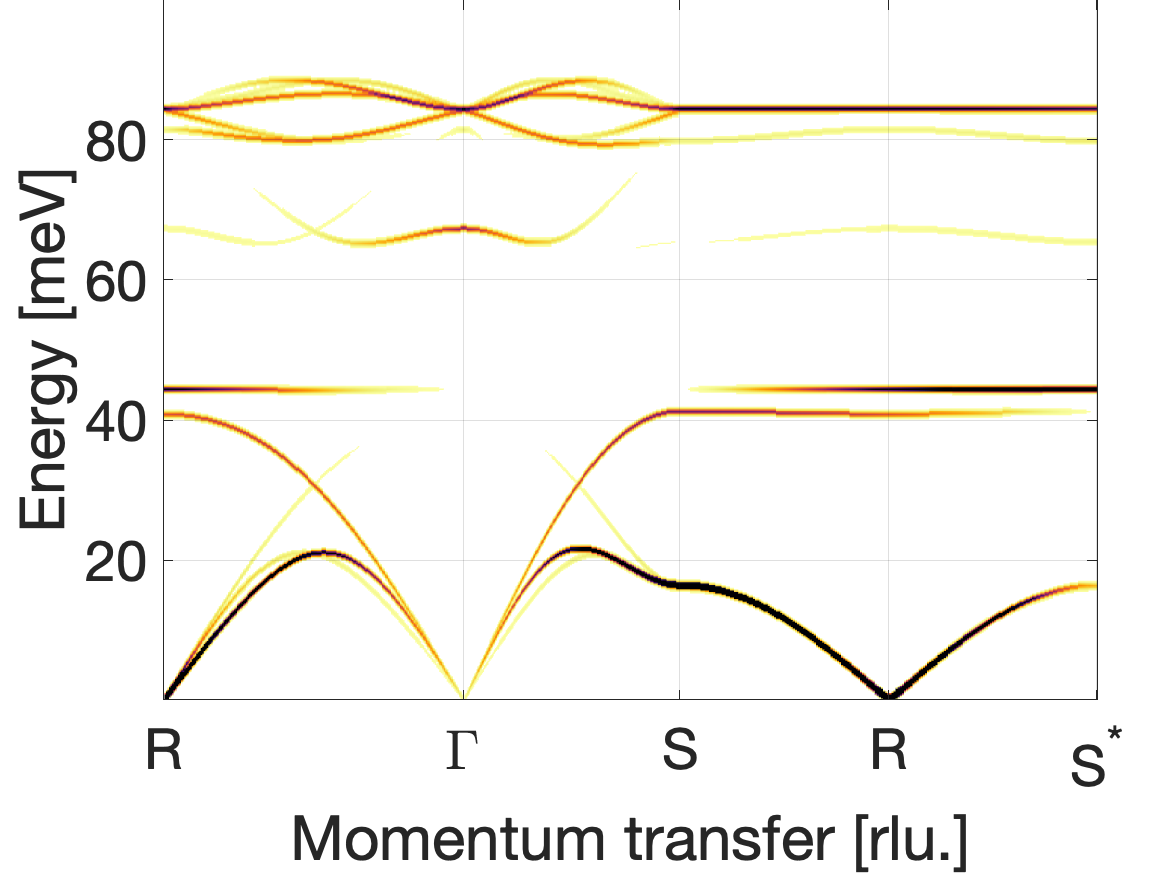}
         \caption{$J_5=\frac{3}{2} J_{5D}=4.65$~meV}
     \end{subfigure}
        \caption{Varying the exchange interaction $J_5$, keeping all other parameters constant.}
        \label{fig:parameter_J5}
\end{figure}

Due to an asymmetry in the positions of the oxygen ligands, there are two different interactions connecting the Fe$_1$ and the Fe$_2$ sites, namely $J_3$ and $J_5$. However, the two interactions are very similar in the effect they have on the spin waves, but $J_3$ is a factor of two larger than $J_5$. To see the difference between the two interactions, the values of the two have been switched. The only noticeable difference is that the intensity is slightly different. \\
Looking at Fig.~\ref{fig:parameter_J3} and \ref{fig:parameter_J5}, the exchange interactions determine the bandwidths of the lower and upper bands. When turning off one of the interactions, there are no splitting of the lower, flat and upper bands. When increasing $J_3$ the lower and upper band splits, thus the energy scale of the system increases. For the lower band, the previous 30~meV mode is seen to split from the acoustic mode at R. As the exchange increases, especially the 30~meV band is lifted in energy. The same is apparent for $J_5$. So the stronger the interaction, the larger the bandwidth. 

\FloatBarrier
\subsubsection{Influence of $J_4$}
\begin{figure}[ht!]
     \centering
     \begin{subfigure}[t]{0.32\textwidth}
         \centering
         \includegraphics[width=\textwidth]{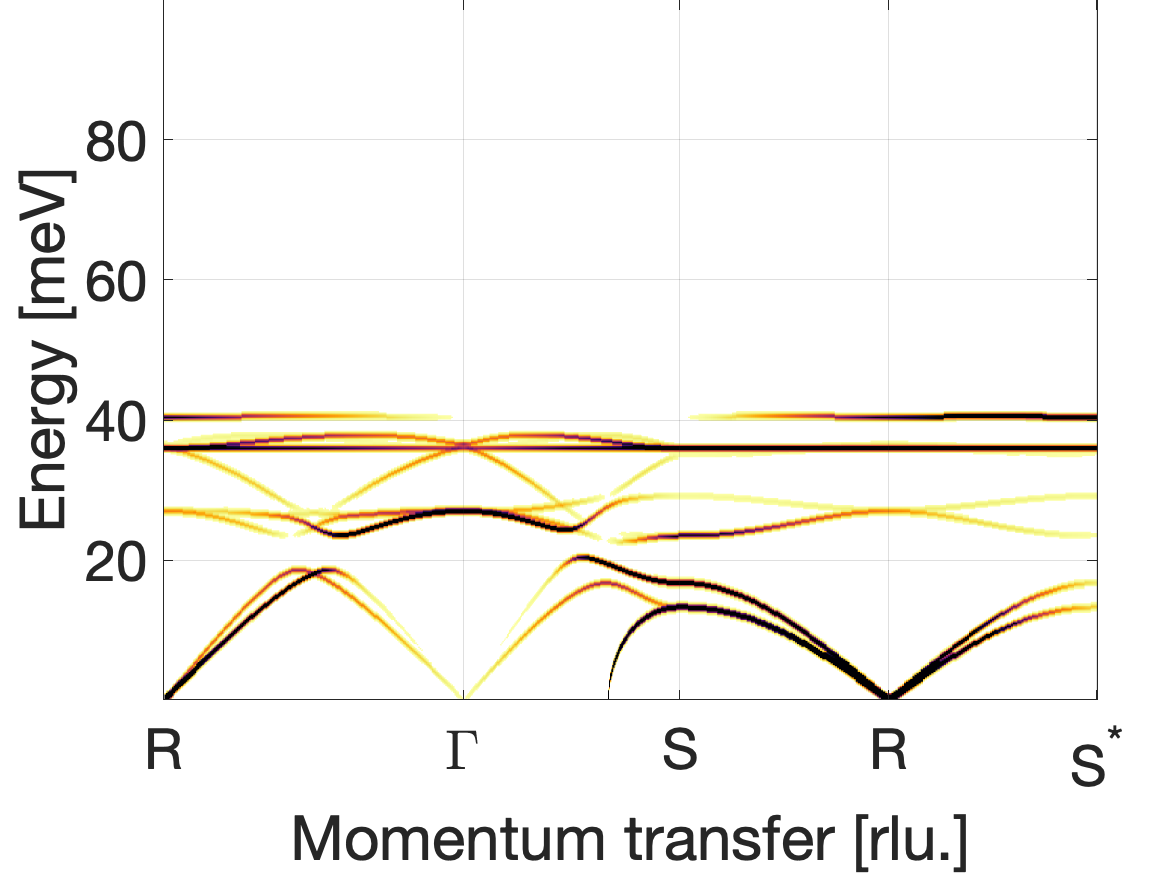}
         \caption{$J_4=0$~meV}
     \end{subfigure}
     \hfill
     \begin{subfigure}[t]{0.32\textwidth}
         \centering
         \includegraphics[width=\textwidth]{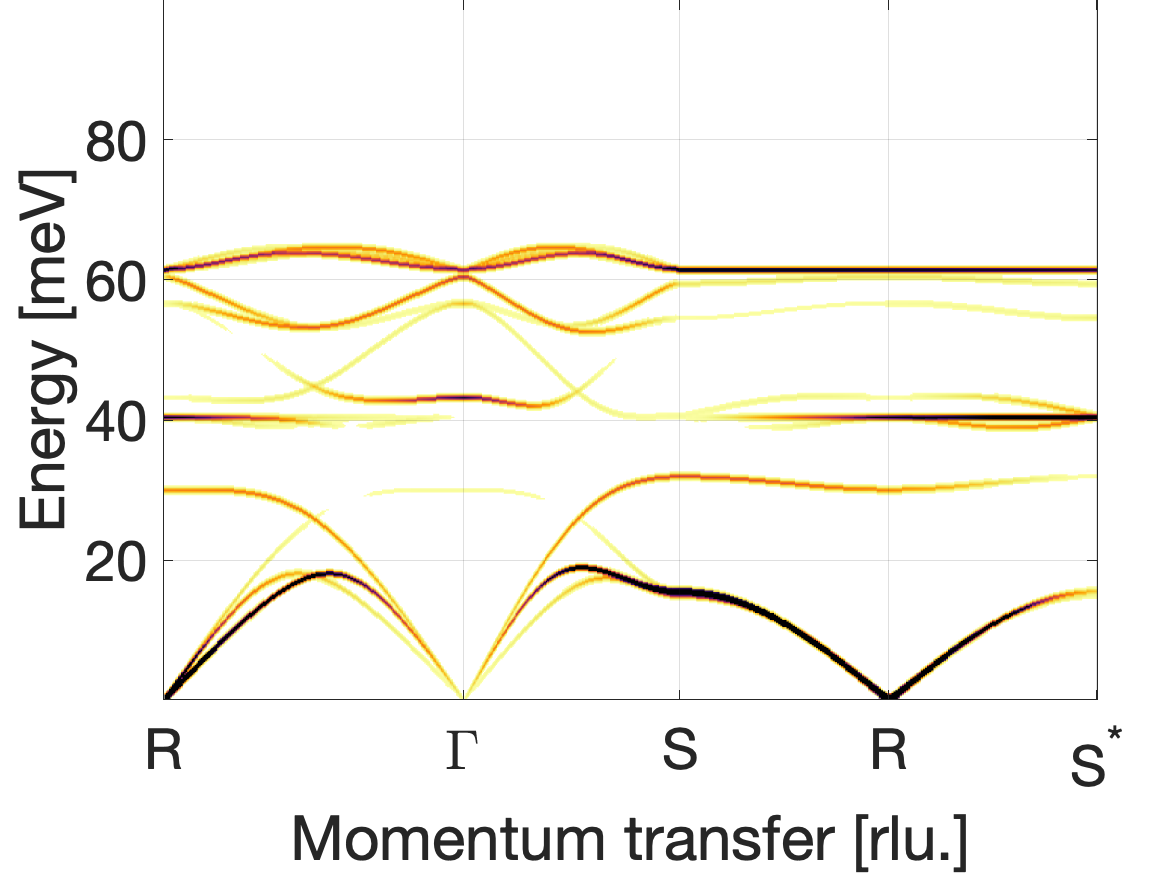}
         \caption{$J_4=\frac{1}{2} J_{4D}=13.8$~meV}
     \end{subfigure}
     \hfill
     \begin{subfigure}[t]{0.32\textwidth}
         \centering
         \includegraphics[width=\textwidth]{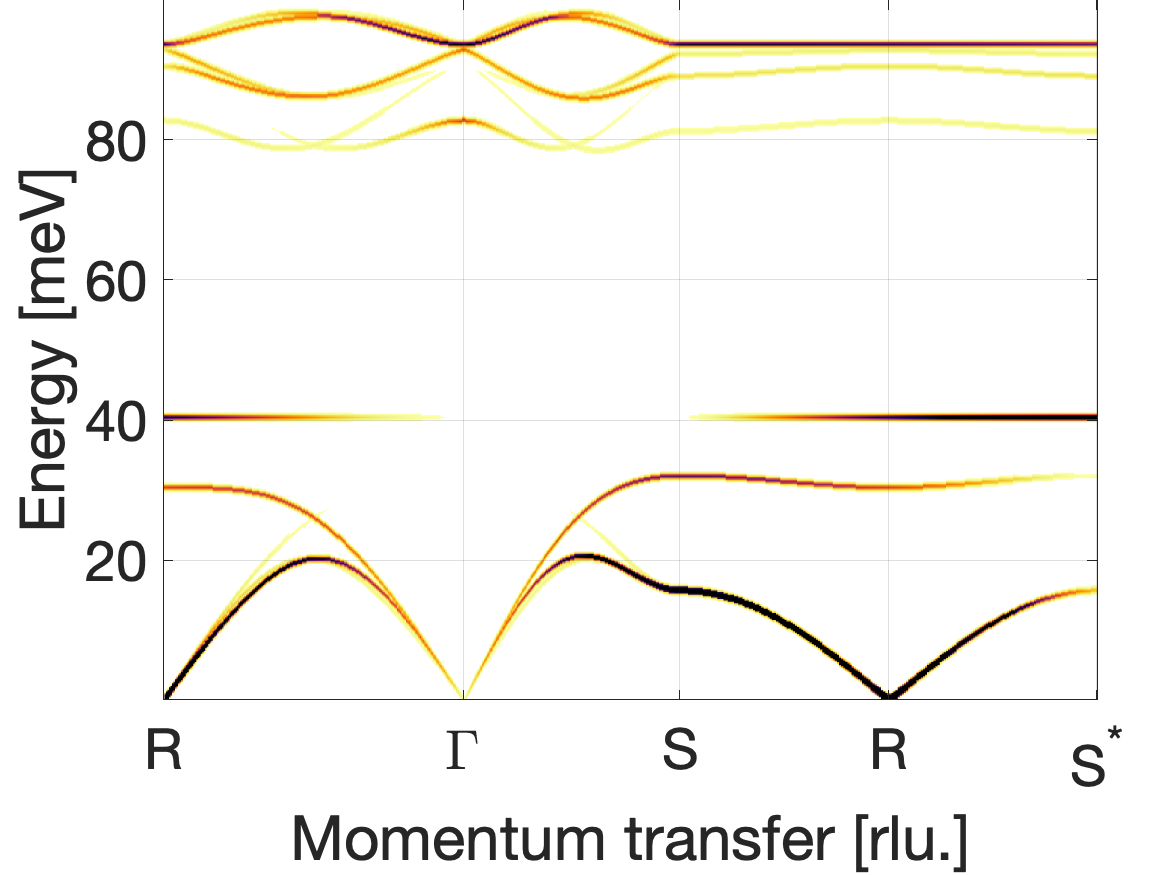}
         \caption{$J_4=\frac{3}{2} J_{4D}= 41.4$~meV}
     \end{subfigure}
        \caption{Varying the exchange interaction $J_4$, keeping all other parameters constant.}
        \label{fig:parameter_J4}
\end{figure}

The $J_4$ interaction is the in-plane Cairo dimer interaction between two Fe$_1$ spins and it is the dominant interaction. This interaction splits the dispersive modes into two band (see Fig.~\ref{fig:parameter_J4}), thus it lifts the upper bands, showing that they mostly involve precession of the spins on the Fe$_1$ sites. The lower band is not affected. Setting the interaction to zero, makes the dispersion non-physical. The observation of the upper bands being between 60 and 80~meV (from Duc Le et al.\cite{LeDuc2021}) points to a surprisingly large antiferromagnetic $J_4$.

\subsection{Test of Anisotropies} 

\subsubsection{Planar and axial single-ion anisotropies of the two iron sites}\label{app:test_SIA}

In the CAMEA data  we observed a double gap, indicating that there are two anisotropies in the system, i.e. the two Fe sites have different SIA. In principle Fe$^{3+}$, in the large spin configuration ($S=5/2$) has $L=0$, but the spin-orbit coupling can still act as a perturbation to the system, thus giving the spins an anisotropy.
\begin{figure}[H]
    \centering
    \includegraphics[width=0.65\linewidth]{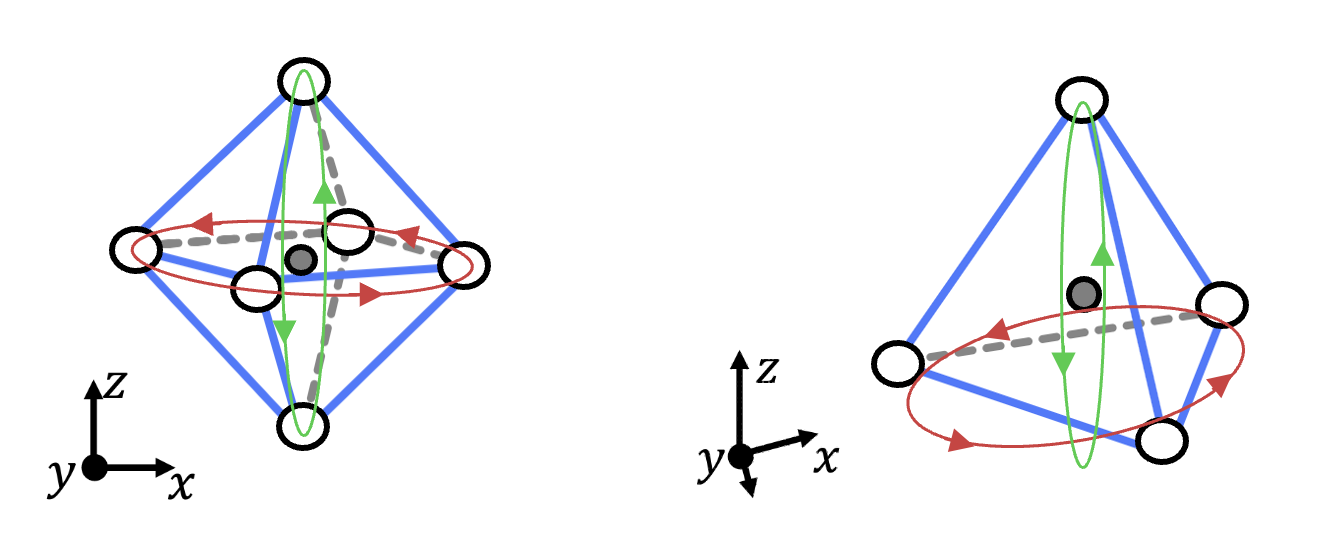}
    \caption{Left is the Fe$_2$ complex; octahedral and right is the Fe$_1$ complex; tetrahedral. The Fe is in the middle of the complex in grey, surrounded by oxygen atoms in white. They are in their respective local coordinate system, such that $z$ is the uniaxial axis. Easy axis is indicated in green and easy plane is shown in red.}
    \label{fig:octahedral_tetrahedral_guess}
\end{figure}

It is difficult to give a qualified guess on the direction of the anisotropy of the iron sites, but looking at the individual complex, they contain a symmetry axis. For the octahedral Fe$_2$ in its local coordinate system ($xyz$), we propose the uniaxial axis to be along the $z$-axis, see Fig.~\ref{fig:octahedral_tetrahedral_guess}. This is due to octahedra in Bi$_2$Fe$_4$O$_9$ are distorted along this axis. Thereby, the easy plane will be the perpendicular plane in $xy$. For the tetrahedral complex, it is not as clear. However, from looking at the crystal structure of Bi$_2$Fe$_4$O$_9$, we propose the direction of where the two Fe$_1$ (tetrahedrals) are facing each other and coupled through $J_4$, will be the uniaxial axis. This will be along the local $z$-axis and the easy plane in the $xy$-plane, both can be seen in Fig. \ref{fig:octahedral_tetrahedral_guess}.

\begin{figure}[H]
  \centering
  \begin{subfigure}[b]{0.45\textwidth}
    \centering
    \includegraphics[width=\textwidth]{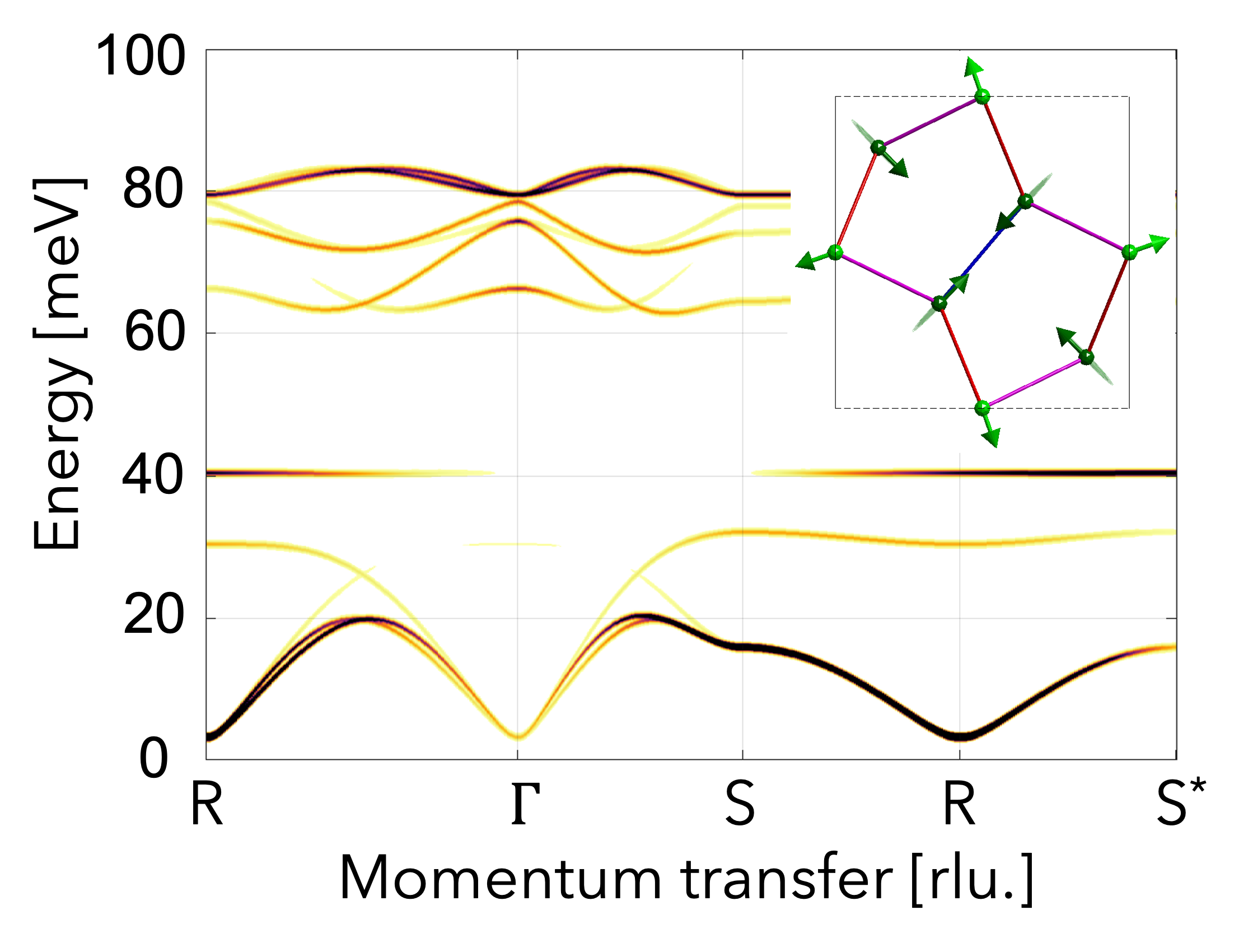}
    \caption{Fe$_1$ axial}
  \end{subfigure}%
  \hfill
  \begin{subfigure}[b]{0.45\textwidth}
    \centering
    \includegraphics[width=\textwidth]{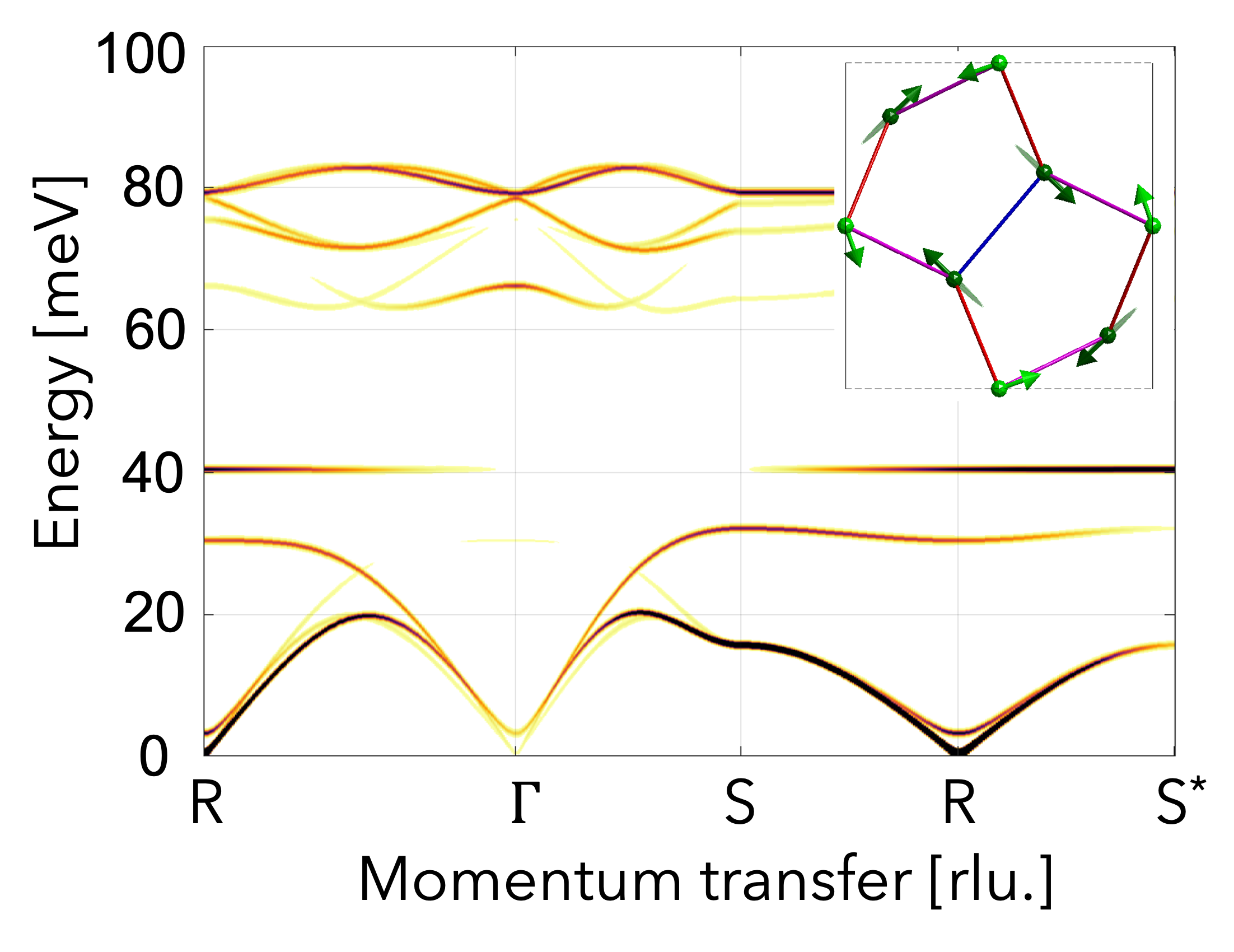}
    \caption{Fe$_1$ planar}
  \end{subfigure}%

  \begin{subfigure}[b]{0.45\textwidth}
    \centering
    \includegraphics[width=\textwidth]{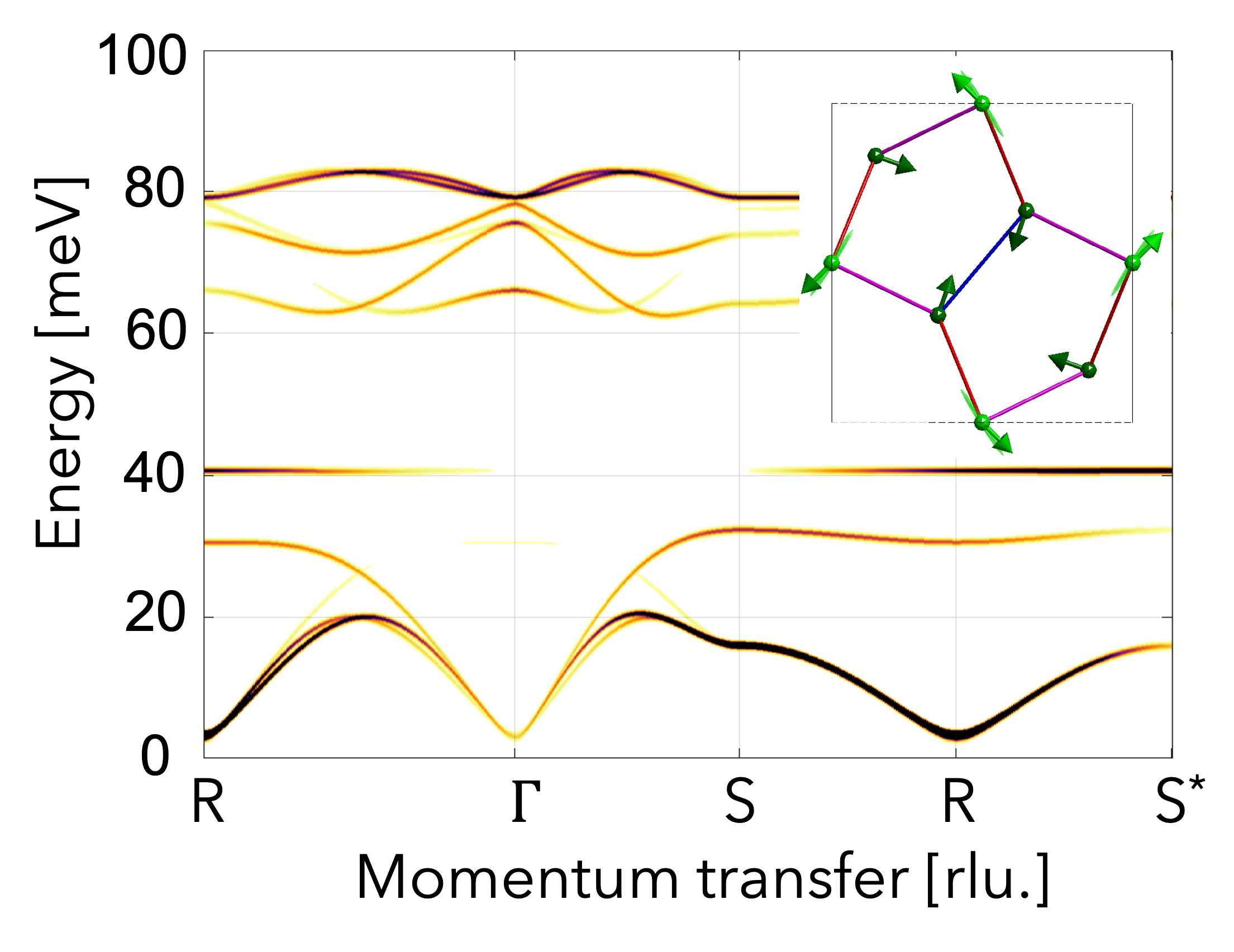}
    \caption{Fe$_2$ axial}
  \end{subfigure}%
  \hfill
  \begin{subfigure}[b]{0.45\textwidth}
    \centering
    \includegraphics[width=\textwidth]{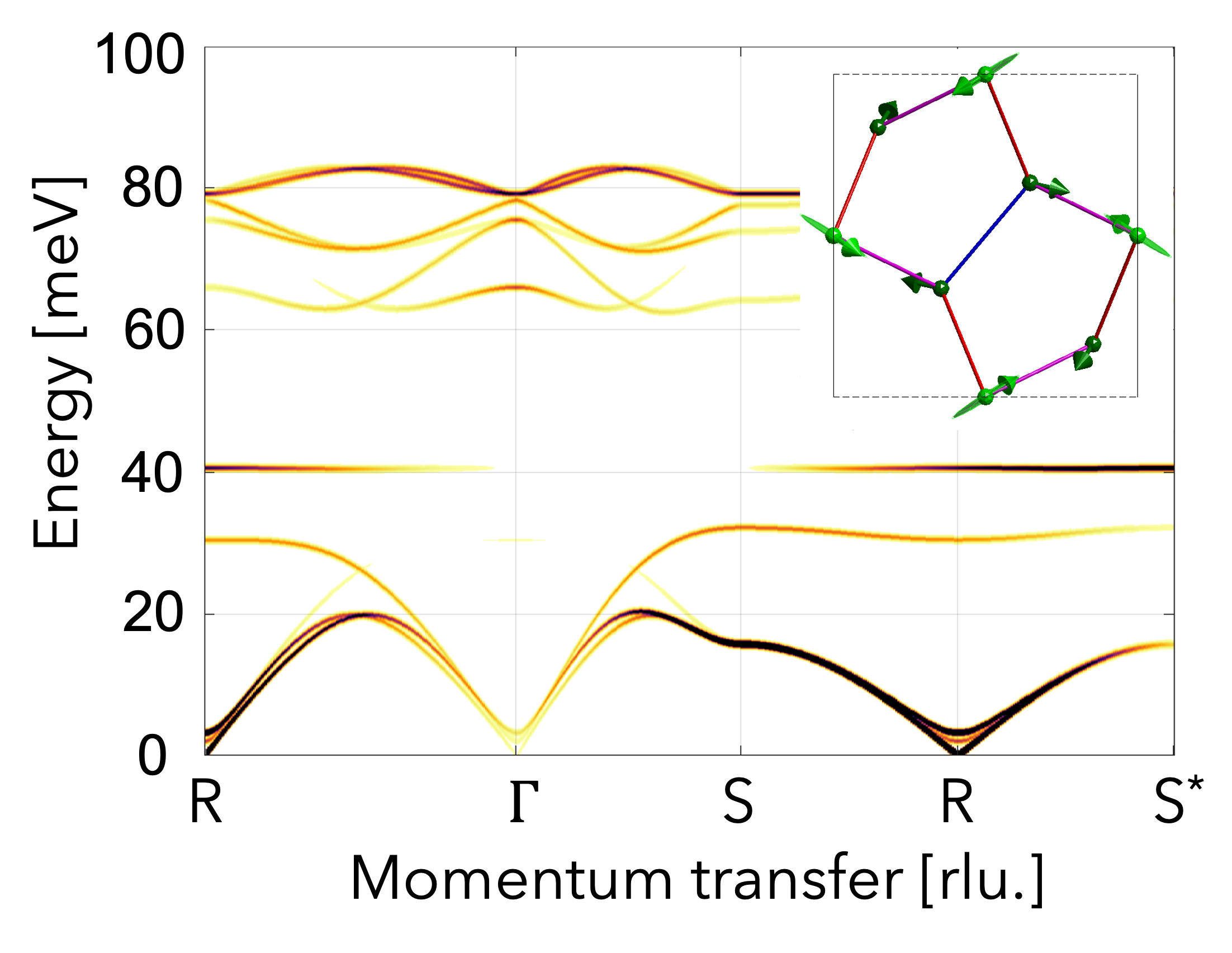}
    \caption{Fe$_2$ planar}
  \end{subfigure}%
  \caption{Only anisotropy of 0.05~meV on one iron atom, with the other being zero. In the top right corner of each figure is the optimized spin structure in one unit cell shown in the $ab$-plane.}
  \label{fig:anisotropy_and_zero}
\end{figure}

\begin{figure}[h]
    \centering
    \begin{subfigure}[t]{0.45\textwidth}
        \centering
        \includegraphics[width=\textwidth]{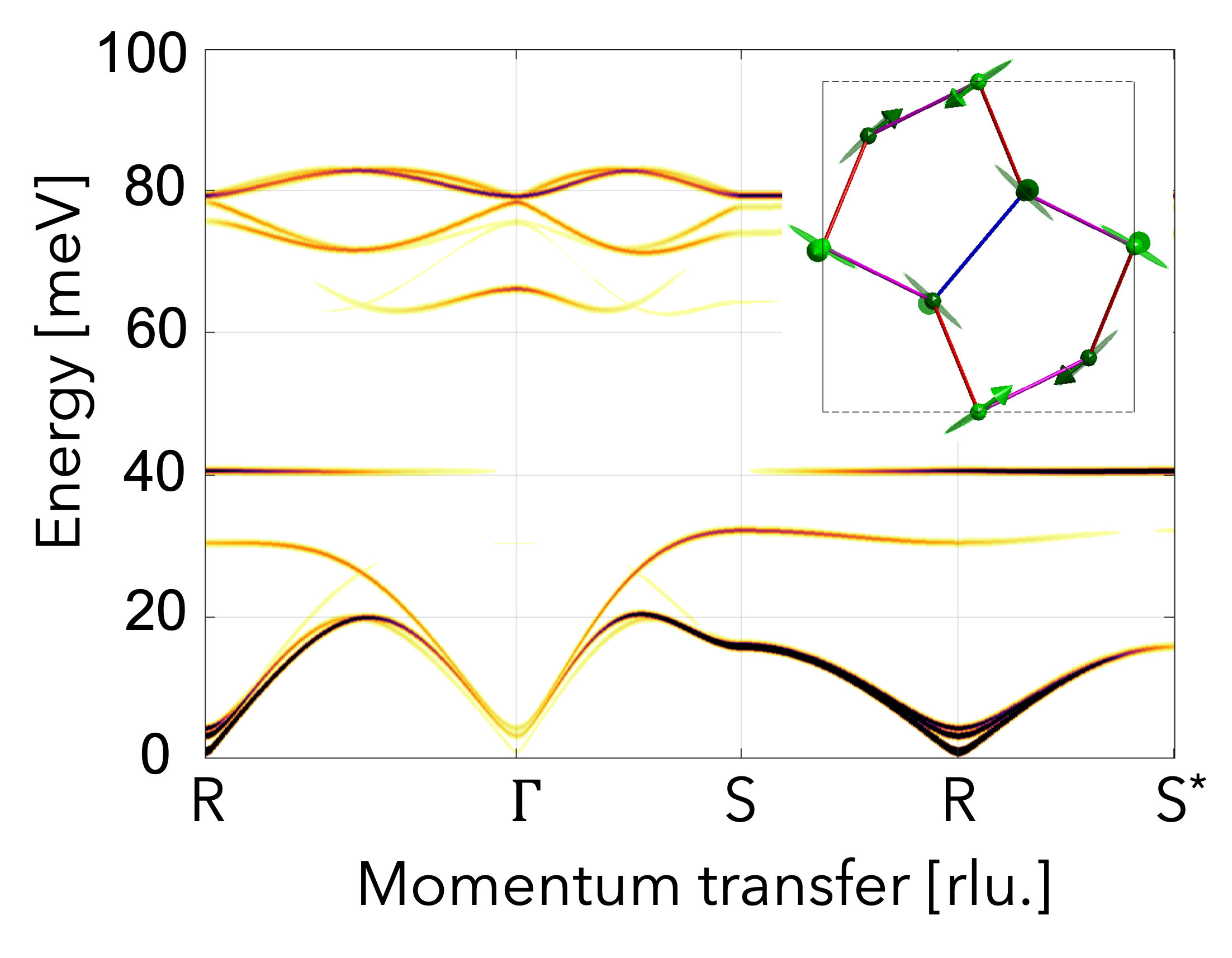}
        \caption{Both planar}
    \end{subfigure}
    \begin{subfigure}[t]{0.45\textwidth}
        \centering
        \includegraphics[width=\textwidth]{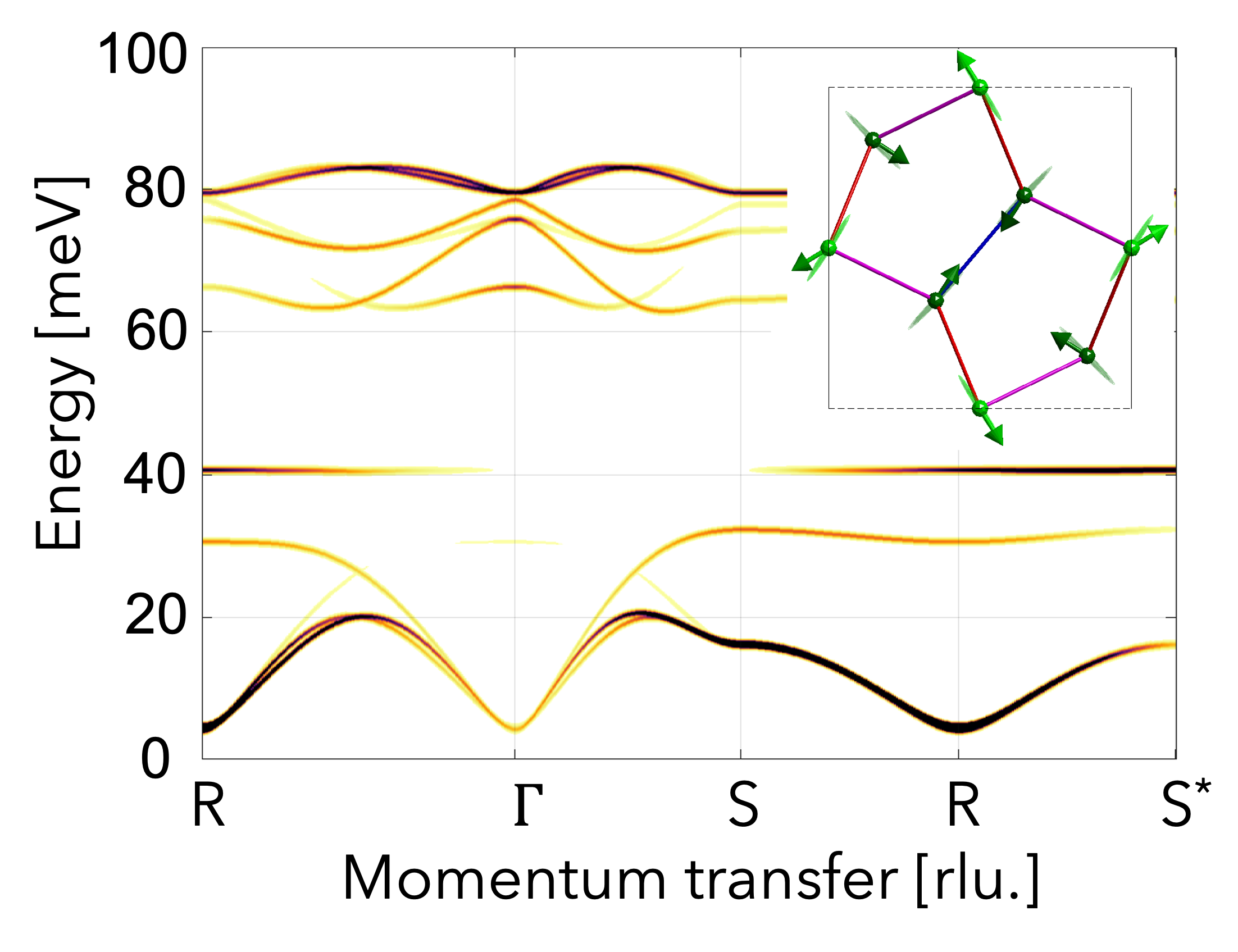}
        \caption{Both axial}
    \end{subfigure}
    
    \begin{subfigure}[t]{0.45\textwidth}
        \centering
        \includegraphics[width=\textwidth]{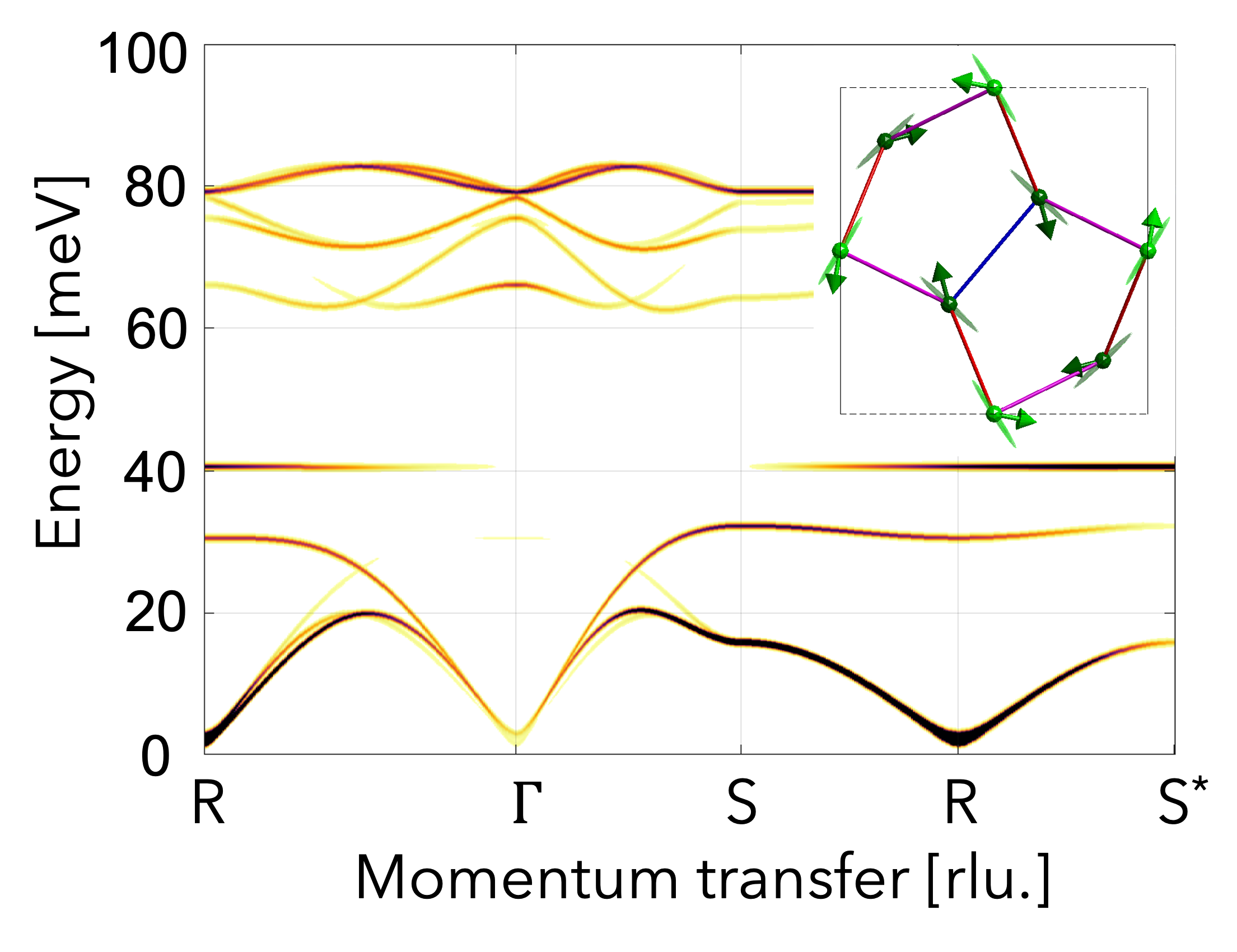}
        \caption{Fe$_1$ planar and Fe$_2$ axial}
    \end{subfigure}
    \begin{subfigure}[t]{0.45\textwidth}
        \centering
        \includegraphics[width=\textwidth]{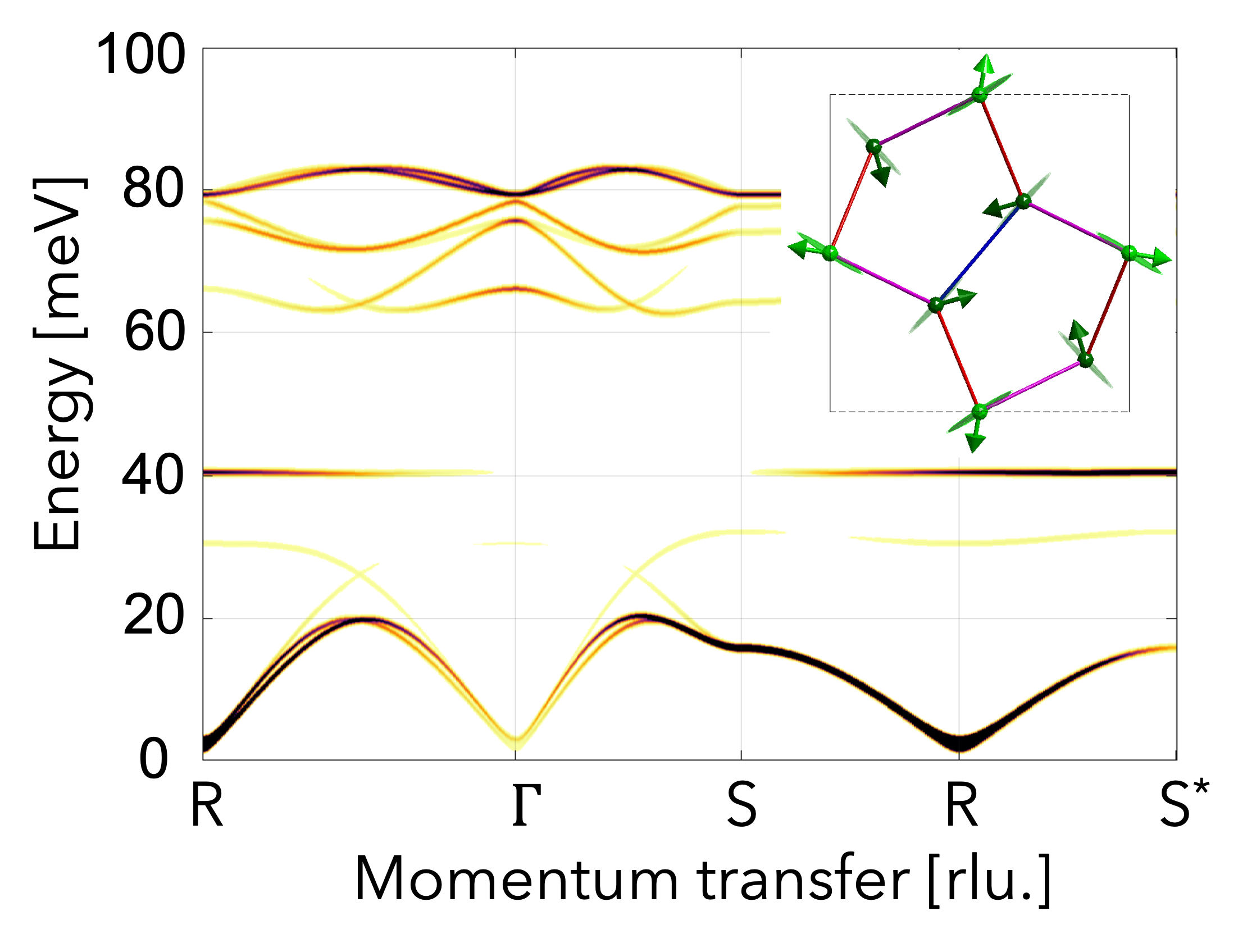}
        \caption{Fe$_1$ axial and Fe$_2$ planar}
    \end{subfigure}
    \caption{Spin wave spectrum with the different combinations of axial and planar SIA on the two Fe sites, both with the value $A = 0.05$~meV. In the top right corner of each figure, the optimized spin structure in one unit cell is shown in the $ab$-plane. }
    \label{fig:different_anisotropies}
\end{figure}

This results in eight different scenarios of zero, axial and planar anisotropy, 4 with one iron having a SIA and the other iron without SIA (Fig.~\ref{fig:anisotropy_and_zero}) and 4 with different combinations of the axial and planar SIA (Fig.~\ref{fig:different_anisotropies}). In these figures, the anisotropies are given the value 0.05~meV for both planar and axial anisotropies. 
From Fig.~\ref{fig:anisotropy_and_zero} the effect of one of the iron sites having an axial anisotropy is to lift the dispersion giving rise to a gap, but the band does not split. In the case of one iron has a planar anisotropy, the band splits, but the lowest band remains gapless. Looking at the optimized spin structures in these 4 cases, we observe that we are far from the ground state found by neutron diffraction\cite{Ressouche2009} (Fig.~\ref{fig:structure}b). All are inconsistent with the data, which means that a combination of the two SIA is required.

We now look at both Fe sites having a combination of both axial and planar anisotropy, see Fig.~\ref{fig:different_anisotropies}. All spectra show small deviations from one another, but looking at the spin structure, there are large differences. When optimizing the spin structure, all converged except the case with both anisotropies being planar. The only combination close to the one found by neutron diffraction\cite{Ressouche2009} is a planar Fe$_1$ anisotropy and an axial Fe$_2$ anisotropy. Looking closer at the dispersion, this combination also has a gap and the modes are split (can only be seen when making the resolution smaller, e.g. dE = 0.2). The double gap does not currently have the correct value, since the arbitrary value of 0.05~meV is chosen for both axial and planar anisotropies.

\subsubsection{Test of DMI}
\begin{figure}[h]
    \centering
    \begin{subfigure}[t]{0.45\textwidth}
        \centering
        \includegraphics[width=\textwidth]{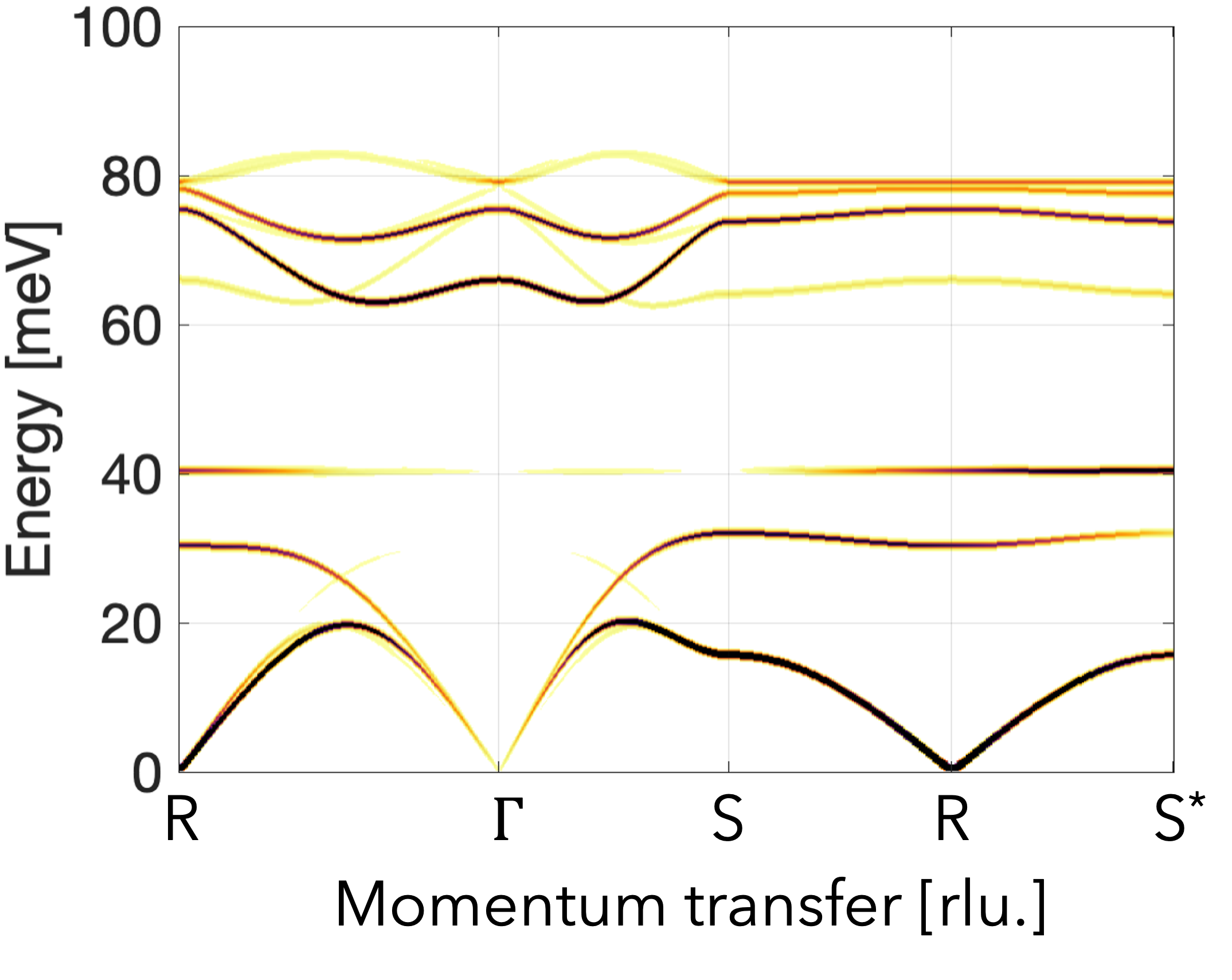}
        \caption{7$\%$ of both $J_{3D}$ and $J_{5D}$}
    \end{subfigure}
    \begin{subfigure}[t]{0.45\textwidth}
        \centering
        \includegraphics[width=\textwidth]{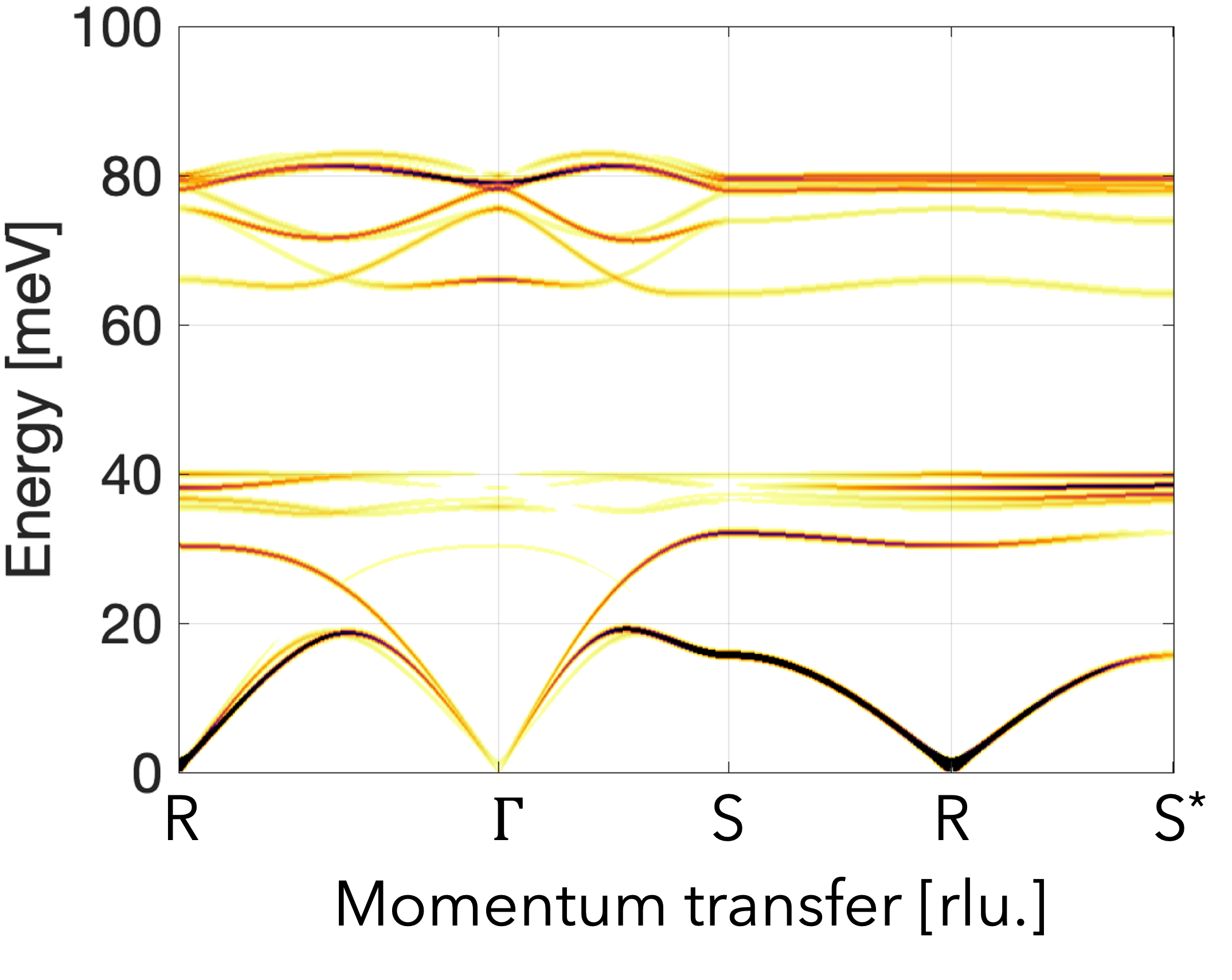}
        \caption{50$\%$ of both $J_{3D}$ and $J_{5D}$}
    \end{subfigure}
    
    \begin{subfigure}[t]{0.45\textwidth}
        \centering
        \includegraphics[width=\textwidth]{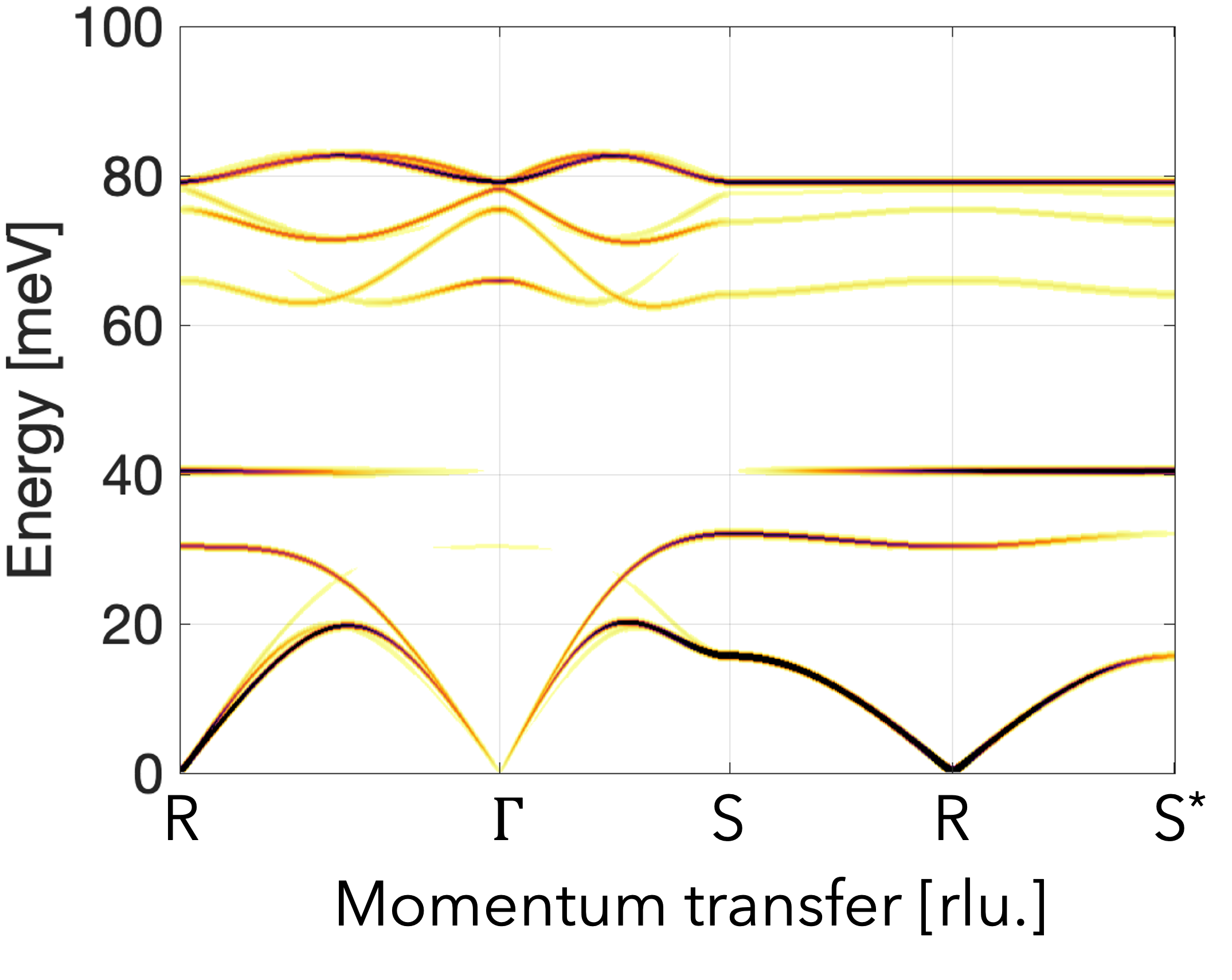}
        \caption{4$\%$ of $J_{3D}$ and 8$\%$ of $J_{5D}$}
    \end{subfigure}
    \begin{subfigure}[t]{0.45\textwidth}
        \centering
        \includegraphics[width=\textwidth]{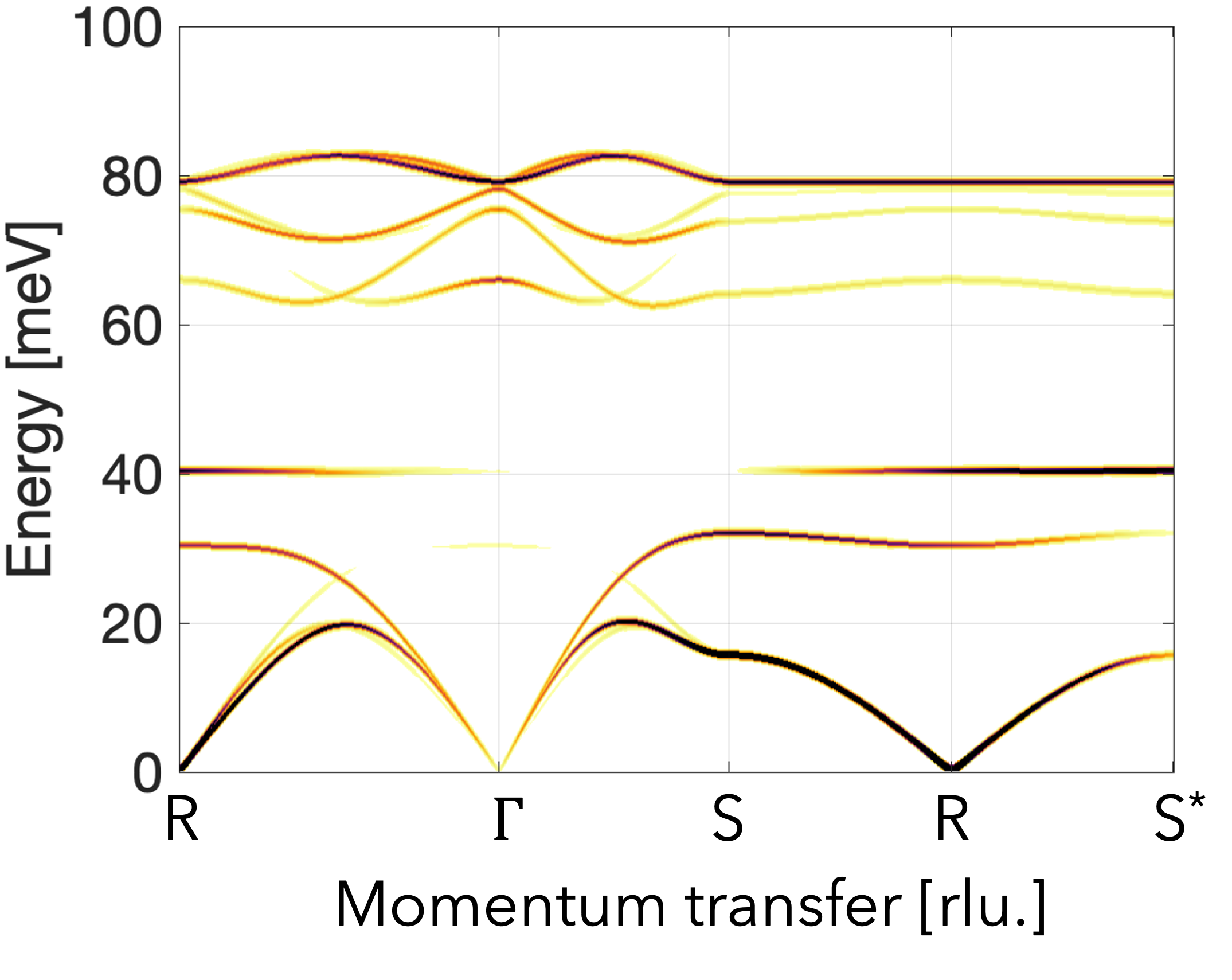}
        \caption{8$\%$ of $J_{3D}$ and 4$\%$ of $J_{5D}$}
    \end{subfigure}
    \caption{Spin wave spectrum with isotropic exchange interactions and DMI on $J_3$ and $J_5$ with varying magnitudes.}
    \label{fig:different_DMI}
\end{figure}

DMI is allowed by symmetry on the $J_3$ and $J_5$ interactions. To test whether DMI can create the double gap observed in the CAMEA data, we fix SIA to zero and calculate the dispersion for the exchange interactions from above and vary the magnitude of the D-vector as a percentage of the exchange interaction. \\
The D-vector is defined from the Fe-O-Fe positions and kept fixed, such that only the magnitude of DMI is varied. \\
For DMI being 7$\%$ of $J_3$ and $J_5$, 0.455 and 0.217~meV respectively, we observed merely minor differences in the dispersion with no anisotropies and no spin gap (Fig.~\ref{fig:different_DMI}a). Even for different percentages for the two DMI (Fig.~\ref{fig:different_DMI}c-d), no gap is observed. Above the 7$\%$ threshold for both exchanges, the ground state does not converge until unrealistic large values of DMI are obtained. E.g. at 50$\%$ of $J_3$ and $J_5$, 3.25 and 1.55~meV respectively, where still no gap at R=(2.5 2.5 0.5) is observed (Fig.~\ref{fig:different_DMI}b). Thus, we can conclude that DMI on $J_3$ and $J_5$ does not create a spin gap. This is because the DMI perturbation in the Hamiltonian adds zero energy, due to the weak FM $J_1$ between the Fe$_2$ pair. However, at larger DMI values, a splitting and softening of the 40~meV mode is observed. 

\subsection{Tested simulations leading to splitting of the dispersionless 40~meV mode} \label{app:40meV_tests}
\subsubsection{SIA and DMI}
Adding both the SIA and the DMI term to the Hamiltonian will create more anisotropy in the system, which might describe the softening and splitting of the 40~meV mode. Using our parameters from Table \ref{tab:interactions} and adding a constant percentage of the DMI of both the $J_3$ and $J_5$ interaction, will create a splitting of the 40~meV mode.

\begin{figure}[h]
    \centering
    \begin{subfigure}[t]{0.45\textwidth}
        \centering
        \includegraphics[width=\textwidth]{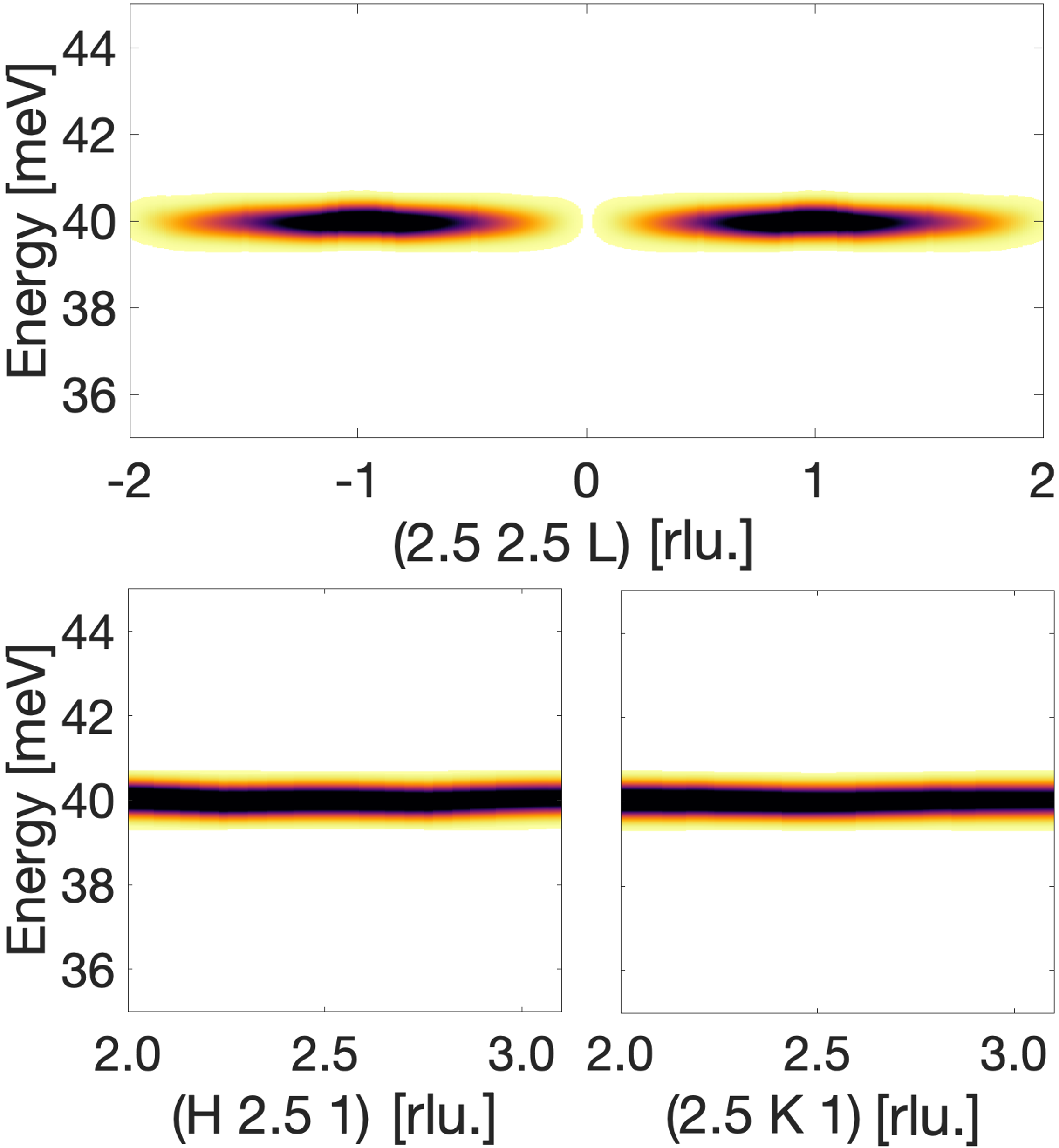}
        \caption{DMI of 10$\%$ of both $J_3$ and $J_5$}
    \end{subfigure}
    \hspace{1 cm}
    \begin{subfigure}[t]{0.45\textwidth}
        \centering
        \includegraphics[width=\textwidth]{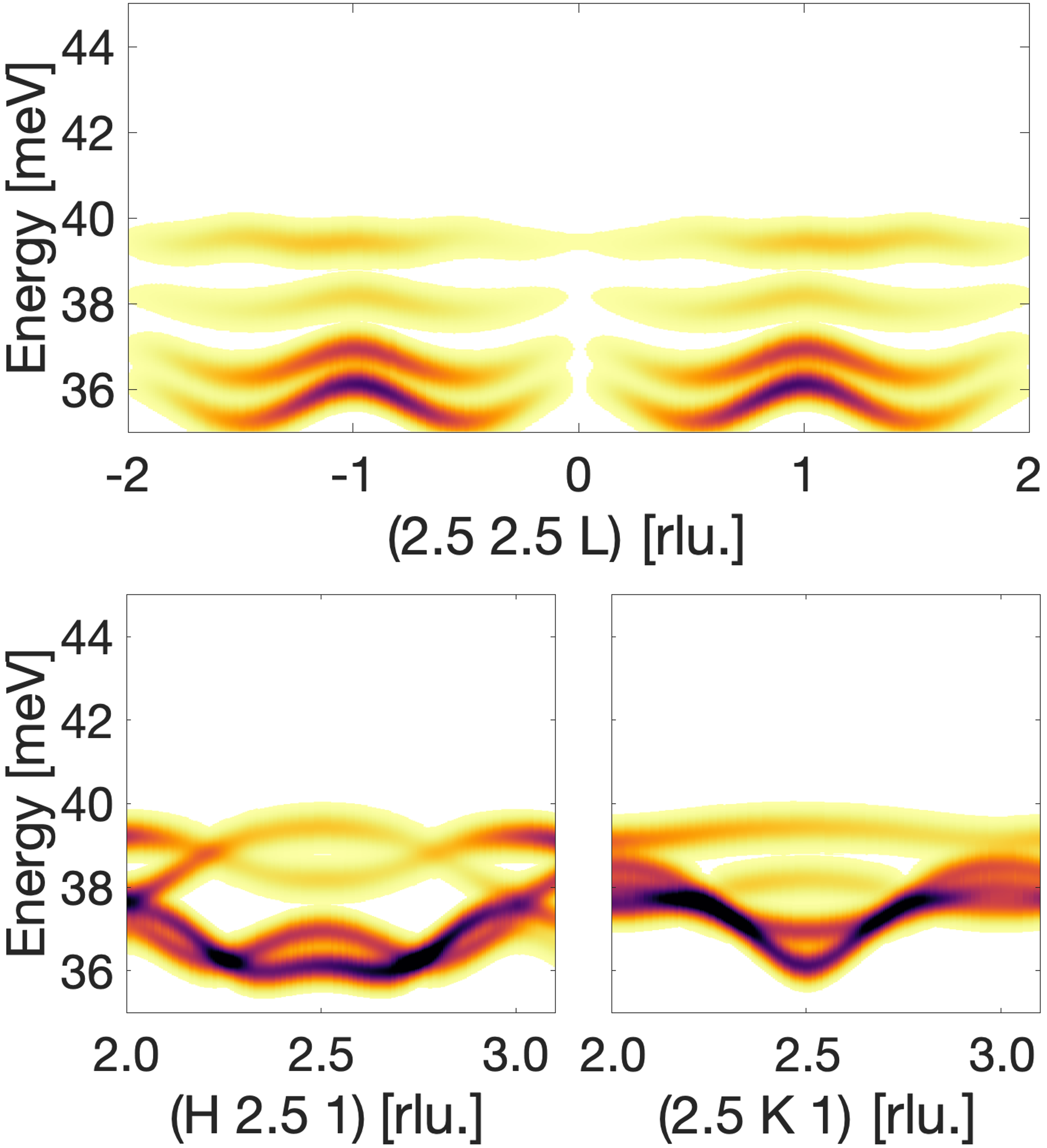}
        \caption{DMI of 50$\%$ of both $J_3$ and $J_5$}
    \end{subfigure}
    
    \caption{Spin wave spectrum of the 40~meV mode with isotropic exchange interactions, SIA from Table \ref{tab:interactions} and DMI on $J_3$ and $J_5$ with varying magnitudes.}
    \label{fig:different_DMI_SIA}
\end{figure}

SIA stabilizes the magnetic structure when adding DMI. In Fig.~\ref{fig:different_DMI_SIA} is the 40~meV mode, on the same energy scale as shown in the data in Fig.~\ref{fig:40meV_mode}b-c, with a magnitude of DMI of 10$\%$ and 50$\%$ of $J_3$ and $J_5$, respectively. The 10$\%$ DMI magnitude (normally considered a large DMI) show no obvious splitting or softening of the 40~meV mode with energy resolution of $dE=0.5$~meV. At 50$\%$ DMI magnitude, we observe both a splitting and a softening of the modes in all three \textbf{Q}-directions, however, it is not representing the data. We would expect the mode to soften along L with a maximum splitting of the mode at L=1, while the dispersion along H and K remains dispersionless. Thus, we conclude that DMI does not describe the 40~meV behaviour observed in the data. 

\subsubsection{Anisotropic exchange and higher order SIA}\label{app:Frida}
\begin{figure*}[h]
    \centering
    \includegraphics[width=1\linewidth]{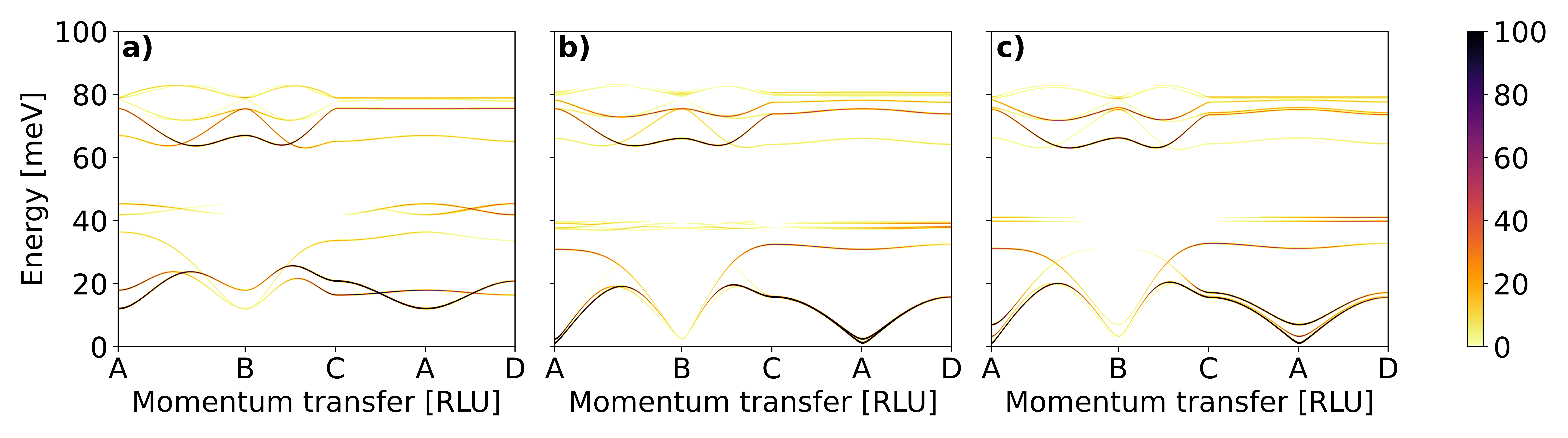}
    \caption{Three examples of simulations with non-Heisenberg exchange interactions or higher order single ion anisotropy, where a) $J_2\rightarrow J_2\begin{pmatrix}
1 & 1 & 0\\
1 & 1 & 0 \\
0 & 0 & 1
\end{pmatrix}$, b) $J_5\rightarrow J_5\begin{pmatrix}
1 & 0 & 0.7\\
0 & 1 & 0 \\
0.7 & 0 & 1
\end{pmatrix}$, and c) $A_{2} \rightarrow A_{2}-0.03 S_y^4$. Here $J_2$, $J_5$, and $A_2$ refer to the values presented in Table \ref{tab:interactions}. All other exchange and anisotropy parameters are kept constant.}
    \label{fig:test_sim40meV}
\end{figure*}

In our determined exchange interactions and single-ion-anisotropies (see Table \ref{tab:interactions}), only isotropic exchange and second order anisotropy (easy-plane and easy-axis) are included. LSWT simulations using these parameters do, however, not replicate the splitting and dispersing of the $40$~meV mode observed in the 4SEASONS data (see Fig.~\ref{fig:40meV_mode}). Therefore, we test whether more complicated descriptions of the exchange interactions and anisotropies are necessary, by performing simulations with anisotropic exchange and higher order anisotropy. Three typical examples of the resulting spin wave dispersions can be seen in Fig.~\ref{fig:test_sim40meV}, where a splitting of the $40$~meV mode is obtained. These LSWT simulations are made in Sunny\cite{dahlbom_sunnyjl_2025}.

We observe that a splitting and dispersing of the $40$~meV mode can be obtained by including more complicated exchange interactions and anisotropies. By applying these modifications to the Hamiltonian, we are, however, not able to simulate the observed behaviour of the $40$~meV mode, while also replicating the shape and energy of the lower modes in the spin wave dispersion. For this reason, the final Hamiltonian includes only isotropic exchange and easy-plane and easy-axis anisotropy.

\subsection{Steps of fitting dispersion and final model}
\begin{figure*}[h]
    \centering
    \includegraphics[width=0.9\linewidth]{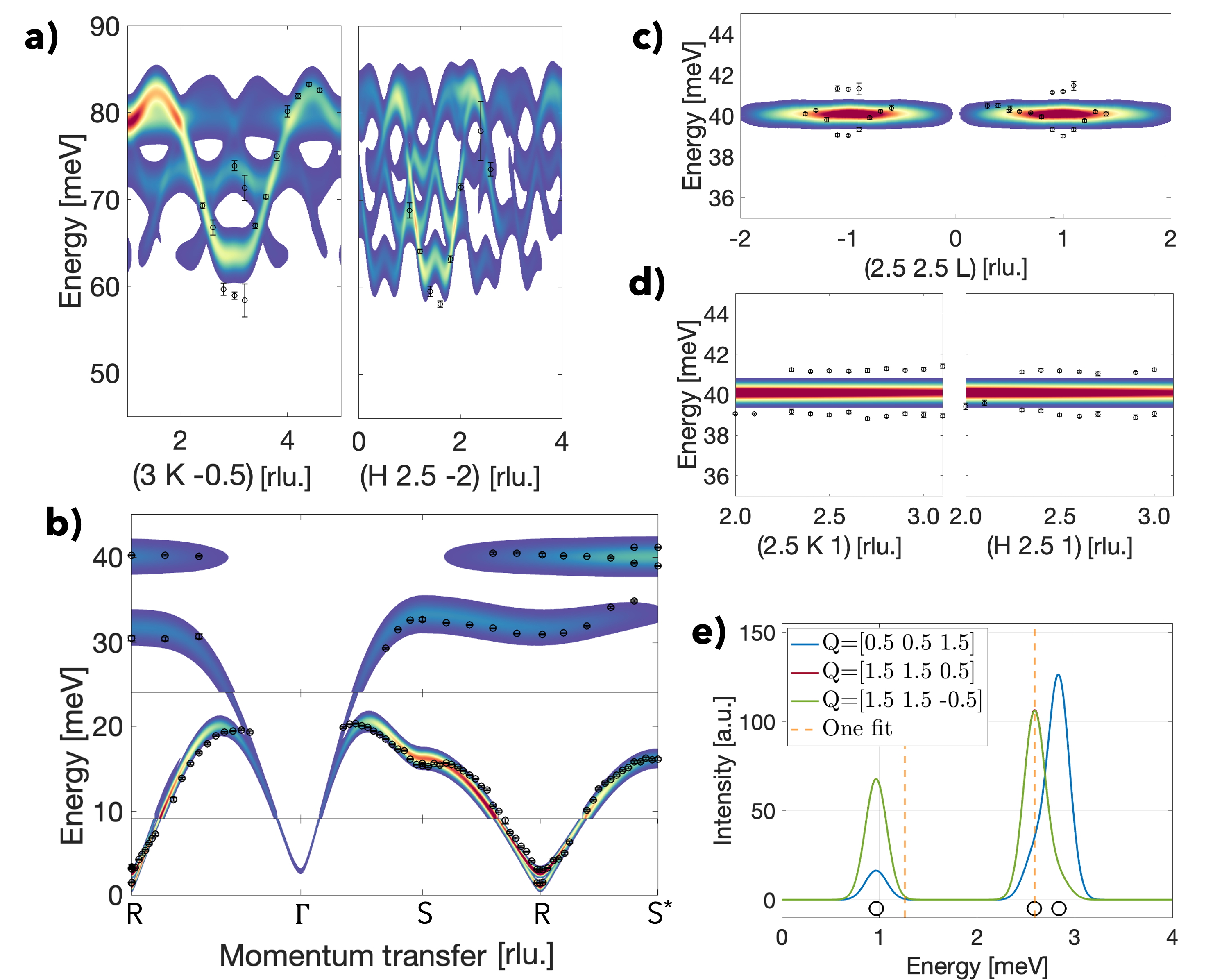}
    \caption{Fitting all parameters; five exchange interactions and the two SIA, reported in Table \ref{tab:interactions_full}.}
    \label{fig:spinW_all_param}
\end{figure*}

To obtain the final model in Table \ref{tab:interactions} with five exchange interactions and SIA different for the two iron sites, we performed the following fitting procedure: We iteratively fit the model parameters to the dataset; with the new parameters we optimize the magnetic ground state and fit again, until convergence is reached. First, all 7 parameters were fitted (see Fig.~\ref{fig:spinW_all_param}). The model fits the data points very well, except the magnitude of the gaps; the low energy gap is underestimated, while the splitting of the higher energy gap is much larger than the data. Interestingly, the relative intensity between the gaps matches that observed in the data better. The fitting algorithm in SpinW weighs all data points equally. However, we believe that our data points for the gap values are much more consistently determined than the other data points, because we take the resolution tail into account (see section \ref{sect:data_double_gap} and appendix \ref{app:double_gap_fit}). Therefore, the gap size is very well determined from the data, and we use the fitted parameters to optimize the magnitude of the anisotropies to get the best possible values of the gap sizes. These are found to be; $A_1=0.034$ and $A_2=0.046$. The anisotropies are fixed, and the exchange interactions are fitted until convergence, giving the model in Table \ref{tab:interactions} and plotted in Fig.~\ref{fig:spinW} and \ref{fig:spinW_40meV}.\\

Table \ref{tab:interactions_full} has all the information of the fitted parameters and the quality of the fits. The reduced $\chi^2$ becomes surprisingly large for the fixed anisotropy, but this is due to the fewer degrees of freedom. When fixing the anisotropy, the fitting routine punishes some data points with very small statistical error bars that systematically adds to a larger reduced $\chi^2$ value. Judging the fit by eye, the fixed anisotropy fit is better, especially for fitting the magnitude of the gaps.

\begin{table*}[h!]
    \centering
    \setlength{\tabcolsep}{12pt}
    \renewcommand{\arraystretch}{1.2}
    \begin{tabular}{lccc}
        \toprule[0.1pt]
        Parameter & Fixed anisotropy & All parameters & With DM \\
        \midrule[0.1pt]  
        \makecell{$J_1$ [meV]}  & -0.2099  & -0.0876   & -0.2654  \\
        \makecell{$J_2$ [meV]} & 1.3987   & 1.4618   & 1.4219   \\
        \makecell{$J_3$ [meV]} & 6.3812   & 6.4373   & 6.2569    \\
        \makecell{$J_4$ [meV]} & 27.8724   & 27.2095  & 27.7470   \\
        \makecell{$J_5$ [meV]} & 3.1513   & 3.2573 & 3.2095    \\
        \makecell{$A_1$ [meV]} & 0.0344   & 0.0444 & 0.0442  \\
        \makecell{$A_2$ [meV]} & -0.0458   & -0.0436 & -0.0497  \\
        \makecell{$D_{J_3}$ [meV]} &          &            & -0.004   \\
        \makecell{$D_{J_5}$ [meV]}&          &            &  0.013  \\
        \makecell{Reduced $\chi^2$} &  419 & 163 & 155\\
        \makecell{Gap sizes [meV] \\  \\  \\} & \makecell{1.2963 \\ 2.6370\\ 2.6370\\ 2.7189} &  \makecell{0.9655 \\ 2.5882 \\ 2.8366 \\ 2.8366} & \makecell{1.2044 \\ 2.7836 \\ 2.8839 \\ 2.8839} \\
        Ground state energy [meV/spin]  & -90.613 &-90.084 & -90.011\\
        \bottomrule[0.1pt]
    \end{tabular}
    \caption{Spin wave exchange parameters fitted to experimental data. Positive values indicate antiferromagnetic exchange. $A_1$ denotes easy-plane for Fe$_1$ sites and $A_2$ is easy-axis anisotropy for Fe$_2$ sites (described in the text). $D$ is the magnitude of the DMI vector of $J_3$ and $J_5$, respectively. }
    \label{tab:interactions_full}
\end{table*}

\begin{figure*}[h]
    \centering
    \includegraphics[width=0.9\linewidth]{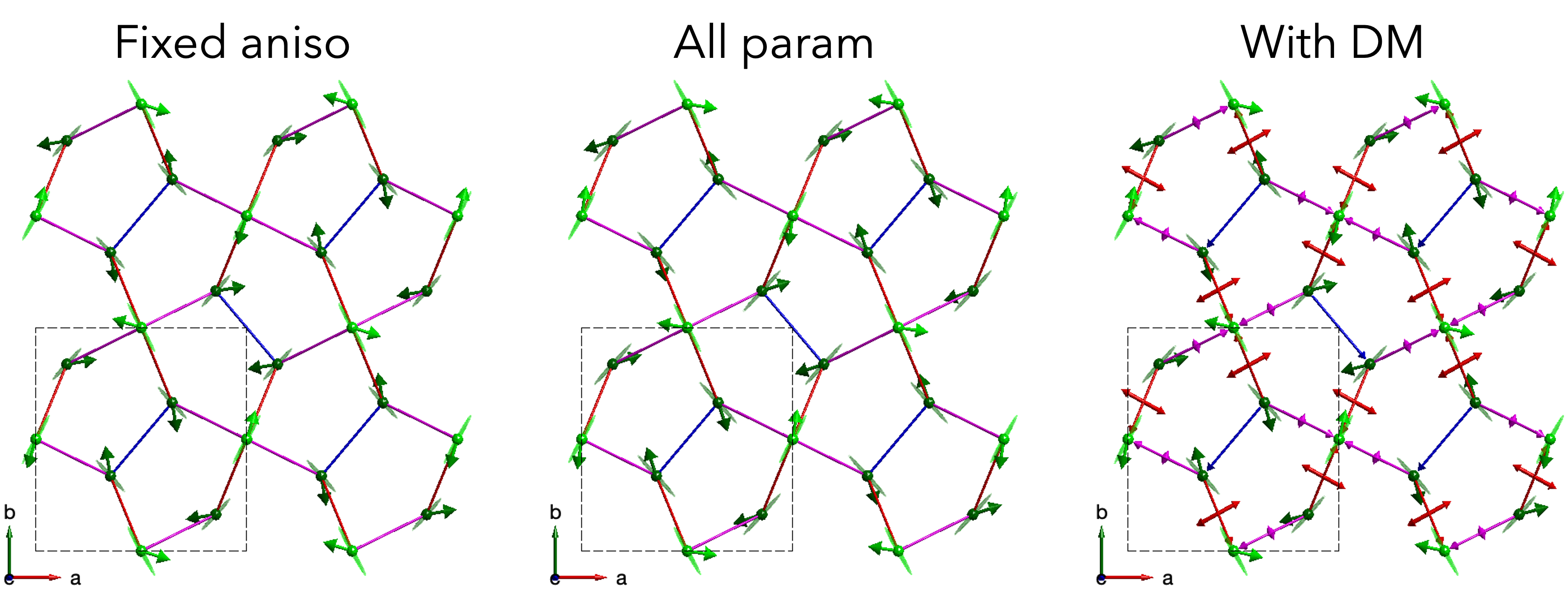}
    \caption{Optimized magnetic ground states for the three scenarios of parameters in Table \ref{tab:interactions_full}.}
    \label{fig:ground_states}
\end{figure*}

\begin{figure*}[h]
    \centering
    \includegraphics[width=0.9\linewidth]{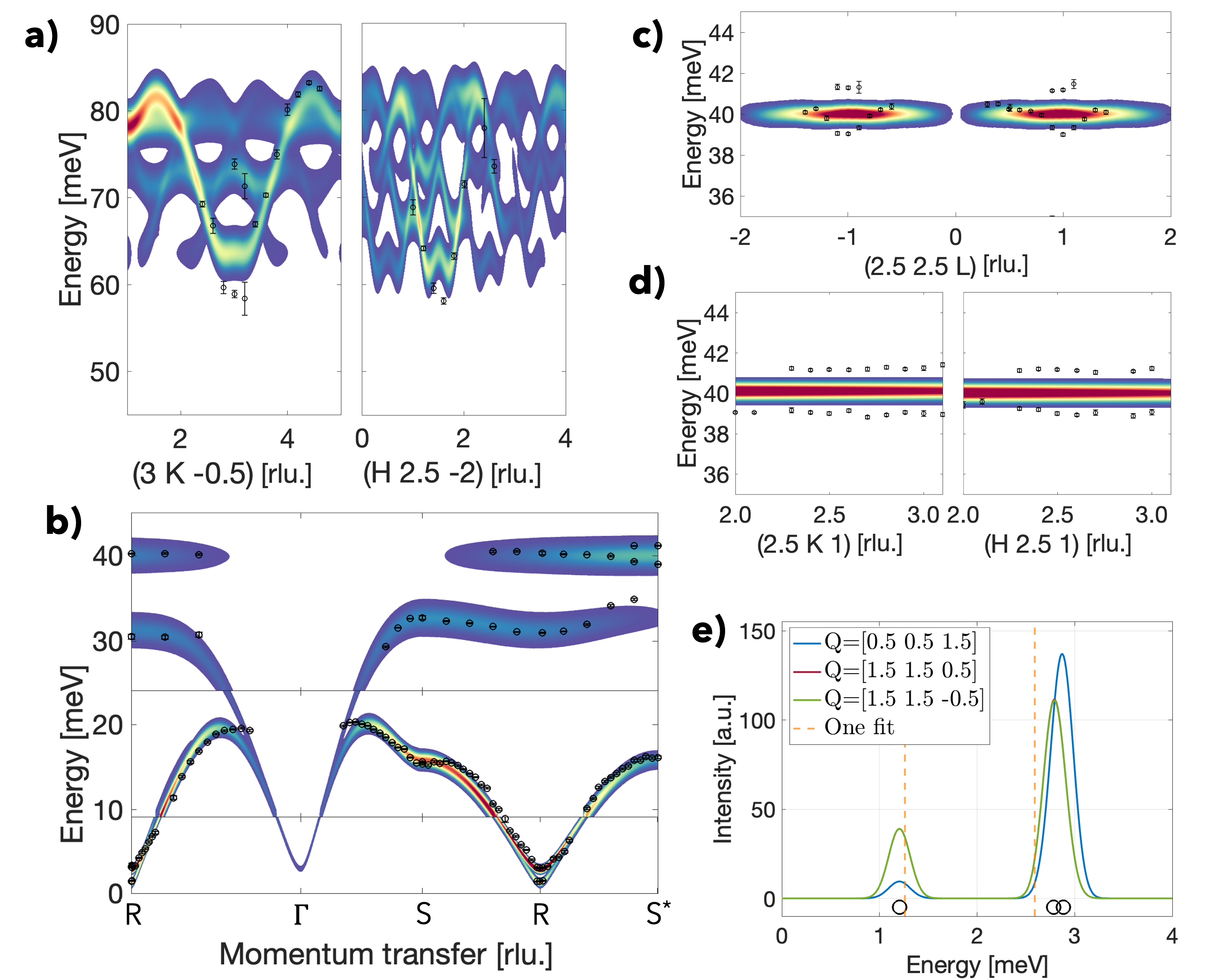}
    \caption{Fitting all parameters with DM; five exchange interactions, the two SIA and DMI on $J_3$ and $J_5$, reported in Table \ref{tab:interactions_full}.}
    \label{fig:spinW_all_param_DMI}
\end{figure*}

Additionally, we have also fitted the data with DMI on $J_3$ and $J_5$, see Fig.~\ref{fig:spinW_all_param_DMI} and the parameters in Table \ref{tab:interactions_full}. Here the higher energy gap is overestimated, but overall it is a good fit. The fitted $D$ magnitudes are close to zero, indicating that the dispersion is not very dependent on DMI on $J_3$ and $J_5$. Thus, we can not justify adding the additional 2 parameters to the fit, so the best fit and thus the one reported, is decided to be the one with the fixed anisotropy. \\
The magnetic structures for all 3 cases of the fitted parameters, shown in Fig.~\ref{fig:ground_states}, have very similar ground states and are in agreement with the magnetic structure found with neutron diffraction by Ressouche et al. Ref. \onlinecite{Ressouche2009}.

\newpage
\section{Polarisation }
\subsection{Polarisation directions in the model}
\label{app:pol}
\begin{figure}[H]
    \centering
    \includegraphics[width=1\linewidth]{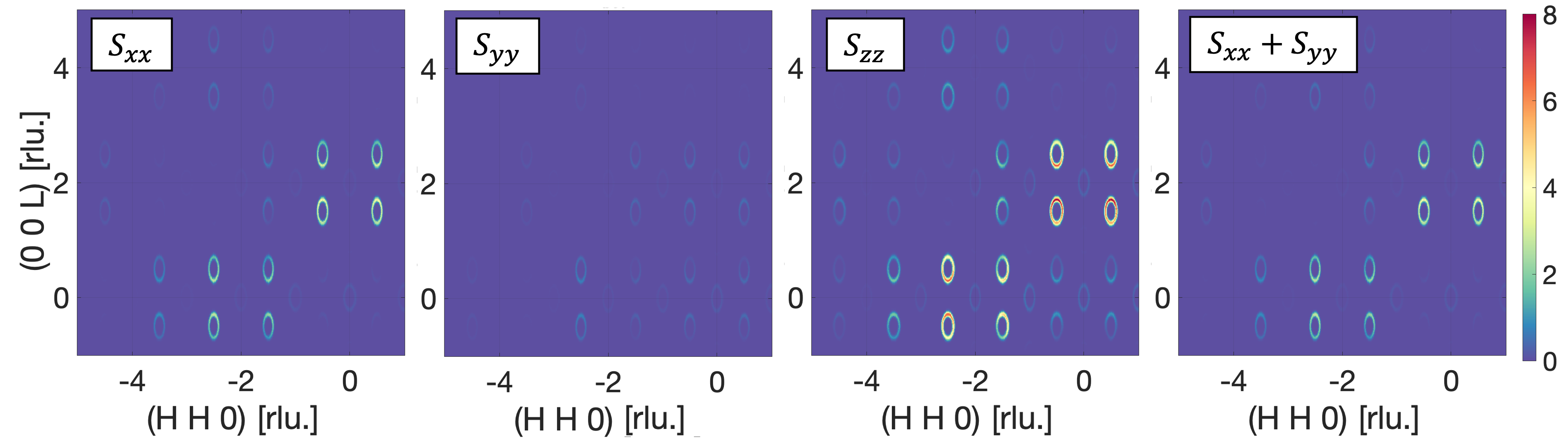}
    \caption{Calculated magnetic polarised maps at 10~meV energy transfer in the scattering plane [HH0]-[00L] (as measured with INS) separated into:  $S_{xx}$ is magnetic amplitudes along the $a$-axis, $S_{yy}$ is along $b$ and $S_{zz}$ is along $c$. The fluctuations in the $ab$-plane are given: $S_{xx}+S_{yy}$. All have the same colorscale.}
    \label{fig:pol_S}
\end{figure}

Since we work in the [HH0]-[00L] scattering plane, it has proven difficult to completely separate the magnetic amplitudes in the $ab$-plane and along the $c$-axis, especially in the $M_{yy}$ channel the contribution differs for each \textbf{Q}-position (described in the main text). Since the model reproduces the behaviour of respectively the $M_{yy}$ and $M_{zz}$, we have trust in our model. This can then be used to separate fluctuations respectively along the $a$, $b$ and $c$-axis, plotted in Fig.~\ref{fig:pol_S}. Here we are at 10~meV energy transfer and in the scattering plane as measured with INS, but for the polarised amplitudes, we have separated it such that $S_{xx}$ is amplitudes along the $a$-axis, $S_{yy}$ is along $b$ and $S_{zz}$ is along $c$. Comparing them, it is very clear that at 10~meV, the main fluctuations are along the $c$-axis. This is consistent with our expectations of the fluctuations being transverse of the ordered spin directions. Surprisingly, we see a large difference between fluctuations along $a$ and $b$, with $a$ being very dominant at most \textbf{Q}-positions. Our best explanation of this, is due to the direction of the Fe$_2$ symmetry axis, which points mostly along $b$, and thus, giving rise to precessions along $a$ and $c$. To understand this, one would need to do an analysis of the eigenmodes to describe the fluctuations in this complex structure.\\ 
Summing, the calculated intensities coming from amplitudes in the $ab$-plane, $S_{xx}+S_{yy}$, at 10~meV energy transfer (in the low-energy range, Fig. \ref{fig:pol_S} right), we find that they are half the size of the amplitudes along $c$; $S_{xx}+S_{yy} \approx \frac{1}{2}S_{zz}$. This is to be expected for a spin wave calculation. 

\subsection{Polarisation of the continuum}
\label{app:pol_con}
\begin{figure}[H]
    \centering
    \includegraphics[width=1\linewidth]{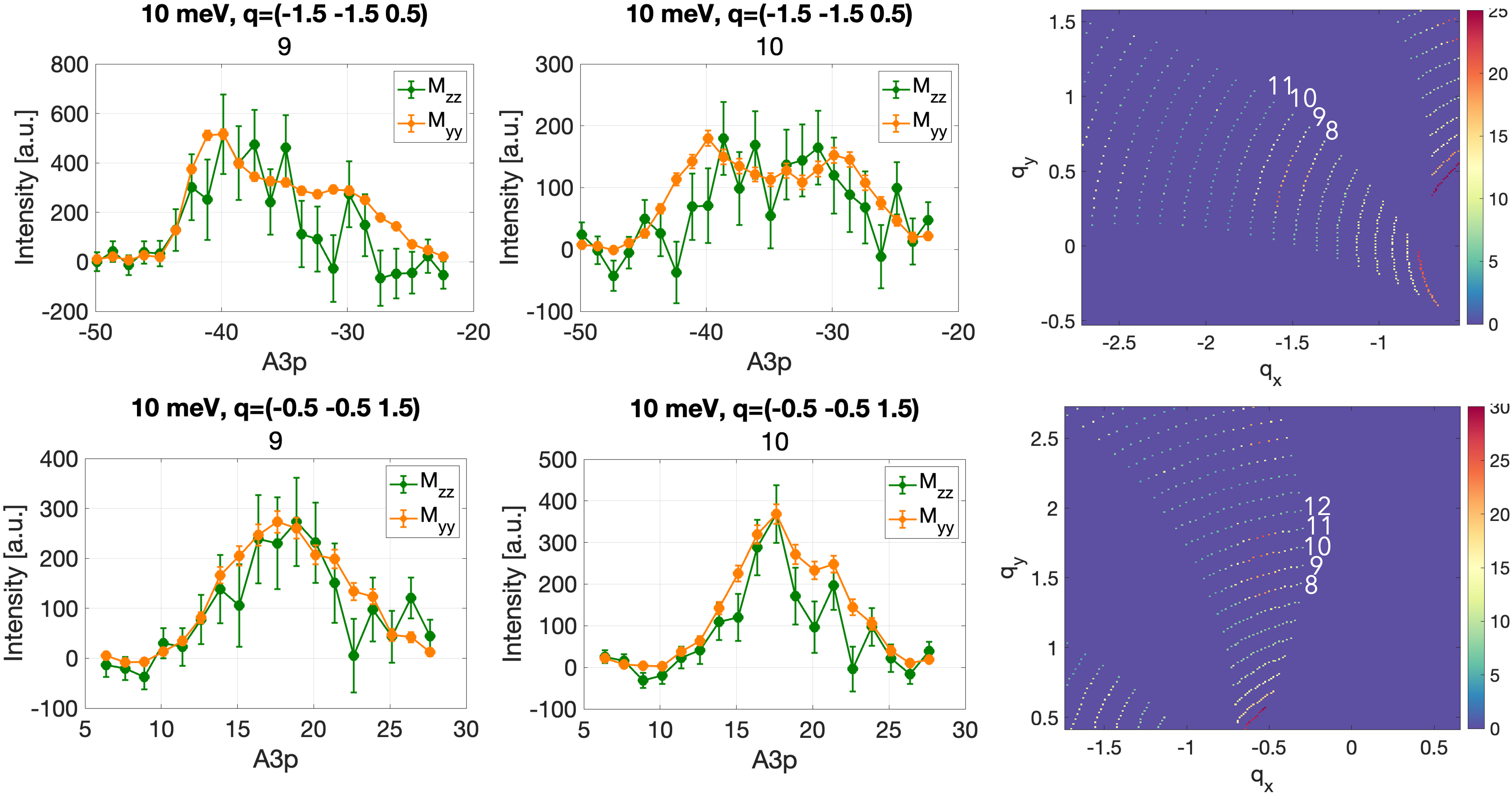}
    \caption{Detector scans over the Bragg peaks; (-1.5 -1.5 0.5) (top row) and (-0.5 -0.5 1.5) (bottom row) in the polarised IN20 data. The detectors 9 (1st column) and 10 (2nd column) both shows the magnetic amplitudes of $M_{yy}$ and $M_{zz}$. The 3rd column indicates detector scans in a colormap in scattering plane $q_x=$[HH0] and $q_y=$[00L] at 10~meV energy transfer.}
    \label{fig:pol_continuum}
\end{figure}

If there is a continuum of scattering present above the spin wave dispersion, then with polarised INS we should be able to differentiate the spin wave dispersion and the 2-magnon scattering. They will behave opposite in the spin wave channel, such that if the spin wave fluctuates out of plane (along $c$), the 2-magnon would fluctuate transverse to this, so in the $ab$-plane. However, the scattering from a 2-magnon is only about 10$\%$ of the spin wave signal and then spread out across a larger area in \textbf{Q}. This means that the signal will be very small in comparison. Looking at the polarised IN20 data at 10~meV (Fig. \ref{fig:pol_maps}b and d), the spin wave scattering is mostly present in the $M_{yy}$ channel, which, as discussed above, is mainly fluctuations along $c$ (but also partly in the $ab$-plane). Assuming that we have 2-magnon scattering, fluctuations in the $ab$-plane should show an effect on the shape of the signal seen in the $M_{zz}$-channel. Plotting detector scans over the Bragg peak positions at 10 meV; (-1.5 -1.5 0.5) and (-0.5 -0.5 1.5), we should observe a difference, see Fig.~\ref{fig:pol_continuum}. The intensities are normalised to the maximum peak intensity. From this, the signal in $M_{zz}$ might tend to have less intensity in the second peak. However, with the low statistics, it is inconclusive whether the shape of the peak is different in the two channels, and we cannot conclude whether a 2-magnon signal is present. We would need better statistics and better energy resolution. 

\FloatBarrier

\end{document}